\documentclass[aps,prl,amssymb,twocolumn]{revtex4}%
\usepackage{graphicx}

\begin{document}

\title{Superconductor-insulator quantum phase transition}

\author{V.F. Gantmakher, V,T. Dolgopolov}

\affiliation{\mbox{Institute of Solid State Physics,  Chernogolovka 142432,
Russia}}

\begin{abstract}
The current understanding of the superconductor--insulator transition is
discussed level by level in a cyclic spiral-like manner. At the first level,
physical phenomena and processes are discussed which, while of no formal
relevance to the topic of transitions, are important for their implementation
and observation; these include superconductivity in low electron density
materials, transport and magnetoresistance in superconducting island films and
in highly resistive granular materials with superconducting grains, and the
Berezinskii--Kosterlitz--Thouless transition. The second level discusses and
summarizes results from various microscopic approaches to the problem, whether
based on the Bardeen--Cooper--Schrieffer theory (the disorder-induced reduction
in the superconducting transition temperature; the key role of Coulomb blockade
in high-resistance granular superconductors; superconducting fluctuations in a
strong magnetic field) or on the theory of the Bose--Einstein condensation. A
special discussion is given to phenomenological scaling theories. Experimental
investigations, primarily transport measurements, make the contents of the
third level and are for convenience classified by the type of material used
(ultrathin films, variable composition materials, high-temperature
superconductors, superconductor--poor metal transitions). As a separate topic,
data on nonlinear phenomena near the superconductor--insulator transition are
presented. At the final, summarizing, level the basic aspects of the problem
are enumerated again to identify where further research is needed and how this
research can be carried out. Some relatively new results, potentially of key
importance in resolving the remaining problems, are also discussed.
\end{abstract}

\maketitle

\section{1. Introduction}

As temperature decreases, many metals pass from the normal to the
superconducting state which is phenomenologically characterized by the
possibility of a dissipationless electric current and by the Meissner effect.
As a result of a change in some external parameter (for example, magnetic field
strength), the superconductivity can be destroyed. In the overwhelming majority
of cases, this leads to the return of the superconducting material to the
metallic state. However, it has been revealed in the last three decades that
there are electron systems in which the breakdown of superconductivity leads to
the transition to an insulator rather than to a normal metal. At first, such a
transition seemed surprising, and numerous efforts were undertaken in order to
experimentally check its reality and to theoretically explain its mechanism. It
was revealed that the insulator can prove to be quite extraordinary; moreover,
upon breakdown of superconductivity with the formation of a normal metal, the
metal can also be unusual. This review is devoted to a discussion of the state
of the art in experiment and theory in this field.

\subsection{1.1 Superconducting state, electron pairing}

By the term `superconducting state', we understand the state of metal which, at
a sufficiently low temperature, has an electrical resistance exactly equal to
zero at the zero frequency, thus indicating the existence of a macroscopic
coherence of electron wave functions. This state is brought about as a result
of superconducting interactions between charge carriers. Such an interaction is
something more general than superconductivity itself, since it can either lead
to or not lead to superconductivity.

According to the Bardeen--Cooper--Schrieffer (BCS) theory, the transition to
the superconducting state is accompanied by and is caused by a rearrangement of
the electronic spectrum with the appearance of a gap with a width of $\Delta$
at the Fermi level. The superconducting state is characterized by a complex
order parameter

\begin{equation}\label{OrderParameter}
  \Phi{(\bf r)}=\Delta\exp(i\varphi(\bf r)),
\end{equation}
in which the value of the gap $\Delta$  in the spectrum is used as the modulus.
If the phase $\varphi ({\bf r})$ of the order parameter has a gradient,
$\varphi(\bf r)\neq\mathrm{const}$, then a particle flow exists in the system.
Since the particles are charged, the occurrence of a gradient indicates the
presence of a current in the ground state.

The rearrangement of the spectrum can be represented as a result of a binding
of electrons from the vicinity of the Fermi level (with momenta ${\bf p}$ and
$-{\bf p}$ and oppositely directed spins) into Cooper pairs with a binding
energy $2\Delta$. The binding occurs as a result of the effective mutual
attraction of electrons located in the crystal lattice, which competes with the
Coulomb repulsion.

A Cooper pair is a concept that is rather conditional, not only since the pair
consists of two electrons moving in opposite directions with a velocity $v_{\rm
F}$, but also since the size of a pair in the conventional superconductor,
$\zeta\sim\hbar v_F/\Delta\sim10^{-4}$~cm, is substantially greater than the
average distance between pairs, $s \sim (g_0\Delta )^{-1/3}\sim 10^{-6}$~cm
($g_0$ is the density of states in a normal metal at the Fermi level):
\begin{equation}\label{1}
  \zeta\gg s.
\end{equation}
In fact, the totality of Cooper pairs represents a collective state of all
electrons. It has long been known that superconductivity also arises in systems
with an electron concentration that is substantially less than that
characteristic of conventional metals, for example, in SrTiO$_3$ single
crystals with an electron concentration of about $n\sim 10^{19}\,$cm$^{-3}$
[1]. Furthermore, the parameter $\zeta$ in type-II superconductors can be less
than 100\,\AA . Therefore, inequality (\ref{1}), which is necessary for the
applicability of the BCS model, can prove to be violated. The materials in
which $\zeta\lesssim s$ are referred to as `exotic' superconductors; these also
include high-temperature superconductors in which the superconductivity is
caused by charge carriers moving in CuO$_2$ crystallographic planes. As in any
two-dimensional (2D) system, the density of states $g_0$ in the CuO$_2$~planes
in the normal state is independent of the charge carrier concentration and,
according to measurements, is $g_0=2.5\times 10^{-4}$~K$^{-1}$ per one CuO$_2$
crystal plane [to approximately one and the same magnitude in all families of
the cuprate superconductors (see, e.g., Ref.~[2])]. Assuming, for the sake of
estimation, that $\Delta$ is on the order of the superconducting transition
temperature $T_{\rm c}$, we obtain the average distance between the pairs in
CuO$_2$ planes: $s\approx(g_0T_c)^{-1/2}\approx25$\,\AA\ at $T_c\approx100\,$K.
This value is comparable with the typical coherence length $\zeta \approx
20$\,\AA\  in high-temperature superconductors.

The existence of exotic superconductors, for which inequality (2) is violated,
induced to turn to another model of superconductivity --- the Bose--Einstein
condensation (BEC) of the gas of electron pairs considered as bosons with a
charge 2e [3] --- and to investigate the crossover from the BCS to the BEC
model (see, e.g., the review [4]).One of the essential differences between
these models consists in the assumption of the state of the electron gas at
temperatures exceeding the transition temperature. The BEC model implies the
presence of bosons on both sides of the transition. An argument in favor of the
existence of superconductors with the transition occurring in the BEC scenario
is the presence of a pseudogap in some exotic superconductors. It is assumed
that the pseudogap is the binding energy of electron pairs above the transition
temperature (for more detail, see the end of Section 4.3 devoted to
high-temperature superconductors).

In the BCS model, the Cooper pairs for $T>T_{\rm c}$ appear only as a result of
superconducting fluctuations; the equilibrium concentration of pairs exists
only for $T<T_{\rm c}$. The crossover from the BCS to the BEC model consists in
decreasing gradually the relative size of Cooper pairs and appearing the pairs
on both sides of the transition, which are correlated in phase in the
superconducting state and uncorrelated in the normal state. The conception that
in superconducting materials with a comparatively low electron density the
equilibrium electron pairs can exist for $T>T_{\rm c}$ began to be discussed
immediately after the discovery of these materials [5].

For the problem of the superconductor--insulator transition, the question of
the interrelation between the BCS and BEC models is of large importance, since
near the boundary of the region of existence of the superconducting state it is
natural to expect a decrease in the density of states $g_0$ and an increase in
$s$, so that inequality (2) must strongly weaken or be completely violated. In
any case, the problem of a phase transition that is accompanied by localization
makes sense within the framework of both approaches.

By having agreed that the superconductivity of exotic superconductors can be
described using the BEC model, we adopt that for a temperature $T>T_{\rm c}$
there can exist both fluctuation-driven and equilibrium electron pairs. Then, a
natural question arises: since the electron pairs can exist not only in the
superconducting but also in the dissipative state, can it happen that pair
correlations between the localized electrons can be retained as well on the
insulator side in superconductor--insulator transition? Below, we shall
repeatedly return to this question.

\subsection{1.2. Superconductor--insulator transition as a quantum phase
transition}

It is well known that in the ground state the electron wave functions at the
Fermi level can be localized or delocalized. In the first case, the substance
is called an insulator, and in the second case a metal. As was already said
above, it has long been considered that superconductivity can arise only on the
basis of a metal, i.e., the coherence of the delocalized wave functions can
arise only as an alternative to their incoherence. We now know that with the
breakdown of superconductivity all electron wave functions that became
incoherent can immediately prove to be localized. In this case, it is assumed
that the temperature is equal to zero, so that on both sides of the transition
the electrons are in the ground state.

The phase transition between the ground states is called the quantum
transition. This means that it is accompanied by quantum rather than thermal
fluctuations. The transition can be initiated by a change in a certain control
parameter $x$, for instance, the electron concentration, disorder, or magnetic
field strength. Superconductivity can also be destroyed by a change in the
control parameter $x$ at a finite temperature, when thermodynamic thermal
fluctuations are dominant. It can be said that in the plane $(x,T\,)$ there is
a line of thermodynamic phase transitions $x(T\,)$, which is terminated on the
abscissa $(T=0)$ at the point $x=x_0$ of the quantum transition.

\begin{figure}
\includegraphics{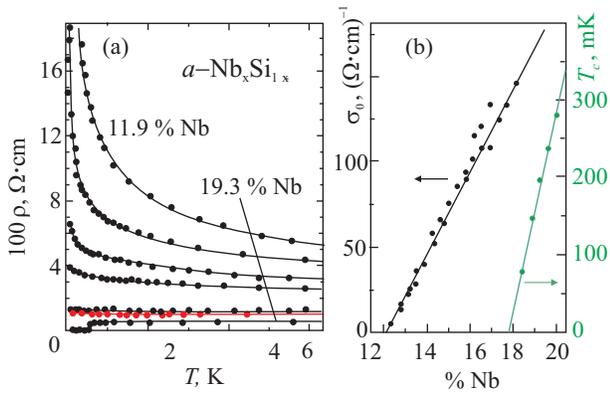}
\caption{((a) Temperature dependence of resistivity $\rho (T\,)$ of films of
the amorphous ${\rm Nb}_x{\rm Si}_{1-x}$ alloy at various concentrations of Nb
[6]. (b)~Dependence of the superconducting transition temperature $T_{\rm c}$
and of the extrapolated value of the low-temperature conductivity $\sigma _0=
\lim \sigma (T\rightarrow 0)$ on the Nb concentration in films of the amorphous
${\rm Nb}_x{\rm Si}_{1-x}$ alloy [6]. Here and below, the curves nearest to the boundary which separates  $R(T)$ curves for superconducting states are shown in red}
 \label{SiNb}
\end{figure}

Let us shift the state of the superconducting metal toward the region of
insulating states by changing a certain parameter $x$. Under the effect of this
shift, it can happen that, first, superconductivity will disappear, and then
the normal metal--insulator transition will occur. It is precisely according to
this scenario that the events develop with a decreasing concentration of Nb in
the amorphous ${\rm Nb}_x{\rm Si}_{1-x}$ alloy [6]: at an Nb concentration of
approximately 18\%, the temperature of the superconducting transition drops to
zero and the alloy becomes a normal metal, and the metal--insulator transition
occurs only at an Nb concentration of 12\% (Fig.~1). The
superconductor--insulator transition is split into two sequential transitions.
This example is instructive in the sense that though in the set of $\rho (T\,)$
curves (Fig.~1a) the boundary between the superconducting and
nonsuperconducting states is clearly visible, to prove the existence of an
intermediate metallic region and to reveal the metal--insulator transition, it
is necessary to perform extrapolation of the $\sigma (T\,)=1/\rho (T\,)$
dependence as $T\rightarrow 0$ in a certain interval of concentrations. The
quantity $\sigma _0$ presented in Fig.~1b as a function of the Nb concentration
is the result of this extrapolation.

Of greater interest is the case of unsplit transition, where the superconductor
directly transforms into the insulator, possibly passing through a bordering
isolated normal state. This survey is mainly devoted to precisely such
transitions, which are, as we will see, sufficiently diverse.
\begin{figure}[b]
\includegraphics{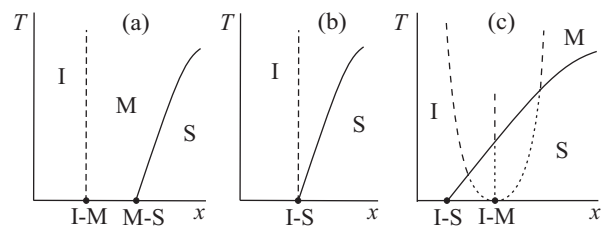}
\caption{Variants of the phase diagram with transitions between three different
states of the electron system: insulator (I), normal metal (M), and
superconductor (S). The quantum transitions are shown by dots lying on the
abscissa; the thermodynamic transitions, by solid curves; the crossovers and
the boundaries of the critical region, by dashed lines, and the virtual
boundaries in figure (c), by dotted lines. }
 \label{diagr-ab}
\end{figure}

Let us schematically depict the phase diagrams of these phase transitions in
the plane $(x,T\,)$ (Fig.~2), assuming for the sake of certainty the
three-dimensional nature of the electronic system. As is known, the
metal--insulator transition is depicted on this plane in the form of an
isolated point on the x-axis, because the very concept of an `insulator' is
strictly defined only at $T=0$ (see, e.g., the review [7]). Therefore, the
vertical dashed straight lines in Fig.~2 do not mark real phase boundaries.

In the diagram presented in Fig.~2a, which corresponds to a split transition,
the dashed straight line issuing from the point $\rm I\!-\!M$ shows that in the
region I the extrapolation of the conductivity to $T=0$ will give zero, and in
the region M it will give a finite value. According to Fig.~1, the alloy ${\rm
Nb}_x{\rm Si}_{1-x}$ has precisely such a phase diagram. In Section 4.4, we
shall return to ${\rm Nb}_x{\rm Si}_{1-x}$ type substances and shall see that
the diagram presented in Fig.~2a has, in turn, several variants.

In the diagram shown in Fig.~2b, for any state to the right of the dashed line
a temperature decrease will lead to emergence of superconductivity; therefore,
to determine whether the state is metallic or insulating, it is necessary to
measure the temperature dependence of resistivity in the region that lies above
the superconducting transition, with the extrapolation of this dependence to
$T=0$. Such a disposition appears to be realized, for example, in ultrathin
films of amorphous Bi (see Fig.~18 in Section 4.1).

Finally, one more variant of the phase diagram, which was for the first time
proposed in Ref.~[8], is given in Fig.~2c. In this diagram, the
metal--insulator transition is completely absent, since it should have to be
located in the superconducting region. From this transition, only part of the
critical region is retained, which lies higher than the region of
superconductivity. This phase diagram has been observed for TiN (see Fig.~31 in
Section 4.2).

\subsection{1.3.  Role of disorder. Granular superconductors}

From the before-studied theory of the normal metal--insulator quantum
transition, it is known that this transition can be initiated by two
fundamentally different reasons: growing disorder in the system of
noninteracting electrons (Anderson transition) or decreasing electron
concentration in the presence of a Coulomb electron--electron interaction in an
ideal system without random potential (Mott transition). In this review (in any
case, in its experimental part), we shall assume that the
superconductor--insulator transition occurs in a strongly disordered Anderson
type electron system. Even when the control parameter $x$ is the electron
concentration, it is assumed that the latter changes against the background of
a sufficiently strong random potential.

In order to answer the question concerning in which case and which of the
diagrams shown in Fig.~2 can be realized, it is necessary to study the
influence of disorder on the superconductivity. The first result in this area
was obtained by P.W.~Anderson as early as 1959. In Ref.~[9] he showed that if
the electron--electron Coulomb interaction is ignored, then the introduction of
nonmagnetic impurities does not lead to a substantial change in the
superconducting transition temperature. The allowance for Coulomb interactions
changes the situation. As was shown by Finkel'shtein [10, 11] for
two-dimensional systems, the Coulomb interaction does suppress
superconductivity in so-called dirty systems, the mechanism of suppression
being caused by the combination of electron--electron interaction with impurity
scattering (see Section 2.1).
\begin{figure}[b]
\includegraphics{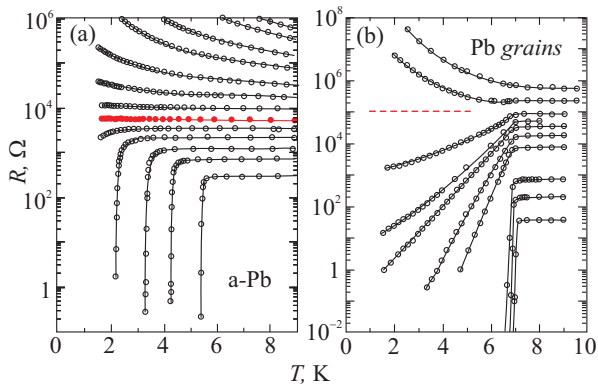}
\caption{Resistance variations with temperature in Pb films upon increasing
their thickness (from top to bottom) [12]. (a)~Superconductor--insulator
transition in finely dispersed quasihomogeneous films deposited on an SiO
surface over an intermediate thin layer of amorphous Ge. In the superconducting
region, the $R(T\,)$ curves demonstrate a correlation between the normal
resistance and the superconducting transition temperature. (b)
Superconductor--insulator transition in granular films deposited directly onto
the SiO surface. In such a method of deposition, the lead atoms coalesce into
granules. The temperature of the superconducting transition in the film becomes
constant at a film thickness exceeding the critical one. } \label{Frydman}
\end{figure}

From the variety of random potentials that describe disorder, let us single out
two limiting cases: systems with a potential inhomogeneity on an atomic scale,
which are subsequently considered as uniform, and systems with inhomogeneities
that substantially exceed atomic dimensions. We shall call the latter systems
granular, assuming for the sake of certainty that they consist of granules of a
superconductive material with a characteristic dimension $b$, which are
separated by interlayers of a normal metal or an insulator. A control parameter
in such a granular material can be, for example, the resistance of the
interlayers.

There exist both theoretical and experimental criteria which make it possible
to relate a real electronic system to one of these limiting cases. The
theoretical criterion is determined by the possibility of the generation of a
superconducting state in one granule taken separately, irrespective of its
environment. For this event to occur, it is necessary that the average spacing
between the energy levels of electrons inside the granule be less than the
superconducting gap $\Delta$:
 \begin{equation}\label{g0}
  \delta\varepsilon=(g_0b^3)^{-1}<\Delta
\end{equation}
where $g_0$ is the density of states at the Fermi level in the bulk of the
massive metal, and $b^3$ is the average volume of one granule. The relationship
$\delta\varepsilon\approx\Delta$ specifies the minimum size of an isolated
granule:
 \begin{equation}\label{g00}
b_{sc}=(g_0\Delta)^{-1/3},
\end{equation}
for which the concept of the superconducting state makes sense. When the
inequality $b<b_{\rm SC}$ is fulfilled, no granules that could be
superconducting by themselves exist. Such a material, from the viewpoint of the
superconductive transition, is uniformly disordered; in it, the transition
temperature $T_{\rm c}$ is determined by the average characteristics of the
material and can smoothly change together with these characteristics.

The experimental criterion which makes it possible to distinguish between the
superconductor--insulator transitions in granular and quasihomogeneously
disordered systems is illustrated in Fig.~3. Here, the control parameter is the
thickness $b$ of a lead film deposited on the surface of an SiO substrate. The
curves shown in Fig.~3a were obtained for lead films deposited on an
intermediate sublayer of amorphous Ge. The temperature of the superconductive
transition decreases with decreasing $b$ in this series of films; at a zero
temperature, an increase in the thickness $b$ leads to a direct transition from
the insulating to the superconductive state. No macrostructure was revealed in
these films by the structural analysis performed simultaneously. The
intermediate thin layer of amorphous Ge appears to prevent the coalescence of
atoms into granules in the deposited material (see also Section 4.1). In any
case, if the granules exist, their size should be lower than the critical size
(4).

The curves shown in Fig.~3b were obtained for lead films deposited directly
onto a mirror surface of SiO cooled to liquid-helium temperature. With this
method of deposition, the lead atoms are collected into droplet-like granules,
which reach a diameter of 200\,\AA\  and a height of 50--80\,\AA\ before they
start coalescing. A film in which no coalescence has yet occurred is called an
island film: it represents a system of metallic islands between which the
conductivity is achieved via tunneling. In all the films, the superconducting
transition begins, if it occurs at all, at one and the same temperature $T_{\rm
c}\approx 7$\,K. This means that the granule sizes are sufficiently large, so
that in them $\delta \varepsilon <\Delta$, the superconducting transition in
the granules occurs at the same temperature as that in the massive metal, and
the behavior of the entire material on the whole depends on the interaction
between the granules.

As can be seen from Fig.~3, the transition in the granular system possesses one
more specific feature. Near the transition, on the superconductor side, the
temperature dependence of the resistance $R(T\,)$ for $T<T_{\rm c}$ follows a
very strange formula [13] $$R=R_0e^{T/T_0},$$ which can be called the
`inverse-Arrhenius law'. In quasi-homogeneous systems, to which this survey is
devoted, such relation have not been observed.

In an isolated particle with the size of $b<b_{\rm SC}$, no superconducting
state exists, in the sense that there is no the coherent state of all electrons
with a common wave function. However, a superconducting interaction through
phonons is retained, which causes effective attraction between the electrons.
The superconducting interaction gives rise to the parity effect: the addition
of an odd electron to the electron system leads to a greater increase in the
total electron energy than the addition of a subsequent even electron. The
difference is equal to $2\Delta _{\rm p}$, where $\Delta _{\rm p}$ is the
binding energy per electron:
\begin{equation}\label{Delta_p}        
 \Delta_p=E_{2l+1}-\frac{1}{2}(E_{2l}+E_{2l+2})
\end{equation}
The parity effect was examined experimentally when studying the Coulomb
blockade in superconducting grains [14, 15]. A theoretical treatment [16]
showed that, because of strong quantum fluctuations of the order parameter, the
binding energy in small grains,
\begin{equation}\label{smallGrains}     
 b\ll b_{sc},\qquad\mbox{i.e.}\qquad\delta\varepsilon\gg\Delta,
\end{equation}
not only is retained, but, in general, becomes greater:
\begin{equation}\label{Delta_p-small}       
 \Delta_p=\frac{\delta\varepsilon}{2\ln(\delta\varepsilon/\Delta)}>\Delta.
\end{equation}
 The magnitude of
$\Delta _{\rm p}$ is much less than the level spacing $\delta \varepsilon$, but
it is by no means less than the superconducting gap $\Delta$.

\subsection{1.4 Fermionic and bosonic scenarios for the transition}

There are two scenarios for a superconductor--insulator transition. The
foundation of the theory of the fermionic scenario of the
superconductor--insulator transition was laid by  Finkel'shtein [10, 11]. Its
essence lies in the fact that, due to various reasons, the efficiency of the
superconducting interaction in a dirty system at a zero temperature gradually
drops to zero, and Anderson localization occurs in the arising normal fermionic
system. However, this scenario is by no means unique. As a result of the rapid
development of theoretical and experimental studies in this field, it was
revealed that there is one more scenario, the bosonic scenario, for this
transition. The difference between the scenarios can conveniently be formulated
using the complex order parameter (1). The phase $\varphi$ of the order
parameter inside the massive superconductor is constant in the absence of
current; this reflects the existence of quantum correlations between the
electron pairs. In the presence of fluctuations, the superconducting state of a
three-dimensional system is retained until the correlator $G({\bf r})$,
\begin{equation}\label{Correl}
  G({\bf r})=\langle\Phi({\bf r})\Phi(0)\rangle\rightarrow G_0\neq0\quad
  \mbox{при}\quad|{\bf r}|\rightarrow\infty
\end{equation} tends
to a finite value with increasing $|{\bf r}|$. The angular brackets in formula
(8) indicate averaging over the quantum state of the system, and $\Phi({\bf
r})$ is the complex order parameter.

The consideration given in Refs~[10, 11] is based on the BCS theory. In the BCS
and related theories, the energy gap $\Delta$, i.e., the modulus of the order
parameter $|\Phi |$, becomes zero at the phase-transition point and the phase
automatically becomes meaningless. However, the superconducting state can be
destroyed by another way as well: the correlator (8) can be made vanishing at a
nonzero modulus of the order parameter by the action of phase fluctuations of
the order parameter. This is exactly the bosonic scenario for the transition.
This name comes from the fact that the finite modulus of the order parameter at
the transition indicates the presence of coupled electron pairs, i.e., the
concentration of bosons during transition does not become zero. The realization
of the bosonic scenario is favored by the fact that the superconductors with a
low electron density are characterized by a weaker shielding and a
comparatively small `rigidity' relative to phase changes, thus raising the role
of the phase fluctuations [17, 18].

The bosonic scenario was mainly developed for the case of uniform disordered
superconductors [8]. However, it should be noted that in granular
superconductors this scenario is realized quite naturally in the framework of
the BCS theory. Indeed, if we move from one curve to another in Fig.~3b from
bottom to top, assuming for simplicity that the difference between the states
arises as a result of a gradual increase in the resistance of the interlayers
between the unaltered granules, we shall see that even when the
superconductivity of the macroscopic sample disappears (upper curves in
Fig.~3b), the granules remain superconducting. However, the Cooper pairs in
them prove to be `localized,' each in its own granule.

The word localized is put in quotation marks, since if the size $b$ of the
granules is macroscopic, then the limitation on the displacement of Cooper
pairs will not agree with the conventional understanding of the term
`localization'. Let, however, $b\lesssim b_{sc}$. Relationship (4) determines
the applicability boundary of the concept of granular superconductors: below
this boundary they transform into so-called dirty superconductors with
characteristic atomic lengths describing disorder. The boundary of a granule
with parameters (6) can already be considered simply as a defect, and the
electrons located inside it, as being localized on a length $b<b_{\rm SC}$,
irrespective of the structure of the wave function inside this region.
According to the parity effect [14--16], pair correlations with a finite
binding energy are retained between the electrons localized on such a defect.

Thus, granular superconductors prove to be a natural model object for studying
the bosonic scenario for superconductor--insulator transitions. It is
interesting that some manifestations of this scenario were discovered
experimentally in granular two-dimensional systems at a time when the problem
of superconductor--insulator transitions had not yet appeared [19, 20].

The tunneling current between two superconducting granules, in fact, consists
of two components: the superconducting Josephson current of Cooper pairs, and a
single-particle dissipative current. The Josephson current in the junction can
for various reasons be suppressed; in particular, it is suppressed by
fluctuations in the case of too high a normal resistance $R_{\rm n}$ of the
junction [21]. Then, even through contact with the superconducting banks of the
junction, only a normal single-particle current $j_{\rm n}=V/R_{\rm n}$ can
flow, and then only if a potential difference $V$ is applied across the
junction. This gives rise to a paradoxical behavior of the granular
superconductor with decreasing temperature. The concentration of
single-particle excitations in superconducting granules diminishes
exponentially with a decrease in the temperature: $n\propto \exp {(-\Delta
/T\,)}$ and, correspondingly, the resistance of all junctions grows
exponentially: $R\propto \exp {(\Delta /T\,)}$. As a result, the resistivity
$\rho$ of the entire material increases rather than decreases with temperature
for $T<T_{\rm c}$. This exponential increase in the resistivity,
 $$\ln(\rho/\rho_0)\propto T^{-1},$$
starting at a temperature equal to the temperature of the superconducting
transition $T_{\rm c}$, was experimentally examined in island films [19, 20]
(Fig.~4a) and, later, in granular films with superconducting granules (Fig.~4b
[22]) and in a three-dimensional (3D) material [23].

\begin{figure}
\includegraphics{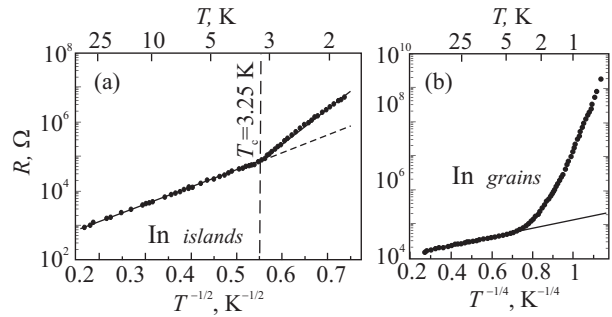}
\caption{(a) Temperature dependence of the resistance of an In island film
exhibiting a significant increase in the resistance at temperatures less than
the temperature of the superconducting transition in indium granules. (b)~The
same for a granular film consisting of In granules separated by insulating
oxide layers [22].} \label{Adki}
\end{figure}

If we destroy (by an external magnetic field) the superconducting gap in the
granules, making them normal, then the number of quasiparticles at the Fermi
level on the superconducting sides of the junction will grow and the junction
resistance will return to the normal resistance $R_{\rm n}$. In other words, a
system of metallic granules in an insulating matrix over a certain interval of
parameters can have a finite resistivity $\rho$  at $T=0$ if the granules are
normal, but becomes an insulator, with $\rho =\infty$, if the granules are
superconducting. A specific feature and, at the same time, an attribute of such
a system is negative magnetoresistance, which becomes stronger as the
temperature lowers:
\begin{equation}\label{OMR}     
  \rho(B,T)/\rho(0,T)\approx\exp(-T/\Delta),
  \qquad B>B_c,
\end{equation}
where $\Delta \approx T_{\rm c}$ (it is everywhere assumed that the temperature
$T$ is measured in energy units), $T_{\rm c}$ is the critical temperature, and
$B_{\rm c}$ is the magnetic field induction that destroys the superconductivity
of separate granules. In the experiment whose results are presented in Fig.~5,
a magnetic field of 10\,T decreases the resistance by more than two orders of
magnitude at a temperature of 0.5\,K.

\begin{figure}
\includegraphics{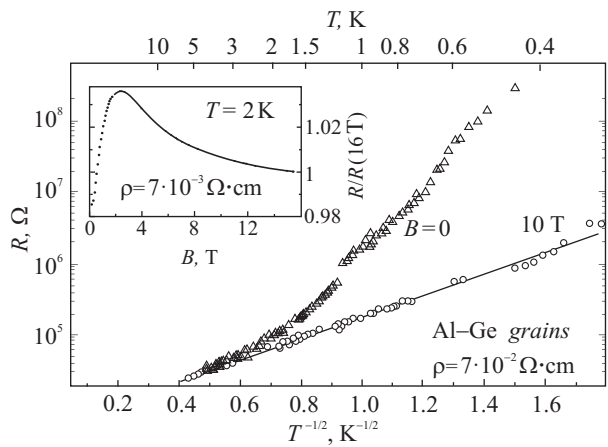}
\caption{(((a) Restoration of the `normal' conductance of a sample with
granular Al in the matrix of amorphous Ge [24]. The sample represents a film
about 2000\,\AA\  thick with granules about 120\,\AA\ in size. In a zero
magnetic field, the granules become superconducting at a temperature of about
2~K. The inset displays the $R(B)$ curve for a sample with the same geometrical
characteristics but exhibiting tenfold less normal resistance and a
superconducting transition. The curve was recorded at a temperature slightly
exceeding $T_{\rm c}$ [24].}
 \label{granOMR}
\end{figure}

An increase in resistance in a zero field and negative magnetoresistance are
possible, even at temperatures that exceed the temperature of the
superconducting transition, due to superconducting fluctuations [25, 26]. As a
result of the absence of a Josephson coupling between the granules, the virtual
Cooper pairs that arise due to fluctuations make no contribution to electron
transport. However, the fluctuation-induced decrease in the density of
single-particle states in the granules strongly increases intergranular
resistance; this resistance decreases if the fluctuations are suppressed by a
strong magnetic field. This is illustrated in the inset to Fig.~5 [curve
$R(B)$] obtained in a sample of amorphous Ge, in which the Josephson couplings
between the Al granules ensure a superconducting state at a low temperature
T=2~K which only slightly exceeds $T_{\rm c}$; the negative magnetoresistance
caused by the suppression of superconducting fluctuations is observed in
magnetic fields of up to 16~T.

Thus, experiments on granular superconductors revealed a new experimental area
of searching for the realization of the bosonic scenario for the
superconductor--insulator transition. {\it If in an insulator that is formed
after the breakdown of superconductivity there exist electron pairs localized
on defects, then in a strong magnetic field we can expect the appearance of a
negative magnetoresistance caused by the destruction of these pairs.}

\subsection{1.5 Berezinskii--Kosterlitz--Thouless transition}

A distinguishing feature of two-dimensional superconducting systems is the
possible existence of a gas of fluctuations in the form of spontaneously
generated magnetic vortices at temperatures smaller than the temperature
$T_{\rm c0}$ of the bulk superconducting transition. A magnetic flux quantum
\begin{equation}\label{Phi0}        
\Phi_0=2\pi\hbar c/2e.
\end{equation}
passes through each vortex. The factor 2 in the denominator of expression (10)
is preserved in order to emphasize that the quantization is determined by
charge carriers with a charge $2e$.

The vortices are generated by pairs with the oppositely directed fields on the
axis (the vortex--antivortex pairs) and in a finite time they annihilate as a
result of collisions. In a zero magnetic field, the concentrations of vortices
with opposite signs are equal, $N_+=N_-$; they are determined by the dynamic
equilibrium between the processes of spontaneous generation and annihilation. A
decrease in temperature to $T_{\rm c}\equiv T_{\rm BKT}<T_{\rm c0}$ leads to a
Berezinskii--Kosterlitz--Thouless (BKT) transition [27, 28]. The generation of
vortex pairs ceases, and the concentration of vortices decreases sharply and
becomes exponentially small.

Thus, in a certain temperature range
\begin{equation}\label{BKT}
 T_c<T<T_{c0}
\end{equation}
in two-dimensional superconductors, the vortices coexist with Cooper pairs. The
modulus of the order parameter, which is the binding energy $\Delta$ of a
Cooper pair in the space between the vortices, decreases to zero on the axis of
the vortex; there is no superconductivity near the axis of the vortex, and the
electrons are normal. The phase of the order parameter in the space between the
vortices fluctuates as a result of their motion. Correspondingly, correlator
(8) on the interval (11) falls off exponentially, and at temperatures below the
temperature of the BKT transition ($T<T_{\rm BKT}$) it diminishes according to
a power law:
\begin{equation}\label{Correl-1}
  G({\bf r})\propto r^{-\eta},\qquad0<\eta<1,
\end{equation}
i.e., at large distances it tends to zero rather than to a finite value. At
large distances, a coherent state with the finite correlator (8) is established
at $T=0$.

The vortices being considered as quasiparticles are bosons. Therefore, it can
be said that the presence of free vortices-bosons leads to energy dissipation
when current flows, in spite of the presence of 2e-bosons (Cooper pairs).

\begin{figure}[t]
\includegraphics{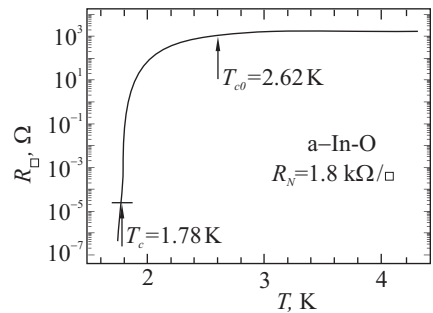}
\caption{Temperature $T_{\rm c0}$ at which an equilibrium concentration of
Cooper pairs appears, and the temperature $T_{\rm c}$ of the BKT transition in
an In-O film 100\,\AA\ thick [30]. $R_\Box$ is the resistance per square
(resistivity of a two-dimensional system).} \label{Hebard0}
\end{figure}

There is a purely experimental problem in determining the temperatures $T_{\rm
c0}$ and $T_{\rm c}$ from the curve of the resistive transition. The resistance
of the system in the temperature range $T_{\rm c}<T<T_{\rm c0}$ was calculated
in Ref.~[29], and a thorough experimental examination was carried out in
Ref.~[30] using a superconducting transition in In-O amorphous films. It is
seen from Fig.~6, in which the result of such an analysis is given for one of
the films, that the temperatures $T_{\rm c0}$ and $T_{\rm c}$ differ strongly:
$T_{\rm c0}$ lies in the high-temperature part of the $R(T\,)$ curve, so that
$R(T_{\rm c0})\approx 0.5R_{\rm N}$, while $R(T_{\rm c})$ is less than the
resistance $R_{\rm N}$ of the film in the normal state by several orders of
magnitude. The relationship between the resistances $R(T_{\rm c0})$, $R(T_{\rm
c})$, and $R_{\rm N}$ changes from film to film, but even more they differ
because of the fact that in various laboratories the $T_{\rm c0}$ and $T_{\rm
c}$ temperatures are usually determined differently. Therefore, when comparing
the results of experiments, it is sometimes more convenient to use the ratio
$R(T\,)/R_{\rm N}$ for determining the characteristic points in the resistance
curve.

\section{2. Microscopic approaches to the problem of the
superconductor--insulator transition}

Among different theoretical models used for the description of
superconductor--insulator transitions, there is no one unconditionally leading
model, such as the BCS model employed for the superconductivity itself.
Approaching the problem from different sides, the existing models emphasize its
different aspects and together create an integral picture, demonstrating at the
same time the existence of different variants of the transition.

\subsection{2.1 Fermionic mechanism for the superconductivity suppression}

As already mentioned in Section 1.4, the fermionic scenario requires the
vanishing of the modulus of the order parameter with increasing the number of
impurities in the system. For the realization of the fermionic scenario, it is
necessary to go beyond the limits of the validity of the Anderson theorem [9],
namely, it is necessary to take into account the Coulomb interaction between
the electrons, together with the disorder. The first idea in this area, which
was formulated in Ref.~[31], was based on the use of formula (3). First, we
shall assume that the system is granular. With increasing impurity
concentration in a granule, the density of states at the Fermi level is
suppressed by the Coulomb interelectron interaction due to the
Aronov--Altshuler effect [32, 33] and, correspondingly, the spacing (3) between
the energy levels grows. In this case, the critical size (4) of a granule
increases, while at a fixed size $b_{\rm SC}$ the gap $\Delta$  and, therefore,
the temperature $T_{\rm c}$ of the superconducting transition decrease. It can
be expected that the temperature $T_{\rm c}$ will become zero at a certain
critical concentration of impurities. The same reasoning is also applicable to
a uniform system if the granular size is replaced by the length of electron
localization in the normal state [34--36].

However, it turned out that the Coulomb interaction suppresses the modulus of
the order parameter in a completely different way, which is not related to the
granular or quasigranular character of the system. In the dirty limit, the
Coulomb interelectron interaction itself is renormalized [10], and the
processes of repulsion of electrons with opposite momenta and spins, which lead
to a low transfer of the momentum, become stronger. The result of Ref.~[10]
resembles the suppression of the density of states $g_0$ at the Fermi level in
a normal dirty metal by the Coulomb interaction [32, 33], with the difference
that in the superconductor it is the temperature $T_{\rm c}$ that decreases
with increasing disorder, rather than the density of states $g_0$ at the Fermi
level.

The effect of $T_{\rm c}$ reduction as a result of a renormalization of the
Coulomb interaction was known earlier [37--39] in the form of a weak correction
to the superconducting transition temperature. For example, in the
two-dimensional case we have
\begin{equation}\label{Ovch}        
   T_c=T_{c0}\left(1-\frac{1}{12\pi^2y}\ln^3(\hbar/T_c\tau)\right),
\end{equation}
where $y$ is the dimensionless conductance:
\begin{equation}\label{conduct}
  y=\hbar/e^2R_\Box
\end{equation}
$R_\Box$ is the resistance per square (resistivity of a two-dimensional
system), and $\tau$ is the relaxation time of the momentum in the normal state.
Formally, expression (13) already reveals the possibility of vanishing $T_{\rm
c}$ with increasing disorder. However, the extrapolation over such a large
distance cannot serve as a serious argument.

The expression for the critical temperature $T_{\rm c}$ of a two-dimensional
system that is valid at low temperatures, $T_{\rm c}\ll T_{\rm c0}$, has been
obtained [40] using the renormalization-group analysis (see also papers [8,
11]):
\begin{equation}\label{Tc-renorm}        
\begin{array}{c}
 \displaystyle\frac{T_c}{T_{c0}}=
 e^{-1/\gamma}\left[\frac{\gamma-\varrho/4+(\varrho/2)^{1/2}}
              {\gamma-\varrho/4-(\varrho/2)^{1/2}}\right]^{1/\sqrt{2\varrho}},\\
 \displaystyle\rule{0pt}{7mm}\gamma=(\ln T_{c0}\tau/\hbar)^{-1}<0,\\
 \varrho=\frac{e^2}{2\pi^2\hbar}R_\Box= \displaystyle\frac{1}{2\pi^2y}.
\end{array}
\end{equation}

\begin{figure}
  \includegraphics{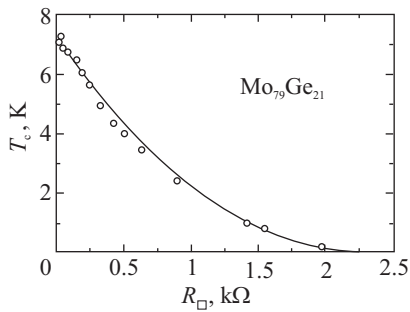}
 \caption{Suppression of superconductivity by a disorder in amorphous $\rm
Mo_{79}Ge_{21}$ films [40]. Circles correspond to experimental data taken from
Refs~[41, 42], and the solid curve was constructed from formula (15).}
 \label{MoGeFink}
\end{figure}

Figure 7 displays experimental data for films of an amorphous $\rm
Mo_{79}Ge_{21}$ alloy with various thicknesses and, consequently, with
different resistances $R_\Box$  [41, 42]. The solid curve was constructed in
Ref.~[40] using formula (15) on the assumption that $\ln {(\hbar /T_{\rm
c0}\tau )}=8.2$.

Thus, the theory correctly describes in the two-dimensional case the decrease
in the temperature of the superconducting transition under the disorder effect.
For the three-dimensional case, there are no exact answers, but we can expect
the same qualitative picture. Depending on which of the situations, i.e.,
Anderson localization in the normal state or vanishing of the superconducting
transition temperature, occurs earlier, one of the three phase diagrams
presented in Fig.~2 is realized.

The theory developed in Ref.~[40] corresponds to the use of a mean field
concept, i.e., an order parameter that is independent of the coordinates. In
recent years, it has been revealed, however, that the possible inhomogeneity of
the order parameter both with and without allowance for the Coulomb interaction
effect can by itself lead to the loss of macroscopic coherence. In the vicinity
of the quantum phase transition, where the conductance (14) is on the order of
unity, mesoscopic effects caused by a nonlocal interference of electron waves
scattered by impurities can become essential [43]. As a result, the originally
uniform system can become nonuniform upon transition. Superconducting droplets
can appear in it.

This possibility is realized in the two-dimensional case if the Coulomb
interelectron interaction is taken into account, i.e., when using the model
[40] beyond the framework of the mean-field approximation [44]. The mesoscopic
effects in a wide temperature range of $T>T_{\rm c}$ generate a nonuniform
state of the system with superconducting droplets embedded into the normal
regions. According to Ref.~[44], the temperature interval $\delta T_{\rm c}$ in
which the superconducting droplets can appear is specified by the relationship
\begin{equation}\label{Skvortsov}       
  \delta T_c/ T_c \simeq \frac{0.4 \pi^2 \varrho^2 }{(1-\varrho/\varrho_c)},
\end{equation}
where $\rho _{\rm c}$ is the critical value of the dimensionless resistivity at
which $T_{\rm c}$ calculated according to formula (15) becomes zero. As can be
seen from formula (16), the width of the region of the nonuniform state can be
on the order of $T_{\rm c}$.

\subsection{2.2 Model of a granular superconductor}

The first analytically solvable model with a phase transition to the insulating
state was constructed by Efetov [45] for a granular superconductor with a
superconducting gap $\Delta$, a granule size $b$, and the frequency $1/\tau _b$
of electron hopping between adjacent granules. It was assumed that $\tau _b$
falls in the range assigned by the following inequalities:
\begin{equation}\label{limitsEf}
 \delta\varepsilon\ll\hbar/\tau_b\ll\Delta
\end{equation}
where the energy $\hbar /\tau _b$ is less than the superconducting gap, but
more than the level spacing in the granules. The left-hand inequality means
that in the absence of superconducting interaction the localization effects can
be neglected and the granular material can be considered as a normal metal.

The granule size $b$ is assumed to be smaller than the coherence length $\xi$.
The left-hand inequality (17) chosen as the bound from below for the size $b$
is more strict than the above-considered condition (4). As a result, the
following interval was assumed for $b$:
\begin{equation}\label{limits-1-Ef}
 (\hbar g_0/\tau_b)^{-1/3}< b<\xi.
\end{equation}

The effective Hamiltonian describing the system is written out as follows:
\begin{equation}\label{Ham-Ef}
 \begin{array}{c}
 H_{\bf eff}=\sum\limits_{ij}\frac12 B_{ij}\widehat{\rho_i}\widehat{\rho_j}+
 \sum\limits_{ij}J_{ij}(1-\cos(\varphi_i-\varphi_j)),\\
 \displaystyle\rule{0pt}{5mm}\widehat{\rho_i}=
  -{\rm i}\frac{\partial}{\partial\varphi_i}.
 \end{array}
\end{equation}
Here, $\hat {\rho \,}\!_i$ are the operators of the number of Cooper pairs in
the i-th granule (with the integers as the eigenvalues), and the quantities
$B_{i\,j}$ at low temperatures are proportional to the elements of the matrix
that is inverse to the capacitance matrix. On the order of magnitude, for
example, for granules with thin interlayers of thickness $\tilde {b}$, we have
\begin{equation}\label{Bij-Ef}     
 B_{ij}\sim (e^2/b) (\widetilde{b}/\kappa_0b),
\end{equation}
where )$\kappa _0$ is the dielectric constant of the insulating interlayer. The
first term under the summation sign in Hamiltonian (19) describes the
electrostatic energy arising upon the generation of pairs on the granules. The
second term contains the Josephson energy $J_{i\,j}$, which is nonzero only for
the nearest neighbors and is expressed through the normal contact resistance
$R_{i\,j}^{\,\rm n}$ as
\begin{equation}\label{joseph}      
  J_{ij}=\frac\pi4\left(\frac{\hbar/e^2}{R_{ij}^{(n)}}\right)\Delta(T).
\end{equation}
It is assumed for simplicity that all the granules and the insulating
interlayers are identical and arranged regularly, so that $B_{i\,j}$ and
$J_{i\,j}$ depend only on the difference $|i-j\,|$.

The solution was obtained by the self-consistent field method. To this end, the
interaction in the Hamiltonian was replaced by a mean effective field:
\begin{equation}\label{cos-Ef}
  \cos(\varphi_i-\varphi_j)\rightarrow
  \langle\cos\varphi_i\rangle\cos\varphi_j
\end{equation}
The phase transition point is found from the condition of phase coherence in
different granules, i.e., from the condition that $\langle \cos \varphi
_i\rangle$  is nonzero in the equation for self-consistent solution of the
problem with Hamiltonian (19). In this way, a critical value is obtained of the
ratio between the Josephson and Coulomb energies, at which a phase transition
at a zero temperature occurs. In the simplest case, one has
\begin{equation}\label{epsilon_c-Ef}        
 J_{ic}=\left(\sum_jJ_{ij}\right)_c=\frac12 B_{ii}(0) .
\end{equation}

For $J_i>J_{i{\rm c}}$, a macroscopic superconducting state is realized in the
granular superconductor. In order to understand the properties of the
incoherent phase in which $\langle \cos \varphi _i\rangle =0$, it is necessary
to solve the kinetic problem of the response of a granular superconductor in
the incoherent state to a static electric field for
\begin{equation}\label{Ef-7}
J_i\ll J_{ic}.
\end{equation}
The current between separate granules is equal to the sum of normal and
Josephson currents. Owing to the first of these terms, the conductivity at the
zero frequency proves to be finite at a nonzero temperature and, to an accuracy
of a numerical coefficient, is expressed in the form
\begin{equation}\label{Ef-8}        
 \sigma(0) = R^{-1}\exp(-\Delta/T).
\end{equation}
The exponential dependence on the temperature indicates that the system resides
in the insulating state.

\begin{figure}
\includegraphics{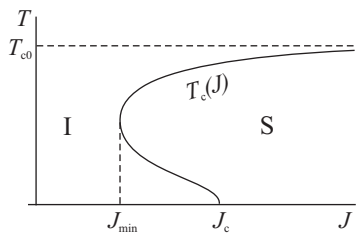}
\caption{Phase diagram on the $(J,T\,)$ plane, corresponding to the possibility
of the occurrence of a reentrant superconductor--insulator transition with
decreasing temperature [45].} \label{Efet}
\end{figure}

The case of a finite temperature is rather interesting. In this case, for
$\tilde {b}\ll b$ it is necessary to take into account the contribution from
the off-diagonal elements $B_{i\,j}$, but this means the possibility of the
appearance of charges in two adjacent granules rather than in only one granule.
The critical value of the Josephson energy is additionally increased under
these conditions. However, with increasing temperature the spaced charges will
be screened by the normal excitations of adjacent granules and the critical
value of the Josephson energy will decrease. Irrespective of this, an increase
in temperature leads to an increase in the spread of the phases of separate
granules. The resulting dependence of the superconducting transition
temperature on the Josephson energy is illustrated qualitatively in Fig.~8.
This dependence indicates that under specific conditions the granular
superconductor can pass into an insulating state upon a decrease in
temperature. This transition is called reentrant.

The theory of reentrant transitions was developed in many studies (see, e.g.,
Refs~[46--48]), mainly within the framework of the ideas presented above.
Experimentally, the reentrant transitions are manifested in the fact that the
rapid decrease in resistance with decreasing temperature in the process of the
superconducting transition is changed by its rapid growth. The reentrant
transition is usually considered to be a specific property of granular
superconductors. Frequently, the presence of such a transition was assumed to
indicate that the sample had a granular structure and served as a criterion for
the selection and classification of samples. However, as we shall see in
Section 2.5, a reentrant transition in the presence of a magnetic field can
occur even in the absence of a granular structure.

The upper branch of the phase diagram in Fig.~8 is also very informative. It
shows that the temperature of the superconducting transition can decrease when
approaching the critical value of the control parameter not only in a uniformly
disordered superconductor but also in a granular superconductor with granules
of a small size (18), if some additional conditions are fulfilled [in
particular, if inequalities (17) are valid and the interlayers between the
granules are relatively narrow, $\tilde {b}\ll b]$.

Thus, Efetov [45] has constructed a strict microscopic theory of the
superconductor--insulator transition for a single specific case of a granular
superconductor for which inequalities (17) and (18) are fulfilled. Some results
of this theory were later obtained based on phenomenological considerations in
Ref.~[49].

The model of the transition constructed in Ref.~[45] occupies an intermediate
place between the fermionic and bosonic scenarios. On the one hand, this model
proceeds from the BCS theory and deals exclusively with Cooper pairing. On the
other hand, because of the coordinate dependence of the order-parameter
modulus, which is due to the very formulation of the problem (difference in the
magnitude of $\Delta$ inside and outside the granules), this model allows the
existence of regions with $\Delta \neq 0$ for temperatures $T>T_{\rm c}$. Note
in conclusion that Efetov's model does not require the presence of disorder in
the granular system (if the very existence of the granules is not considered as
disorder) for the implementation of the transition. There is no doubt that the
existence of disorder does not prevent the transition of a superconductor to an
insulating state, but neither is it a driving force for such a transition: the
latter could occur even on a regular lattice of granules. In this respect, the
transition considered is more likely analogous to a Mott--Hubbard
metal--insulator transition than to an Anderson transition.

\subsection{2.3 Bose--Einstein condensation of a bosonic gas}

As was already noted above, in some cases it is more convenient to employ the model of Bose--Einstein condensation in a gas of bosons for describing the behavior of a superconductor. Recall that according to the statistics of Bose particles at a temperature lower than a certain critical value, a macroscopic number of particles find themselves at the lower quantum level and form the so-called Bose condensate. In the general case, the lower quantum level is not separated by a spectral gap from the excited states of the system. At a zero temperature, all Bose particles prove to be in the ground state. The assertion about the existence of a Bose condensate is correct both for a gas of charged Bose particles [50], i.e., particles with interaction, and for a gas of noninteracting Bose particles which are scattered by the short-range field of impurities [51]. The presence of a Bose condensate by itself by no means implies that the particles will demonstrate superfluidity (or ideal conductivity in the case of a gas of charged particles). The problem of the dynamic low-frequency response of the interacting gas of Bose particles in the field of impurities was posed and solved by Gold [52, 53] for two concrete cases: a Bose gas with a weak repulsion in the field of neutral impurities, and a charged Bose gas in the field of charged impurities.

The problem was set up as follows. The dependence of the kinetic energy of bosons on the momentum is assumed to be parabolic, $\varepsilon (k)=k^{\,2}/2m$, and the Hamiltonian comprises three terms:
\begin{equation}\label{Ham-Gold}
 H=H_0 +H_I +H_D.       
\end{equation}
The first term describes the kinetic energy of free bosons:
\begin{equation}\label{Ham-Gold-0}      
 H_0=\sum_{\bf k}\varepsilon(k) a^+_{\bf k} a_{\bf k};
\end{equation}
where the operators $a^{\,+}_{\bf k}$ and $a_{\bf k}^{}$ correspond, as usual, to the creation and annihilation of a boson with a momentum {\bf k}. The second term describes the interaction between the bosons:
\begin{equation}\label{Ham-Gold-I}
 H_I=\frac12\sum_{\bf q}\varrho({\bf q})V_{\bf q}\varrho^+({\bf q}),
\end{equation}
where $V_{\bf q}$ is the Fourier component of the interaction potential, and $\varrho ({\bf q})$ and $\varrho ^{\,+}({\bf q})$ are the operators of the density fluctuations:
$\varrho ({\bf q})=\sum _{\bf k}a^{\,+}_{{\bf k}-{\bf q}/2}a_{{\bf k}+{\bf q}/2}$ and
$\varrho ^{\,+}({\bf q})= \sum _{\bf k}a^{\,+}_{{\bf k}+{\bf q}/2}\,a_{{\bf k}-{\bf q}/2}$.
The last term in sum (26) corresponds to the interaction of bosons with impurities:
 \begin{equation}\label{Ham-Gold-D}         
 H_D=\sum_{\bf q}U_{\bf q}\varrho^+({\bf q}),
\end{equation}
where $U_{\bf q}$ is the Fourier component of the scattering potential.

It is necessary to calculate the dynamic response of a system with such a Hamiltonian. Let us first examine a weakly interacting gas of repulsive Bose particles having the radius of interaction $q_0^{\,-1}$ with the impurities:  $$V_{\bf q}=V,$$ and
\begin{equation}\label{Ham-Gold-5}      
\langle|U_q|^2\rangle=6\pi^2q_0^{-2}U^2\theta(q_0-q),\quad
    \begin{array}{cc}
        \theta(x)=1,& x\geqslant0,\\
        \theta(x)=0,& x<0,
    \end{array}
\end{equation}
where $V$ and $U$ are the constants. According to Ref.~[54], the gas of interacting particles in question possesses a gapless spectrum of excitations, and the introduction of scatterers with a small interaction radius does not lead to the critical behavior of spectral characteristics. Nevertheless, the kinetic characteristics of system (26)--(30) radically change, depending on the relationship between the scale of the interparticle interaction and the scattering potential. For the formal description of this relationship, a dimensionless parameter $A$ was introduced in'Ref.~[52]:
\begin{equation}\label{Gold-6}      
A=3n^2\sum_{\bf q}\langle|U_q|^2\rangle(\hat g(q))^2,
\end{equation}
where $n$ is the density of bosons, and $\hat g(q)$ is the compressibility of the interacting boson gas, which can be expressed through $V_q$. An increase in disorder brings about an increase in $A$.

It turned out that the transport properties of the system radically change at $A=1$. The last condition always corresponds to an increase in the critical value of the effective scattering potential with strengthening interaction between the bosons, and/or with increasing the density of bosons; this fact corresponds to the concept of the collective wave function of the Bose condensate.

For the active response of the system at low frequencies, the following result was obtained:
\begin{equation}\label{Gold-7}      
\sigma'(\omega)\approx\left\{
  \begin{array}{cl}
  (1-A)\delta(\omega),&\:A<1,\\
  0,&\:A>1,
  \end{array}
  \right.
\end{equation}
where $\delta (\omega )$ is the $\delta$ function. For $A<1$, the system possesses an infinite active component of conductivity at $\omega =0$ and is superfluid. At $A=1$, a quantum phase transition from the superfluid state to the state of localized bosons (Bose glass) occurs.

An analogous behavior is characteristic of the gas of charged bosons in the field of charged impurities. The corresponding harmonics of the potentials take on the form
\begin{equation}\label{Gold-8}      
  V_q =4\pi e^2 /q^2,\qquad \langle|U_q|^2\rangle =N(4\pi e^2/q^2)^2,
\end{equation}
where $N$ is the density of scattering centers. For the system to be stable, it is necessary to assume the existence of a uniform background which compensates for the charge of bosons. In expression (31), not only the potential of interaction with the scatterers but also the compressibility $\hat g(q)$ is changed. The ground state proves to be separated by a gap from the excited states; however, the main result described by expressions (31) and (32) remains unaltered.

Bose condensation means the existence of superconductivity with a London penetration depth $\lambda _0^2=mc^{\,2}/4\pi n$ (where $m$ and $n$ are the effective mass and the density of bosons, respectively) and with the conductivity $\sigma (\omega )$ as $\omega \rightarrow 0$:
\begin{equation}\label{sc-sigma}
  \sigma(\omega)=i\frac{c^2}{4\pi}\,\frac{1}{\lambda^2}\,\frac1\omega,\qquad
  \lambda=\lambda_0\frac{1}{\sqrt{1-A}}\;.
\end{equation}
The occurrence of a transition follows from the divergence of $\lambda$ as $A\rightarrow 1$.

Now, the condition $A=1$ connects the concentrations of the scattering impurities $(N)$ and the bosons $(n)$. At the critical concentration of impurities $N_{\rm c}$, namely
\begin{equation}\label{Gold-9}     
  N_c\propto n^{5/4}
\end{equation}
a transition from the state of an ideal conductor to an insulating state occurs.
The Efetov model, which was discussed in Section 2.2, allows a periodic arrangement of granules, so that the transition in this model resembles the Mott transition. In the Gold model, an important feature is precisely the randomness of the arrangement of impurities, and the transition from the superconducting to the insulating state rather resembles the Anderson transition to the Bose-glass phase. However, in the case of a bosonic system it is impossible to assume the complete absence of interaction, unlike the case of the Anderson transition in the electron system. The need to take into account the interaction between the bosons can be explained as follows.

Let us assume that there is only one impurity and only one localized state near it. In the absence of interaction, all the bosons will be condensed into this localized state, i.e., we obtain an insulator. It can be said that the superconducting state of noninteracting bosons is unstable with respect to an arbitrarily weak random potential and that the interaction between bosons stabilizes superconductivity. Hence, relationship (35) appears: a decrease in the boson concentration weakens interaction and, therefore, the critical concentration of impurities decreases, as well.

\subsection{2.4 Bosons at lattice sites}

Fisher et al. [55] suggested in their study a model which partially inherits properties of the two previously considered models [45, 52]. The authors of Ref.~[55] investigated the properties of a system of bosons arranged at sites in the lattice, which possess weak repulsion and are characterized by a finite probability of hopping between the sites and by a chaotically changing binding energy at a site. This model is especially interesting for us, since a general scaling scheme for a superconductor--insulator transition was constructed on the basis of ideas developed in Ref.'[55].

The Hamiltonian of the system in question takes on the following form
\begin{equation}\label{Ham-Fisher}      
 \begin{array}{l}
 \widehat{H}=\widehat{H}_0 +\widehat{H}_1, \\
 \widehat{H}_0=
 -\sum\limits_i(-J_0+\mu+\delta\mu)\widehat{n}_i
 +\frac12\sum\limits_iV\widehat{n}_i(\widehat{n}_i-1),\\
\widehat{H}_1=-\frac12\sum\limits_{ij}J_{ij}(\widehat{\Phi}_i^
+\widehat{\Phi}_j+\mbox{H.c.}),\\
\end{array}
\end{equation}
where $\hat n_i$ is the operator of the number of particles at site $i$; $J_{i\,j}$ is proportional to the frequency of hoppings between the sites $i$ and $j$; the sum $\sum _{\,j} J_{i\,j}=J_0$ is assumed to be identical for all the sites; $\mu$ is the common chemical potential; $V$ is the interaction energy of two bosons at one site, and H.c. denotes the Hermitian conjugate. Randomness in the system is introduced with the aid of variations $\delta\mu _i$ in the chemical potential from site to site (an average of $\delta\mu _i$ over the system is equal to zero). The field operators of the bosonic field, $\hat {\Phi }_i^{\,+}$ and $\hat {\Phi }_j$, in the Hamiltonian $\hat H_1$ can be expressed through the operators of creation and annihilation of particles, $a^{\,+}_{\bf k}$ and $a_{\bf k}^{}$, that were used in Hamiltonian (27):
\begin{equation}\label{Ham-Fisher2}
 \widehat{\Phi}_i\equiv\widehat{\Phi}({\bf r}_i)
=\sum_{\bf k}\psi_{\bf k}({\bf r}_i)a_{\bf k}, \quad
\widehat{\Phi}^+_i=\sum_{\bf k}\psi^*_{\bf k}({\bf r}_i)a^+_{\bf k},
\end{equation}
where $\psi_{\bf k}({\bf r})$ is the wave function of a particle in the state with a wave vector {\bf k}. The field operators can be considered as the operators of particle annihilation or creation at a given point of space; their commutator is $[\hat {\Phi }_i,\hat {\Phi}_j^{\,+}]=\delta _{i\,j}$, and $\hat {\Phi }_i^{\,+}\hat {\Phi }_i=\hat {n}_i$.
\begin{figure}[b]
  \includegraphics{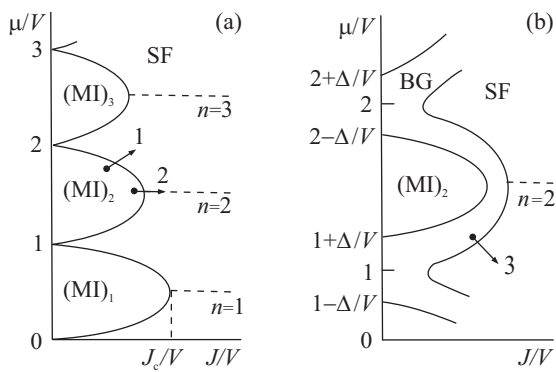}
\caption{Phase diagrams at $T=0$ for a system of bosons interacting at the sites: (a) in the absence of disorder, and (b) in the presence of disorder. $({\rm MI})_n$ is a Mott insulator with $n$ bosons at each site; SF is a superfluid phase. Arrows indicate transitions from the state of the Mott insulator to the superfluid phase: (arrow 1) transitions with a change in the density of bosons; (arrow 2) transitions at the constant density of bosons. BG is a Bose glass with a different number of bosons at different sites. Arrow 3 indicates the transition from the Bose insulator to the superfluid phase (taken from Ref.~[51]).} \label{DiagFish}
\end{figure}

Let us first consider a system without disorder and construct a phase diagram on the plane $(J,\mu )$ (Fig.~9a). To start, we take the case of $J_{i\,j}=0.$ Let the potential $\mu$  be determined by the external thermostat and let it be able to change continuously. The number of bosons $n$ at all the sites is one and the same, since all the sites are equivalent, and is an integer. The number $n$ should be found by minimizing the energy of bosons residing at a single site:
\begin{equation}\label{FF3}     
    \varepsilon(n) = -\mu n+ \frac12V n(n-1).
\end{equation}
Since $n$ is discrete, each value of n is realized on a certain interval of $\mu$  values, namely, $n-1<\mu /V<n$. At the boundary of this interval, the values of the energy (38) for two neighboring values of $n$ become the same: at $\mu =nV$, we have
\begin{equation}\label{FF4}         
  \varepsilon(n)=\varepsilon(n+1)=-\frac V2n(n+1).
\end{equation}

The role of an elementary excitation in the system is played by an extra or missing boson at one of the sites. The energy required to add a boson to the system or remove it from the system depends on the position of the chemical potential $\mu$  relative to the boundaries of the interval. If $\mu$ is fixed at a level
$$\mu/V=n-\frac12+\alpha,\qquad-\frac12<\alpha<\frac12,$$
then, to add a boson to the system or remove it from the system, an energy on the order of
\begin{equation}\label{FF5}
  \delta\varepsilon_\pm\sim\pm V(\frac12\mp\alpha).
\end{equation}
is required.

It was assumed above that the interaction with the thermostat ensures the possibility of a smooth change in $\mu$ and that $n$, considered as an average number of bosons at a site, can assume only discrete integer values. If, on the contrary, we can smoothly vary the total number of bosons in the system, then $n$ changes continuously, and the chemical potential takes only discrete values:
\begin{equation}\label{FF5a}
\mu=[n]V,
\end{equation}
where $[n]$ is the integer part $n$, and the number of bosons is equal to $[n]$ at some sites, and to $[n]+1$ at other sites. According to formula (39), the energies $\varepsilon ([n])$ and $\varepsilon ([n]+1)$ are equal at values of the chemical potential equal to those defined by formula (41). In Section 5.2, we shall consider the experimental realization of precisely such a case.

Now, let us return to the system with smoothly changing $\mu$ and integer $n$, and include weak hopping $J>0$ into the examination, i.e., require that, during the determination of the equilibrium state, the kinetic energy be taken into account, as well. This will influence the state of the system only if $J$ proves to be larger than at least one of the energies specified in estimate (40). In particular, at integer values of $\mu /V$ this will occur at arbitrarily small $J$, and the critical value $J=J_{\rm c}$ will be maximum at half-integer $\mu /V$. Hence, the phase plane $(J,\mu )$ will be divided into two regions (Fig.~9a). To the left of the solid line, in the interval of the values of the chemical potential,
\begin{equation}\label{FF6}     
  V(n-1)<\mu<Vn
\end{equation}
the system resides in the state of an insulator \emph{with equal number n of bosons at all sites}. Since there is no disorder whatever in the system, this insulator is called the Mott insulator (MI). Thus, to the left of the solid line we obtained a set of Mott insulators (MI)$_n$ that differ in the number of bosons $n$ at the sites.

To the right of the phase boundary, it is possible to introduce a boson into the system by supplying it only with kinetic energy $J$, without assigning it to a specific site. Such bosons will be delocalized. They can freely move around the system, and at $T=0$ they, through the Bose condensation, provide superfluidity.

On the upper part of the boundary of an (MI)$_n$ region, for $\mu /V>n$, the potential energy required for an additional boson to appear at some site is compensated for by its kinetic energy. Therefore, the additional boson can freely jump over sites and go into the Bose condensate. For any point $\mu /V<n$ of the lower part of the phase boundary, the same reasoning is valid for the hole (one boson missing from a site). After the intersection of the boundary, the number of bosons ceases to be fixed and an integer, and begins smoothly changing as $\mu$  varies. In contrast to these transitions caused by a change in the density of bosons, at the points $\mu /V=n$ on the boundary, a transition at a constant density can occur, when the kinetic energy of the bosons grows so that they obtain the possibility of moving across the sites, overcoming intrasite repulsion.

Now, let us introduce disorder into the system of bosons, suggesting that $\delta \mu _i$ are distributed uniformly inside the interval $(-\Delta ,\Delta )$, with $\Delta <V/2$. Let us again first exclude the hopping between the sites, assuming $J=0$. Then, we are obliged to minimize the energy for each of the sites separately:
\begin{equation}\label{FF7}     
  \varepsilon(n_i)=-(\mu+\delta\mu_i)n_i+ \frac12Vn_i(n_i-1).
\end{equation}
If we `smear' the quantity $\mu$ in inequality (42) over an interval $\pm\Delta$, then, to retain condition (42), we should correspondingly shift the boundaries of the interval:
\begin{equation}\label{FF8}
  V(n-1)+\Delta<\mu_i=\mu+\delta\mu_i<Vn-\Delta.
\end{equation}
As a result, we obtain the diagram presented in Fig.~9b: the ordinate axis is divided into intervals centered at half-integral values of $\mu/V$, inside which, as before, an equal number of bosons is located at each of the sites. Inside these intervals, the Mott insulator is retained. On the remaining part of the ordinate axis, disorder prevails and the number of bosons at the sites proves to be different. Here, we are dealing with an insulator of another type---a Bose glass.

The introduction of a finite probability of a boson hopping between the sites, $J\neq 0$, leads to appearance of a layer of Bose-glass states on the $(J,\mu)$-plane, so that the transition to the superfluid state occurs from the disordered insulator (arrow 3 in Fig.~9b). Moreover, in the case of a strong disorder, $\Delta >V/2$, the Mott-insulator regions disappear at all.

The above qualitative picture of phase transitions in the system of bosons on a lattice of sites can naturally be extended to insulator--superconductor transitions if we assume the bosons to be charged. The transitions to the superconducting state can occur both upon a change in the concentration $n$ with the chemical potential as the control parameter, $\delta x=\mu -\mu _{\rm c}$, and upon an increase in the hopping frequency, $\delta x=J-J_{\rm c}$. In the above-considered model, the transitions can occur both from the MI state and from the BG state. However, since we are discussing the superconductivity in Fermi systems, the existence of the Bose-glass state, i.e., of localized pairs, should first be proved.

\subsection{2.5 Superconducting fluctuations in a strong magnetic field in the framework
of the Bardeen--Cooper--Schrieffer model}

In the BCS model, Cooper pairs appear only via the fluctuation mechanism at temperatures exceeding $T_{\rm c}$ or, for $T<T_{\rm c}$, in the magnetic field with $B>B_{\rm c2}(T\,)$. Nevertheless, their effect on conductivity is considerable. We here are first interested in the question of whether there is an anomalous component of this influence, i.e., is it possible to observe, in a certain domain of parameters, an increase in resistance under the effect of superconducting fluctuations, as occurs in granular superconductors [25, 26, 45].

In the plane $(T,B)$, the region of existence of fluctuations is that where $B>B_{\rm c2}(T\,)$, including
\begin{equation}\label{B0}      
 T>T_c(B=0)\equiv T_{c0}\qquad\mbox{при}\qquad B=0.
\end{equation}
The fluctuations in a zero magnetic field, i.e., in region (45), were studied sufficiently long ago [56--58]; however, at low temperatures,
\begin{equation}\label{T0}
  T\ll T_{c0},\qquad B>B_{c0}
\end{equation}
such studies were possible to conduct only comparatively recently [59], and only for two-dimensional systems (see also monograph [60]). A positive answer to the question that is of interest for us can be found in the results of Ref.~[59] in the dirty limit $T_{\rm c0}\tau\ll1$ (where $\tau$ is the mean free time) for two-dimensional superconductors at low temperatures in fields near $B_{\rm c2}(0)$ in the region
\begin{equation}\label{Bc2}     
  \begin{array}{l}
 t=T/T_{c0}\ll1, \\
 \beta(T)=(B-B_{c2}(T))/B_{c2}(0)\ll1\rule{0pt}{5mm},
  \end{array}
 \end{equation}

Three forms of quantum corrections exist for conductivity, which are caused by superconducting fluctuations (they are also called corrections in the Cooper channel). These are the Aslamazov--Larkin correction caused by the contribution to the conductivity from fluctuation-induced pairs; the Maki--Thompson correction connected with the coherent scattering of paired electrons by impurities, and the correction caused by a decrease in the density of states of normal electrons at the Fermi level as a result of the appearance of Cooper pairs [60]. In region (47), the contributions from all these corrections are of the same order. The resulting correction $\delta\sigma$ to the conductivity calculated in the first (single-loop) approximation in this region takes on the form
 \begin{equation}\label{GL}     
  \delta\sigma=
  \frac{2e^2}{3\pi^2\hbar}\left[-\ln\frac r\beta-\frac{3}{2r}+\psi(r)
  +4(r\psi'(r)-1)\right],
\end{equation}
where $\psi (x)$ is the logarithmic derivative of the $\Gamma$ function, $r=(1/2\gamma^{\,\prime})(\beta /t)$, and $\gamma ^{\,\prime }=\exp \gamma =1.781$ is expressed through the Euler constant $\gamma$.

\begin{figure}
\includegraphics{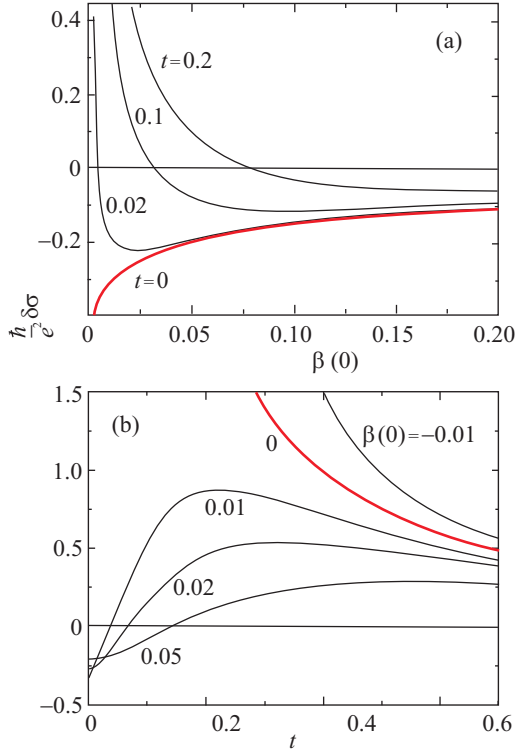}
\caption{Fluctuation correction (48) to the conductivity of a two-dimensional dirty superconductor as a function of (a) magnetic field at four different temperatures, and (b) temperature at five different strengths of the magnetic field [59]. Red curves are the separatrices of both families of the curves.}
 \label{GalLark}
\end{figure}

Formula (48) is illustrated in Fig.~10. The most important thing, from the viewpoint of the problem that is of interest for us, is that the corrections to the conductivity arising as a result of superconducting fluctuations can be not only positive but also negative. In the low-temperature limit of $t\ll \beta$ in fields $B>B_{\rm c2}(0$), formula (48) acquires the form
\begin{equation}\label{GL1}     
  \delta\sigma=
  \frac{2e^2}{3\pi^2\hbar}\ln \beta.
\end{equation}
The correction to the conductivity is negative and becomes quite large as $\beta \rightarrow 0$ (curve $t=0$ in Fig.~10a).

The curves corresponding to very small positive $\beta$ in Fig.~10b describe a reentrant transition, in spite of the absence of the granular structure in the superconductor (cf. Fig.~8). These curves are first held up against the curve $\beta =0$, increase along with it, and then return to the level of $\delta \sigma \sim 0$, so that the resistance first decreases and then returns to the level corresponding to the resistance in the normal state.

\begin{figure}[t]
\includegraphics{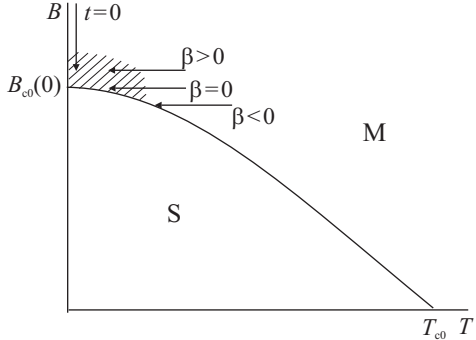}
\caption{Phase plane $(T,B)$ for a superconductor--dirty normal metal transition. The region of increasing resistance due to superconductor fluctuations is hatched (according to the results of calculations [59]). Arrows show different trajectories in the phase plane corresponding to different curves in Fig.~10.} \label{GalLark2}
\end{figure}

The calculation of fluctuation corrections has been done in the dirty limit of the BCS theory. Although the dirty limit means the presence of disorder, so that the mean free path is assumed to be less than the coherence length, both in the BCS theory and in Ref.~[59] a normal metal--superconductor transition is considered. The curve $B_{\rm c2}(T\,)$ in the phase plane $(T,B)$, used in paper [59] (presented in Fig.~11), implies just such a transition. The curve $t=0$ in Fig.~10a demonstrates the behavior of the fluctuation correction upon a decrease in the magnetic field strength, i.e., upon motion downward along the vertical arrow on the phase plane in Fig.~11. It turned out that the superconducting fluctuations in this region lead to an increase in the resistance. Strictly speaking, the results of calculations [59] are valid only in the region where $\Delta \sigma \ll \sigma$. However, based on the results of analogous calculations in the theory of normal metals, the weak localization is assumed to precede the strong localization [61]. If we, analogously to the above case, extend the tendency of an increase in resistance onto the region of $\Delta \sigma \sim \sigma$, we shall see that now the transition to the superconducting state upon a decrease in the field strength is preceded by the transformation of the normal metal into an insulator (or, at least, into a high-resistance state). In Fig.~11, the region in which this transformation occurs is hatched. As can be seen from the curves $\beta >0$ in Fig.~10b, this region is very narrow.

Notice that the conductivity in the vicinity of the critical point $(\beta =0, t=0)$ depends on the way we approach this point. According to the curve $\beta =0$ in Fig.~10b, the conductivity $\sigma$ tends to infinity as $t$ tends to zero. This means that along this path in the phase plane, which is arbitrarily indicated in Fig.~11 by the middle horizontal arrow, the system approaches a superconducting state.

As we shall see when examining experiments on films of different materials in Sections 4.1 and 4.2, an important factor, which is established quite clearly, is the character of the slope of the separatrix $R_{B_{\rm c}}(T\,)$ of the family of curves $R_B(T\,)$ in the limit $T\rightarrow 0$ (in experiment, it is usually the resistivity that is measured rather than the conductivity). In the calculation performed in paper [59], such a separatrix is the curve $\beta=0$ in Fig.~10b:
\begin{equation}\label{GLseparatr}          
 \sigma(T,\beta(0)=0)=\sigma(T,B=B_{c2}(0))\rightarrow\infty
 \quad\mbox{as}\quad T\rightarrow0.
\end{equation}
In the region where the results of calculation [59] are valid, the derivative $\partial (\delta \sigma )/\partial t$ of this curve grows in absolute value with decreasing temperature.

The intersection of the curves in Fig.~10b at low temperatures indicates the presence of a negative magnetoresistance. It turns out that the increase in resistance as a result of superconducting fluctuations and the presence of a negative magnetoresistance are characteristic not only of granular superconductors (see Fig.~5) but also of dirty quasi-homogeneous superconductors, and inequality (6) is not a fundamental limitation for the occurrence of these effects.

\subsection{2.6 Fermions at lattice sites. Numerical models}

Within the framework of the fermionic model, the role of the superconducting interaction in the presence of disorder was also studied by numerical methods. In Refs~[62, 63], the authors investigated the behavior of a system of $N$ fermions with spin $\sigma =\pm 1/2$ on a planar lattice with a model Hamiltonian
\begin{equation}\label{HamiShepel}      
 H=-t\sum_{\langle{\bf ij}\rangle,\sigma}
 c_{{\bf i}\sigma}^\dagger c_{{\bf j}\sigma}+
 \sum_{{\bf i},\sigma}(W_{\bf i}-\mu)n_{{\bf i}\sigma}+
 U\sum_{\bf i}n_{{\bf i}\uparrow}n_{{\bf i}\downarrow},
\end{equation}
where the probability $t$ of an electron hopping to a nearest adjacent site is assumed as the natural scale of all energies; $c_{{\bf i}\sigma }^{\,\dagger }$ and $c_{{\bf i}\sigma }$ are the operators of creation and annihilation of a fermion, respectively; the operator $n_{{\bf i}\sigma }=c_{{\bf i}\sigma }^{\,\dagger }c_{{\bf i}\sigma }$ corresponds to the occupation numbers of states, and the representation of the subscripts ${\bf i}$ and ${\bf j}$ in the vector form implies that the summation is extended over the lattice. The energy of electrons at the sites, $\varepsilon _{\bf i}=W_{\bf i}-\mu$, takes on random values on the interval $[-W/2,W/2]$, where $\mu$ is the chemical potential, and the Hubbard energy is assumed to be negative, $U<0$, which should reflect the presence of superconducting interaction.

The basic calculations were conducted on a lattice $L^2= 24\!\times \!24$. The total number of electrons $\langle n\rangle L^2$ with each spin direction was varied on the interval $0.2\le \langle n\rangle \le 0.875$.

Naturally, the number of electrons at a concrete site differs from $\langle n\rangle$ because of the presence of the random potential $W_{\bf i}$. It turned out that with increasing disorder (increase in $W$\,) the amplitude of the local order parameter,
$$\Delta(\bf r)\propto\langle c_{{\bf i}\uparrow}c_{{\bf i}\downarrow}\rangle$$
also suffered strong fluctuations and, at sufficiently large $W$, it was found that $\Delta=0$ on a significant part of the lattice; i.e., the superconductivity disappeared at all. Just as with the allowance for the Coulomb interaction [44], the nominally spatially uniform but strongly disordered system becomes similar to a granular superconductor. The appearance of a spatial modulation of the order parameter is accompanied by increasing phase fluctuations, and all these factors taken in totality lead to the transition from a superconductor to an insulator. Nevertheless, the single-particle gap in the density of states is still long retained. Its evolution at the initial stage of the introduction of disorder is shown in Fig.~12. As can be seen from this figure, it is the coherent peaks that prove to be most sensitive to the random potential.

\begin{figure}[h]
\includegraphics{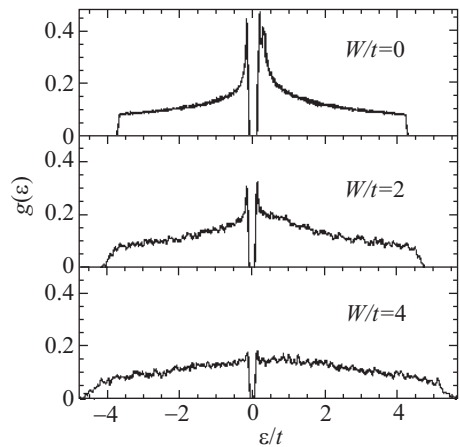}
\caption{Density $g(\varepsilon)$ of single-particle states on a $24\times 24$ lattice at three levels of disorder $W/t$ and an average electron density $\langle n\rangle =0.875$ [63].} \label{Trivedi}
\end{figure}

The results of Refs~[62, 63] can be directly compared with experimental data. First of all, this relates to the dispersion of the local values of the superconducting gap. According to the calculated results, the occurrence of disorder on the scale of the spacing between the adjacent sites (the values of $W_{\bf i}$ are in no way correlated) in the presence of a superconducting attraction leads to the appearance of a macroscopically inhomogeneous structure resembling a granular superconductor. To reveal this inhomogeneity, it was necessary to place the tunnel microscope into a dilution refrigerator. The first similar experiments appeared in 2008 (see Sections 6.2 and 6.4).

A similar problem on a three-dimensional lattice with $L^3$ sites was solved in Ref.~[64], where the same Hamiltonian (51) was investigated, but the problem was formulated somewhat differently. At $U=0$, the Hamiltonian (51) is reduced to the single-particle Anderson model with a metal--insulator transition at $W/t=W_{\rm c}/t\approx 16.5$. The influence of the mutual attraction of electrons at a site on this transition was studied for $U<0$.
\begin{figure}
\includegraphics{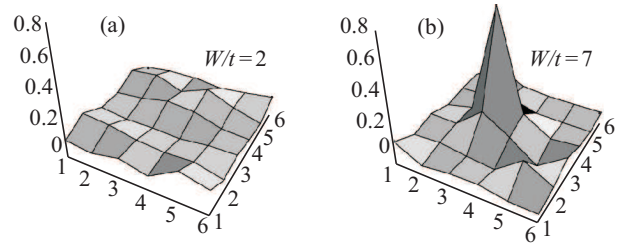}
\caption{Projections onto the $(x,y)$ plane for the probability of the distribution of an extra electron pair introduced at the Fermi level into a lattice of $6\times 6\times 6$ sites at the same attraction energy in the Hubbard model $U/t=-4$, but different levels of disorder: (a) $W/t=2$, and (b)~$W/t=7$ [64]. In the Anderson model without attraction $(U=0)$, the critical level of disorder on such a lattice is $W_{\rm c}/t=16.5$.} \label{Shepelya}
\end{figure}

The occupation number of the lattice sites with electrons, $\langle n\rangle$, was assumed to be about 1/4. The localization properties of the model with attraction $(U/t=-4)$ were determined from the behavior of an additional pair of electrons introduced into the system at the Fermi level. In the case of a small disorder, $W/t=2$, the electrons introduced were uniformly distributed over the lattice (Fig.~13a). However, a localization of the pair occurred already at $W/t=7$, although the disorder remained substantially smaller than that critical for the Anderson model, $W<W_{\rm c}$ (Fig.~13b). Thus, this numerical experiment clearly demonstrates the same tendency that is manifested through an analytical investigation of different models: pairing of electrons favors their localization.

\section{Scaling hypothesis}

\subsection{3.1 General theory of quantum phase transitions}

The general theory of quantum phase transitions [65, 66] is constructed similarly to the theory of thermodynamic phase transitions, but with an inclusion of terms in the partition function $Z$ that reflect the quantum properties of the system. It is desirable that the sum $Z$, in spite of an increase in the number of terms, could, as before, be considered as the partition function of a certain hypothetical classical system. For this to be the case, it is necessary to assume that the dimensionality ${\cal D}$ of the hypothetical system exceeds the real three-dimensional dimensionality $d$ of the system; this is achieved as a result of adding an imaginary time subspace. Thus, the theory of quantum transitions is constructed by mapping a given quantum system in a $d$-dimensional space onto a hypothetical classical system in the
${\cal D}$-dimensional space in such a way that the axes of the imaginary time subspace at a temperature $T$ have a finite length equal to ${\rm i}\hbar/T$ (in more detail, the physical scheme that serves as the basis of this mapping can be found in reviews [7] or [65]).

According to the scaling hypothesis [67], all physical quantities for an equilibrium system in the vicinity of a classical phase transition have a singular part which shows a power law dependence on some variable $\xi$  with a dimensionality of length. In the ${\cal D}$-dimensional space, the ${\cal D}-d$ axes of the imaginary time subspace are nonequivalent to the original spatial axes. Therefore, apart from the correlation length $\xi$  in the subspace of dimensionality $d$, we are obliged to introduce the length $\xi _\varphi$  along the additional axes:
\begin{equation}\label{sca0a}       
  \xi_\varphi\propto\xi^z.
\end{equation}
This length has a dimensionality of inverse energy and cannot be larger than the size ${\rm i}\hbar /T$ of the space in the appropriate direction:
\begin{equation}\label{sca0}
\xi_\varphi\leqslant i\hbar/T.
\end{equation}
The volume element of this fictitious space for the hypothetical classical system can be written out as
$$(d\xi)^d(d\xi_\varphi)\propto(d\xi)^{d+z},\qquad\mbox{т.е.}\qquad{\cal D}=z+d.$$
The correlation length $\xi$, in turn, depends on the proximity to the phase transition point, which is determined by the value of the control parameter x:
\begin{equation}\label{sca00}
 \xi\propto{(\delta x)}^{-\nu},
\end{equation}
and it tends to infinity at the very transition point. The numbers $z$ in formula (52) and $\nu$ in formula (54) are called critical exponents.

The quantity $L_\varphi$ with a dimensionality of length can be put into correspondence with the inverse energy $\xi _\varphi$, by writing, from the dimensionality considerations based on formula (52), that
\begin{equation}\label{Lphi}        
  L_\varphi\propto\xi_\varphi^{1/z}.
\end{equation}
This quantity is called the dephasing length. Upon approaching the transition point, an increase is observed in not only $\xi$, but also in $\xi _\varphi$ and $_\varphi$. However, the last two quantities are bounded in view of inequality (53). As $\delta x\rightarrow 0$ and for $T={\rm const}\neq 0$, the dephasing length $L_\varphi$ ceases to grow at a certain $\delta x_0(T\,)$. A region is formed in which $\xi$ depends only on $\delta x$, and $L_\varphi$, only on $T$:
\begin{equation}\label{Lphi2}
 \xi=\xi(\delta x)\propto(\delta x)^{-\nu},\qquad
 L_\varphi=L_\varphi(T)\propto T^{-1/z}.
\end{equation}
This region is called critical.

Let us examine the application of the above-formulated general postulates of the theoretical scheme using the concrete example of a system of bosons, which was discussed in Section 2.4. The physical quantities characterizing a boson system can contain both a singular part, which depends on $\xi$ and $\xi _\varphi$, and a regular part, which is independent of $\xi$ and $\xi_\varphi$ [55]. As an example, we take the free-energy density of the quantum system, which corresponds to the free-energy density of an equivalent classical system. At $T=0$, it is defined as
\begin{equation}\label{sca1}        
 f(\mu,J)=
\lim_{T\rightarrow0}\lim_{N\rightarrow\infty}(N/T)^{-1}\ln Z,
\end{equation}
where $\mu$ is the chemical potential, $N$ is the number of particles in the system, and $J$ is the frequency of the boson hoppings between the sites.

The singular part $f_{\rm s}$ of the free-energy density builds up on the scale of the correlation length. Therefore, one has
\begin{equation}\label{sca2}
 f_s\propto\xi^{-(d+z)}\propto(\delta x)^{\nu(d+z)}.
\end{equation}

All coordinate axes of the space with dimensionality ${\cal D}$ are, in principle, bounded, and expression (58) for $f_{\rm s}$ can contain, besides the dimensional coefficient, an arbitrary function of the ratio of the correlation lengths to the appropriate sizes. For the length $\xi_\varphi$, the scale is ${\rm i}\hbar /T$, while for the length $\xi$ this is the smaller of the two values --- the size of the sample and the dephasing length:
\begin{equation}\label{sca3}        
 f_s=(\delta x)^{\nu(d+z)}
 F\left(\frac{\xi}{\cal L},\frac{\xi_\varphi}{i\hbar/T}\right),
 \quad{\cal L}=\min(L,L_\varphi).
\end{equation}
It is usually assumed that the system is infinite in space, so that ${\cal L}$ should be replaced by the dephasing length $L_\varphi$.

In the critical vicinity of the transition point, $\xi_\varphi$ acquires a maximum possible value of $\xi _\varphi ={\rm i}\hbar /T$. Therefore, the second argument of the function $F(u_1,u_2)$ in relationship (59) remains constant in the entire critical vicinity, $u_2=1$, so that $F$ becomes a function of a single variable, namely, the ratio between the lengths $\xi$ and $L_\varphi$:
\begin{equation}\label{sca3a}       
\begin{array}{c}
 \displaystyle f_s=(\delta x)^{\nu(d+z)}F\left(\frac{\xi}{L_\varphi}\right)=
 (\delta x)^{\nu(d+z)}F\left(\frac{\delta x}{T^{1/z\nu}}\right),\\
 (\xi<L).
 \end{array}
\end{equation}
The quantity
\begin{equation}\label{scaVar}
    u=\delta x/T^{1/z\nu}
\end{equation}
is called the scaling variable. From the definition of the critical region, it follows that the equation for its boundary takes on the form $\xi =L_\varphi$, or $u=1$, or
\begin{equation}\label{sca3aa}      
 T=(\delta x)^{z\nu}.
\end{equation}
For certainty, we put the constant coefficient in expression (62) equal to unity.

The arbitrary function of the scaling variable enters into the expressions for any physical quantities in the critical region. Subsequently, we shall be interested in the expression for the conductivity, which in the critical region takes the form [55]
\begin{equation}\label{sca3b}
 \sigma\propto
 (\delta x)^{\nu(d-2)}F_\sigma\left(\frac{\delta x}{T^{1/z\nu}}\right)
 \equiv\frac{e^2}{\hbar}\xi^{2-d}
 F_\sigma\left(\frac{\delta x}{T^{1/z\nu}}\right).
\end{equation}
The last form of the representation of expression (63) explains its physical meaning: the coefficient of the arbitrary function has the dimensionality of conductivity.

In expression (63), it is assumed that the system is sufficiently large:
\begin{equation}\label{sca3c}       
 L\gg L_\varphi.
\end{equation}
Since as $T\rightarrow 0$ the dephasing length $L_\varphi \rightarrow \infty$, at low temperatures inequality (64) can be violated. Then, the measurable quantity ceases to depend on temperature. For example, the resistance, instead of tending to zero (superconductor) or infinity (insulator), comes to plateau with lowering temperature. In the experiment, such a situation happens fairly often. Suspicion in this case usually falls, first of all, on the overheating of the electron system relative to the temperature of the bath. However, the reason can also be the violation of inequality (64) (see, e.g., Ref.~[68] and also Ref.~[69] where the effect of finite dimensions was discussed in detail using a concrete example). We shall run into the saturation of resistance curves at low temperatures in the experiments with Be (see Section 4.2) and then return to this issue in Section 5.1 when examining the experiment that concerned precisely the influence of the size of the system (see Fig.~46 and the associated text).
\begin{figure}[h]
\includegraphics{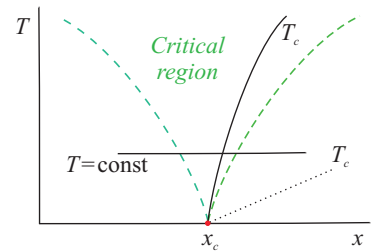}
\caption{Critical region in the vicinity of a quantum superconductor--nonsuperconductor transition (the region boundaries described by Eqn (62) are shown by dashed lines). The solid and dotted lines correspond to two variants of the $T_{\rm c}(x)$ curve, of which one is described by Eqn (65).}
\label{CrRe}
\end{figure}

In the review [7], which was cited at the beginning of this section, it was assumed that the point of a quantum phase transition is an isolated point on the abscissa axis of the phase diagram $(x,T\,)$. This is precisely the case of the metal--insulator transitions. In the case of the superconductor--insulator transitions that are of interest for us here, the point of a quantum phase transition is, on the contrary, an end point of the $T_{\rm c}(x)$ curve of the thermodynamic superconducting transitions at finite temperatures. Let us first assume that in the vicinity of the quantum point $T_{\rm c}(x_{\rm c})=0$ the $T_{\rm c}(x)$ curve finds its way inside the critical region (Fig.~14). Upon intersection of the critical region along the line $T={\rm const}$, the correlation length $\xi$  becomes infinite twice, at points $x=x_{\rm c}$ and $T=T_{\rm c}$. Therefore, the scaling function $F_\sigma (u)$ must exhibit a singularity at a certain critical value $u_{\rm c}$ corresponding to the curve $T_{\rm c}(x)$. Hence it follows that the critical temperature at small $\delta x$ changes in accordance with the equation
\begin{equation}\label{sca3d}       
 T_c=u_c(\delta x)^{z\nu},
\end{equation}
which differs from equation (62) only in a numerical coefficient.

The numerical coefficient in the equation of the boundaries of the critical region has no strict definition. Furthermore, in a sample with infinite dimensions the resistance to the right of the $T_{\rm c}(x)$ line is exactly equal to zero. Therefore, if the curve $T_{\rm c}(x)$ of the thermodynamic superconducting transitions finds its way inside the critical region, then it is expedient to draw the boundary of the critical region precisely along this curve, using Eqn~(65) instead of Eqn~(62).

Generally speaking, the $T_{\rm c}(x)$ curve can pass outside the critical region; this variant is shown in Fig.~14 by a dotted curve. Then, equation (65) is not applicable to this curve.

The application of the above-discussed scheme for describing a critical region to a concrete experiment is given in Section 4.4.

To conclude this section, let us consider the derivative $\kappa =\partial \rho /\partial \mu$  which is frequently called compressibility. Since $\rho=-\partial f/\partial \mu$, the singular part of the compressibility is defined as
\begin{equation}\label{sca4}        
 \kappa_s=-\partial^2f/\partial\mu^2.
\end{equation}
In the transitions corresponding to arrows 1 and 3 in Fig.~9, it is the deviation of the chemical potential of the system from the critical value, $\delta x=\delta \mu$, that can be chosen as the control parameter. Then, using Eqns~(58) and (66), we arrive at the expression for the singular part of the compressibility:
\begin{equation}\label{sca5}
 \kappa_s\propto(\delta x)^{\nu(d+z)-2}.
\end{equation}

For the insulator--superfluid state (and, correspondingly, insulator--superconductor) transitions, we can go further [55] using the condition
\begin{equation}\label{sca6}   
 \delta\mu=\hbar\partial\varphi/\partial t,
\end{equation}
which is equivalent to the well-known Josephson condition. It relates a change in the phase $\varphi$ of the long-wave part of the order parameter for the bosonic system to changes in the chemical potential and suggests that the total compressibility is given by
\begin{equation}\label{sca7}     
 \kappa=-\partial^2f/\partial\mu^2\propto-\partial^2f/\partial\varphi^2.
\end{equation}
Let us expand the free energy into a series in powers of the order-parameter phase. The first term in the series will contain the system density as the coefficient, and the second term the total compressibility, as a result of relationship (69). The third term of the expansion, which is determined by the kinetic energy of the condensate, is proportional to the square of the phase gradient and contains the density of the superconducting component as a coefficient.

Now, let us change the boundary conditions of the system, so that the phase in the space would change by $\pi$, and find the difference between the energy densities of the system after and prior to the change in the boundary conditions:
\begin{equation}\label{sca8}
 \Delta f =f_\pi-f_0.
\end{equation}
The contribution from the first term of the expansion to $\Delta f$ is equal to zero if the boundary conditions are antisymmetric. As the size of the system increases, the third term of the expansion approaches zero more rapidly than the second one. Consequently, one finds
\begin{equation}\label{sca9}        
 \Delta f\propto\kappa/L^2.
\end{equation}
Comparing expressions (59) and (71), we arrive at the following final expression for the total compressibility:
\begin{equation}\label{sca10}
 \kappa\propto(\delta x)^{\nu(d-z)}.
\end{equation}
Changing the phase along the imaginary-time axis, we obtain, using analogous reasoning, the expression for the singular part of the density:
\begin{equation}\label{sca11}
 \rho_s\propto(\delta x)^{\nu(d+z-2)}.
\end{equation}

Since the majority of experimental results for superconductors have been obtained for two-dimensional or quasitwo-dimensional systems, of special importance is the phenomenological theory of superconductor--insulator transitions in two-dimensional superconductors, constructed on the basis of the general theory in the work of Fisher et al. [70, 71]. Its basic ideas will be presented in Section 3.2.

\subsection{3.2 Scaling for two-dimensional systems and the role of a magnetic field}

The superconductor--insulator transitions in two-dimensional superconductors are closely related to the dynamics of magnetic vortices and to the BKT transition. In Section 1.5, we dealt with an ideal system in a zero magnetic field. Now, we introduce disorder and a field, separately or simultaneously. The variety of the variants obtained can be conveniently described using a diagram similar to that given by Fisher [71] (Fig.~15).

\begin{figure}
\includegraphics{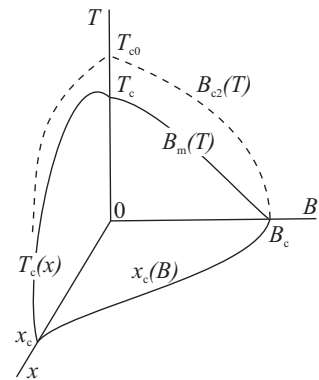}
\caption{Region of superconducting states in the $(x,B,T\,)$ space for a two-dimensional superconductor (the boundaries of the region are shown by solid curves). In an ideal system without a magnetic field $(x=B=0)$, a superconducting transition into a resistive state occurs at a temperature $T_{\rm c0}$; at the temperature $T_{\rm c}$, a BKT transition occurs.} \label{FishDiag}
\end{figure}

Let us first examine the plane $B=0$. A comparatively weak disorder pins (i.e., localizes) the system of vortices-bosons and, according to the Anderson theorem [9], it does not exert a strong influence on the system of 2e-bosons. The fluctuations of the order-parameter phase are suppressed through the pinning of vortices, so that a weak disorder stimulates the establishment of a superconducting state. The temperature $T_{\rm c0}$ is not affected by weak disorder, and the temperature $T_{\rm c}$ can only grow. A strong disorder suppresses the temperature $T_{\rm c0}$ (see Section 2.1). Consequently, $T_{\rm c}$, which is less than or equal to $T_{\rm c0}$, is also suppressed. At a certain critical disorder, $T_{\rm c}$ becomes zero, and the thermodynamic phase transition in the zero magnetic field goes over to a quantum transition.

Let us now return to the region of a weak disorder and switch on a perpendicular magnetic field (plane $x\approx 0$). In a weak field $B\ne 0$, the equilibrium in the vortex--antivortex system is shifted in such a way that the concentration of vortices with the sign corresponding to the direction of the external field prevails: $$N_+-N_-=B/\Phi_0.$$ At a certain strength of the magnetic field, the antivortex concentration $N_-$ becomes zero, and the vortices $N_+$ align into a lattice with a period $\tilde {b}=(B/\Phi _0)^{-1/2}$ [72]. In weak fields at a weak disorder, the vortex lattice is pinned as a whole by pinning centers spaced at a distance $a$, which is much greater than the vortex lattice period [72]:
\begin{equation}\label{2sc1}        
 a\gg\widetilde{b}.
\end{equation}
With an increase in the field strength or disorder, the relative number of pinning centers grows and inequality (74) becomes weaker or is even violated. The disorder `breaks' the vortex lattice, converting it into a vortex glass, and then causes its melting. The melting of the vortex lattice means that the discrete vortices obtain the possibility of moving freely, which leads to dissipation. Thus, the strong magnetic field and strong disorder act on the various types of bosons differently: they suppress the coherent superfluid motion of 2e-bosons, but at the same time delocalize vortices.

Thus, the region of the superconducting states in the diagram shown in Fig.~15 adjoins the origin and is bounded by the surface stretched onto the curves $T_{\rm c}(x)$, $x_{\rm c}(B)$, and $B_{\rm m}(T\,)$. Quantum phase transitions occur along its $x_{\rm c}(B)$ boundary, and thermodynamic superconductor--nonsuperconductor transitions occur on the remaining part of the surface. As can be seen from the curves shown by dashed lines in Fig.~15, a layer of resistive states resides above the region of the superconducting states. In accordance with the BCS theory, 2e-bosons and vortices-bosons coexist in this layer.

The above qualitative picture helps in understanding the origin and meaning of the theoretical model describing the superconductor--insulator transition in two-dimensional superconductors in terms of vortices--2e-bosons duality. The model assumes that the system of 2e-bosons to one side of the phase transition is in the superconductive state, and the vortices are localized, while to the other side it is the electron pairs that are localized and the system of vortices is superfluid [71, 73]. Those who remain unconvinced by the above considerations in favor of duality can find additional arguments in Section 5.1 devoted to Josephson junction arrays (see Fig.~44 and related comments). An additional argument is also the symmetry of the current--voltage characteristics of some systems (connected with superconductor--insulator transitions) relative to the interchange of the current J and voltage V axes. This symmetry is not reduced to the fact that in the superconductor we have V=0, and in the insulator J=0, but is based on more intricate analogies (see Fig.~41 in Section 4.5, and Figs~46b and 46c in Section 5.1).

Let us now turn to a theoretical substantiation of this model. The two-dimensional system under consideration can be described in two alternative languages: the language of charged bosons, and the language of formal quasiparticles (also of the boson type) which carry separate vortices. The Hamiltonian for charged bosons has already been written out above twice, namely, for a regular system of granules [formula (19)] and for bosons at a lattice with a random binding energy at the sites [formula (36)]. Let us now represent this Hamiltonian in a form close to formula (19):
\begin{equation}\label{Ham-boson}       
 \begin{array}{l}
 \widehat{H}=\widehat{H}_0 +\widehat{H}_1,\\
 \widehat{H}_0=
 \sum_{ij}\frac12 V_{ij}(\widehat{n}_i-n_0)(\widehat{n}_j-n_0)+
 \sum_iU_i\widehat{n_i}, \\
 \widehat{H}_1=-J\sum_{ij}\cos(\varphi_i-\varphi_j+A_{ij}^{ext}).
 \end{array}
\end{equation}

Here, the quantity $A_{i\,j}^{\rm ext}$ describes the external magnetic field; it is determined by the difference of the vector potentials of the field at the appropriate sites. This is an additional term with respect to Hamiltonian (19) in which the magnetic field was ignored. The operators $\widehat {n}_i$ of the number of bosons are conjugate to the phase operator: $[\varphi _i\,\widehat {n}_j]={\rm i}\delta _{i\,j}$; $V_{i\,j}$ corresponds to the Coulomb repulsion of bosons at different sites, and $U_i$ corresponds to the random potential changing from site to site, with a zero average value (Hamiltonian (19) contained no corresponding term, since the disorder in that case revealed itself in the spread of the coefficients $J_{i\,j})$. The average number of bosons is equal to $n_0$ and is assumed to be small in comparison with unity. The arrangement of the vortices enters into Hamiltonian (75) through the phase difference $\varphi_i-\varphi _j$ in $\widehat H_1$.

According to Refs~[71, 73], this system can also be described with the aid of a Hamiltonian for an alternative system of quasiparticles:
\begin{equation}\label{Ham-vortex}      
 \begin{array}{l}
 \widehat{H}'=\widehat{H}'_0 +\widehat{H}'_1,\\
  \widehat{H}'_0=
  \sum\limits_{ij}\frac12 G_{ij}(\widehat{\cal{N}}_i-B)(\widehat{\cal{N}}_j-B)+
  \widehat{H}_0[\nabla\times\mathbf{a}]+\rule{0pt}{5mm}\sum_i\wp_i^2, \\
  \widehat{H}'_1=-J'\sum\limits_{ij}\cos(\vartheta_i-\vartheta_j+a_{ij}).
 \end{array}
\end{equation}
Here, $\widehat {\cal N}_i$ is the operator of the number of vortices that is conjugate to the operator of their phase $\vartheta _i$; $G_{i\,j}$ describes the interaction of vortices, and the magnetic field $B$ assigns their average number. The last two terms in the expression for $\widehat {H}^{\,\prime }_0$ contain information on the field {\bf a} created for the alternative quasiparticles by the bare 2e-bosons randomly located at the sites. The first term represents an expression for $\widehat H_0$ entering into Eqn~(75), in which an operator $\nabla\times {\bf a}$ is used instead of the operator $\widehat n_i$, and the `momentum' operators $\wp _i$ are conjugate to the field values ${\bf a}_i$ at the sites.

Certainly, the identity of Hamiltonians (75) and (76) is very conditional. First, the interaction $V_{i\,j}$ between the 2e-bosons occurs according to the Coulomb law, while the interaction between the vortices is logarithmic: $G_{i\,j}\propto \ln r_{i\,j}$. However, this difference can be levelled off by the assumption that the two-dimensional layer possesses a large dielectric constant, so that the electric field of charges is mainly concentrated in this layer. A more essential fact is that the Hamiltonians (75) and (76) do not take into account normal electrons which assist near the vortex axes. These electrons make the motion of vortices in the presence of an external electric field dissipative, which for sure disrupts the possible duality. We shall not discuss under what conditions the difference between the properties of the gas of 2e-bosons and the gas of vortices can be considered unessential, but we shall examine what follows from the application of the theoretical scheme (52)--(63) to the superconductor--insulator transition and what is subject to experimental verification.

Formula (63) for the conductivity in the two-dimensional case takes on the form
\begin{equation}\label{sigma2}      
 \sigma=\frac{e^2}{\hbar}F_\sigma\left(\frac{\delta x}{T^{1/z\nu}}\right).
\end{equation}
This means that the separatrix separating the $\sigma (T\,)$ curves in the regions of the superconductor and insulator is horizontal:
\begin{equation}\label{sigma2a}
 \sigma_c\equiv\sigma(T,x=x_c)=(e^2/\hbar)F_\sigma(0)=\mathrm{const}.
\end{equation}
This assertion is independent of whether the 2e-bosons--vortices duality is realized or not, being a strict consequence of the one-parametric scaling.

Formula (78) looks very simple; nevertheless, a very strong and nontrivial assertion follows from it. The constant $\sigma _{\rm c}$ can be neither zero nor infinity: since the separatrix $\sigma (T\,)=\sigma _{\rm c}$ separates two families of curves in the upper half-plane $\sigma>0$, it must be finite. This assertion is nontrivial for two reasons.

First, the finiteness of $\sigma _{\rm c}$ indicates the presence of a metallic state at the boundary between the superconducting and insulating states. This contradicts the conclusions of the well-known work by Abrahams et al. [74], according to which the system of two-dimensional noninteracting electrons becomes localized at even an arbitrarily low disorder, so that no two-dimensional metal can exist. The problem lies possibly in the fact that the results of Ref.~[74] relate to fermionic systems: it is precisely for these systems that the lowest critical dimensionality, at which the logarithmic corrections lead to localization, is $d_{\rm c}=2$. For bosonic systems, one has $d_{\rm c}=1$.

Second, assertion (78) and the related conclusions do not agree with the results of the calculations of superconducting fluctuations for a two-dimensional superconductor in the dirty limit for $T\ll T_{\rm c}$ and $B\gtrless B_{\rm c2}$ [59]. The absolute value of the separatrix (50) of the set of $\sigma _B(T\,)$ curves that is obtained according to the perturbation theory grows with decreasing temperature (see Fig.~10). The calculations are valid only for $\delta\sigma\ll\sigma$ but, in terms of sense, it is precisely the resistance $1/\sigma _{B_{\rm c2}}$ that must become zero exactly at $T=0$.

The presence of an intermediate metallic state can also be established proceeding from the duality [71]. Let us consider a narrow neighborhood lying to both sides of the transition. In this region, the possibility must exist to write out expressions for the physical quantities in question, relying on any of the two representations. Assuming that both the vortices and the 2e-bosons move in this neighborhood via a diffusive mechanism, we shall use two methods to express the energy that is absorbed by the system of moving 2e-bosons or by the system of moving vortices. The energy $\varepsilon$ absorbed by an individual boson is proportional to the electric field strength $E$ and distance $\xi$  over which the boson preserves the coherence ($\varepsilon \propto E\xi$). Analogously, the expression for the absorbed energy in the case of the vortices contains the product of the current density $j$ determining the Magnus force and the characteristic distance $\xi$  travelled by the vortex in the time of free motion. Let $\varepsilon$ be some function $U$ of this product: $\varepsilon =U(\,j\xi )$. Then, it follows from the identity of the two representations that
\begin{equation}\label{dual1}       
 E\xi\propto U(j\xi).
\end{equation}
When approaching the transition point, where $\xi \rightarrow \infty$, it follows from the condition of preserving the identity that $U(\infty )\rightarrow j\xi$ and
\begin{equation}\label{dual2}
 E\propto j.
\end{equation}
Although both the superconductor and insulator at $T=0$ are nonlinear media exhibiting no linear response, in the boundary state a linear response (80) exists, i.e., the boundary state is metallic.

An even stronger assertion follows from the duality: the conductivity of the boundary state is a universal constant [70, 71, 73] independent of the microscopic structure of the system. Following Ref.~[70], let us write out the dc conductivity in the form of a limiting expression for the frequency-dependent conductivity $\sigma (\omega)$:
\begin{equation}\label{dual2a}      
 \sigma=\lim_{\omega\rightarrow0}\sigma(\omega)=
 (2e)^2\lim_{\omega\rightarrow0}\frac{\rho_s(-i\omega)}{-im\omega}.
\end{equation}
This is a standard trick, which is used, for example, in deriving the Kubo--Greenwood formula. Expression (81) contains only the density of the superconductive (nonlocalized) part of the bosons. The dependence of its limiting value for $\omega \rightarrow 0$ on the changes in the control parameter $\delta x$ is determined by expression (73). Therefore, let us isolate from the density $\rho _{\rm s}(\omega )$ the analogous dependence on $\xi$  explicitly, representing $\rho _{\rm s}(\omega )$ in the form of the product of $\xi ^{\,2-d-z}$ and a certain new function ${\cal R}$ of a dimensionless argument:
\begin{equation}\label{dual2b}      
 \rho_s(\omega)=\xi^{2-d-z}{\cal R}(\omega|\xi_\varphi|).
\end{equation}
Recall that $|\xi _\varphi |^{-1}$ is the characteristic frequency of quantum fluctuations. Relationship (82) for the function $\rho _{\rm s}(\omega )$ indeed goes over into Eqn~(73) as $\omega\rightarrow0$ and at $|\xi_\varphi|^{-1}={\rm const}$ if ${\cal R}(\omega|\xi_\varphi|)$ tends to a constant.

Let us now examine the behavior of the function ${\cal R}(\omega |\xi _\varphi|)$ in the vicinity of the transition point, where $\xi\rightarrow \infty$ and $\xi _\varphi \rightarrow \infty$, while $\omega$ remains constant. Since the density $\rho _{\rm s}(\omega )$ must also remain finite under these conditions, it follows from relationship (82) that in this limit we have
\begin{equation}\label{dual2c}
 {\cal R}(\omega|\xi_\varphi|)=c_d(\omega|\xi_\varphi|)^{(d+z-2)/z},
\end{equation}
where $c_d$ is a universal constant depending only on the dimensionality of the system. Substituting (83) into expression (82) and then into the expression for the conductivity, we obtain the desired formula for the dimensionality $d=2$:
\begin{equation}\label{dual2d}      
 \sigma=c_2(e^2/\hbar).
\end{equation}

Let us also consider the qualitative model reasoning from Ref.~[73]. To this end, we represent a small film with two electrodes in the state close to the superconductor--insulator state as a Josephson element in which phase slips are possible. If the phase at one electrode is assumed to be zero and that at the other electrode is designated as $\varphi (t)$, then, according to the Josephson relation, the potential drop across the electrodes is given by
\begin{equation}\label{dual3}
 V=\frac{\hbar}{2e}\dot\varphi=\frac{\hbar}{2e}\dot n_v,
\end{equation}
where a representation was used, according to which the phase slip is the result of the flow of vortices $\dot n_{\rm v}$ across the electrode line, such that each vortex passed shifts the phase by $2\pi$. Correspondingly, the current $j$ through a film is determined by the flux $\dot n_{\rm c}$ of Cooper pairs from one electrode to another: $j=2e\dot n_{\rm c}$. As a result, the film resistance is defined as
\begin{equation}\label{dual4}
 R=\frac Vj=\frac{2\pi\hbar}{(2e)^2}\left(\frac{\dot n_v}{\dot n_c}\right).
\end{equation}
Since all the vortices in the superconducting state are pinned and $\dot n_{\rm v}=0$, while in the insulator $\dot n_{\rm c}=0$, formula (86) relates only to the boundary state. It remains only to assume that, as a result of the symmetry specified by the duality, in the boundary state the diffusion of each Cooper pair through the system is accompanied by the diffusion of exactly one vortex across the current. Then, the universal resistance of the boundary state is equal to a quantum of the resistance $R_{\rm Q}$:
\begin{equation}\label{dual5}       
 R_Q\equiv
 \frac{2\pi\hbar}{(2e)^2}\approx6.45\;{\rm k}\Omega/\Box\;,
 \quad\mbox{i.e.}\quad c_2=2/\pi.
\end{equation}

As has already been stated, the duality is at best fulfilled only approximately. Therefore, Eqn~(87) should be replaced by the relationship $R_{\rm un}=c_{\rm u}R_{\rm Q}$, where $c_{\rm u}$ is a constant coefficient equal, to an order of magnitude, to unity. This coefficient has repeatedly been calculated. The results obtained depended on the assumptions employed; in particular, $c_{\rm u}=8/\pi$ according to Ref.~[70], 3.51 according to Ref.~[73], 7.1 for short-range repulsive interaction between the bosons, and 1.8 for the Coulomb interaction [75].

The last result, according to which the coefficient $c_{\rm u}$ depends on the nature of interaction, means that $c_{\rm u}$ can depend on the external or internal parameters of an electron system. This means that the magnitude of the resistance $R_{\rm un}$ is not universal, although its value always approximates 10\,k$\Omega/\Box$. In conclusion of this section, let us formulate some questions that the scaling theory of superconductor--insulator transitions in two-dimensional electron systems put to experiment. These questions concern the evolution of the temperature dependence of the dc resistance of thin or ultrathin films with changes in the control parameter $x$. Each subsequent question makes sense only if a positive answer to the preceding one was obtained.

(1) Does there exist for a family of $R_x(T\,)$ curves a separatrix $R_{x_{\rm c}}(T\,)$ that separates the curves for which $R_x(T\,)\,\stackrel {T\,\rightarrow \,0}{\longrightarrow }\,0$ (superconductor) and $R_x(T\,)\,\stackrel {T\,\rightarrow \,0}{\longrightarrow }\,\infty $ (insulator)?

(2) Does the separatrix $_{x_{\rm c}}(T\,)$ have a finite limiting value $R_{\rm un}$ as $T\rightarrow 0$ on the assumption that $R_{\rm un}\neq 0$ and that $R_{\rm un}\neq \infty$ ?

(3) Does the derivative of the separatrix $\partial R_{x_{\rm c}}/\partial T$ tend to zero as $T\to 0$?

(4) What is the magnitude of the coefficient $c_{\rm u}$ and is it stable against changes in the entire family of $R_x$ curves under the effect of some independent parameter $x$?

With the positive answers to questions 1--3, scaling curves (77) can be constructed and the product $z\nu$ of the critical exponents can be determined.

We shall reproduce this procedure using a concrete example in Section 4.1.

\subsection{3.3 Two-parametric scaling}

Everything that was said in Sections 3.1 and 3.2 is directly applicable only to systems with a nonrenormalizable interaction. The meaning of this assertion can be explained for the example of a system of charged particles with a screened Coulomb interaction. If the interaction is independent of the system size $L$ (and, at a finite temperature $T$, of the dephasing length $L_\varphi$), then the system near the transition point follows the laws of one-parametric scaling, i.e., the position of the system in the vicinity of the transition point depends only on one control parameter $x$, so that $\delta x$ in formulas (58)--(60) or (63) depends on neither $L$ nor $T$. In consequence, a separation of the influence of the variables occurs: the correlation length $\xi$  depends only on $x$, and the dephasing length $L_\varphi$, only on $T$. If, on the contrary, it is necessary to take into account the dependence of the interaction on the characteristic dimension of the system, then the scaling becomes two-parametric.

In fact, the Hamiltonians (19), (36), (75), and (76) are constructed in such a way that the renormalizable interaction is neglected in them. Therefore, in all scaling schemes describing superconductor--insulator transitions the formulas of one-parametric scaling are used. The question arises: what will change in the predictions of the scaling theory if we use the scheme of two-parametric scaling?

The difference between one-parametric and two-parametric scaling can be most simply explained with the aid of flow diagrams illustrating the solutions of equations of the renormalization-group theory [76] (see also Refs~[77, 78] for two-parametric scaling). We do this using the example of a metal--insulator transition in two-dimensional systems. The state of the electron system will be characterized by its conductance $y$, considering it to be the only parameter that determines the state of the system in the scheme of one-parametric scaling. For noninteracting electrons, the evolution of the system at a temperature $T=0$ occurs with a change in its size $L$ in accordance with the equation [61, 74]
\begin{equation}\label{2param1}
\frac{d\ln y}{d\ln L}=\beta(\ln y).
\end{equation}
If the system originally has a size $L=\infty$ but is kept at a finite temperature, then the variable $L$ in equation (88) is replaced by the dephasing length $L_\varphi$.

\begin{figure}
\includegraphics{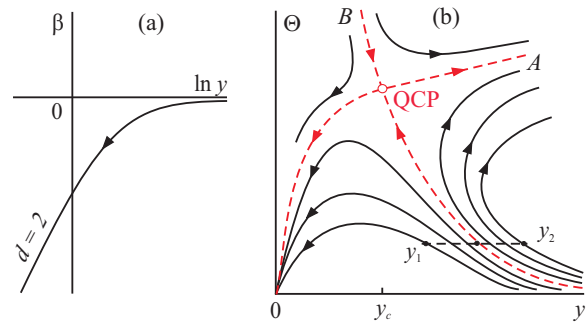}
\caption{(a) Flow diagram of a two-dimensional system of noninteracting electrons described by a single parameter---conductance $y$ [74]. (b)~Part of a flow diagram for a two-dimensional electron gas with interaction constructed in Ref.~[79]; there are two independent parameters: conductance $y$, and interaction $\Theta$; QCP is the quantum critical point (the quantitative relationships calculated in Ref.~[79] have not been retained). The diagram also conditionally shows the new axes A and B directed along the separatrices.}
\label{FlowDiag}
\end{figure}

The function $\beta$ depends on the dimensionality of the system $d$. The fact that at $d=2$ it lies completely in the lower half-plane in Fig.~16a and does not intersect the abscissa axis means that any infinite ($L=\infty$) two-dimensional system of noninteracting electrons at $T=0$ becomes localized. Therefore, no metal--insulator transition exists at all in such a system. For the transition to exist, it would be necessary for the $\beta (y)$ curve to intersect the abscissa axis $\beta =0$ (the interpretation of flow diagrams is considered in more detail in Ref.~[7]).

The possibility of a phase transition appears upon the inclusion of an interelectron interaction $\Theta$, which changes by varying $L$ or $T$. Now, the initial state of the system is determined by two parameters, $y$ and $\Theta$, and two equations appear instead of one equation (88):
(\ref{2param1}) появляются два
\begin{equation}\label{2param2}     
\begin{array}{c} \displaystyle\frac{d\ln y}{d\ln L}=\beta_y(\ln y,\ln\Theta),\:\\
 \displaystyle\frac{d\ln\Theta\rule{0pt}{5mm}}{d\ln L}
  =\beta_\Theta(\ln y,\ln\Theta)\:.
\end{array}
\end{equation}

The solution to equations (89) is sought in the form of some function $\Lambda (\,y,\Theta )$, whose equipotentials $\Lambda (y,\Theta )={\rm const}$ determine the evolution of the system, namely, the changes in the parameters $y$ and $\Theta$ as $L,L_\varphi \rightarrow \infty$; therefore, these equipotentials are called flow lines. It is usually assumed that the motion of the representative point along the flow line begins from the point at which $L=l$ ($l$ is the elastic mean free path). The quantum phase transition is associated with the saddle point of the function $\Lambda$, since the flow lines diverge just near the saddle point. This is clearly seen from the flow diagram shown in Fig.~16b, which was constructed on the basis of the calculated results (see Ref.~[79]) for the model of a two-dimensional system with a multivalley electron spectrum. The abscissa axis in this diagram corresponds to the conductance $y$ of the two-dimensional system of normal electrons; the quantity $\Theta$  plotted along the ordinate axis reflects the effective interaction. The saddle point QCP is the quantum critical point at which the metal--insulator transition occurs. If, initially, the representative point lies in the flow line to the left of the separatrix (point $y_1$), with increasing $L$ or $L_\varphi$ it will approach the line $y=0$, i.e., the system becomes insulating. In the flow lines that lie to the right of the separatrix (for example, beginning from the point $y_2$), the representative point, on the contrary, will move toward larger $y$.
The passage from one flow trajectory to another can be implemented by varying the control parameter. In so doing, the representative point can be placed, in particular, onto the separatrix and then drawn nearer to the QCP by increasing $L$ or $L_\varphi$. As can be seen from the diagram, this motion along the separatrix will change the conductance $y$ of the system: in the case of two-parametric scaling, the finite slope of the separatrix in the set of temperature dependences of $\sigma$ or $R$ of the system of two-dimensional electrons is determined by the angle at which the separatrix in the flow diagram approaches the quantum critical point. This constitutes an essential difference from the case of one-parametric scaling predicting a horizontal separatrix of the temperature dependences of conductivity for quantum transitions in any two-dimensional system, according to relationships (77) and (78).

The occurrence of a metal--insulator transition in the two-dimensional system of normal electrons and the presence of the related inclined separatrix in the set of the temperature dependences of resistance were confirmed in Ref.~[80]. As we shall see in Section 4, the inclined separatrices are also encountered fairly often in the case of superconductor--insulator transitions.

The presence of two independent parameters determining the state of the system changes the entire `system of values' in the vicinity of the transition point. We shall here restrict ourselves to the problem of correlation lengths. By linearizing the set of equations in the vicinity of the saddle point, we can, by replacing the variables $(y,\Theta \rightarrow A,B)$, which involves the rotation and extension of the axes, attain the separation of variables and reduce the set of equations (89) to the form
\begin{equation}\label{2param3}     
\begin{array}{c}
 \displaystyle\frac{d\ln A}{d\ln L}=s_A\ln (A/A_c)\:,\\
 \displaystyle\frac{d\ln B\rule{0pt}{5mm}}{d\ln L}=-s_B\ln (B/B_c)\:,
\end{array}
\end{equation}
where $s_A$ and $s_B$ are the positive numbers.

In the new coordinate system, the motion along the $A$-axis, i.e., along the separatrix $B=B_{\rm c}$, starts from the saddle point $(A_{\rm c},B_{\rm c})$; it is described by the same equation as for the case of noninteracting electrons in a 3D~space [74]. The general solution for the first of equations (90) takes the form
\begin{equation}\label{2param4}
\left|\ln\frac{A}{A_c}\right|=\left(\frac{L}{\xi_A}\right)^{s_A},
\end{equation}
where the correlation length $\xi _A$ depends on the starting point $A_0$ from which the motion along the separatrix toward infinity begins. The nearer $A_0$ is to $A_{\rm c}$, the greater the magnitude of $\xi _A$:
\begin{equation}\label{2param5}
\xi_A\rightarrow\infty\qquad\mbox{при}\qquad
 |\delta A|\equiv|A_0-A_c|\rightarrow0.
\end{equation}
With the motion along the second separatrix $(A=A_{\rm c})$ from the starting point $B_0$ toward the quantum transition point QCP, we have respectively
\begin{equation}\label{2param6} 
\left|\,\ln\frac{B}{B_c}\right| =\left(\frac{L}{\xi_B}\right)^{-s_B}\quad
 \mbox{ и } \quad \xi_B\rightarrow0.
\end{equation}

Thus, since there are two equations in system (90), we had to introduce two correlation lengths. The correlation length $\xi _A$, which corresponds to the effective size of fluctuations, diverges at the transition point, while the corresponding length $\xi _B$, which is connected with interaction, becomes zero:
\begin{equation}\label{2param7}
\begin{array}{l}
 \xi_A\propto|\delta A|^{-\nu_A},\\
 \xi_B\propto|\delta B|^{\nu_B}\rule{0pt}{5mm},
\end{array}
\qquad\nu_A,\;\nu_B>0.
\end{equation}
At an arbitrary point in the vicinity of the point QCP, the physical properties of the system are determined by two correlation lengths, and the function $F$, which was introduced in formula (59), is now a function of four variables, so that the general expression for the conductivity at a zero temperature $T=0$ takes on the form
\begin{equation}\label{2param8}     
\sigma=\frac{e^2}{\hbar}
 F_\sigma\left(\frac{\xi_A}{\cal L},\;\frac{\xi^A_\varphi}{i\hbar/T},\:
 \frac{\xi_B}{\cal L},\:\frac{\xi^B_\varphi}{i\hbar/T}\right)\;,\quad
{\cal L}=\min(L,L_\varphi).
\end{equation}
Here, two additional correlation lengths along the imaginary-time axis appeared:
\begin{equation}\label{2param9}     
\xi^A_\varphi\propto(\xi_A)^{z_A},\qquad\xi^B_\varphi\propto(\xi_B)^{z_B},
\end{equation}
and by the quantity ${\cal L}$, as in Eqn~(59), we imply the size $L$ of the sample at a zero temperature or the temperature-dependent dephasing length $L_\varphi$  for a large-sized system. It is remarkable that the length $L_\varphi$  plays the role of scale with respect to both correlation lengths: $\xi _A$, and $\xi _B$.

\begin{figure}
\includegraphics{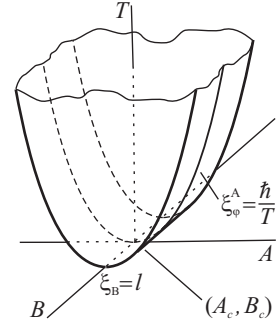}
\caption{Critical vicinity of a saddle point in a flow diagram for two-parametric scaling.}\label{2CritReg}
\end{figure}

When approaching the saddle point $(A_{\rm c},B_{\rm c})$ along the $A$-axis, the length $\xi _A\rightarrow \infty$, and $\xi _\varphi ^{\,A}$ reaches the maximum value $\hbar /T$. Upon approaching the saddle point along the $B$-axis, the length $\xi _B$, which must have been approaching zero, is bounded from below by a certain minimum value, in the simplest case, by the mean free path, $\xi_B\gtrsim l$; simultaneously, $\xi _\varphi ^{\,B}$ also appears to be bounded from below $\xi_\varphi^B\gtrsim l^{z_B}$. Thus, the last three of the four correlation lengths, i.e., $\xi _A$, $\xi _\varphi ^{\,A}$, $\xi _B$, and $\xi _\varphi ^{\,B}$, have constant values in the critical region depicted in Fig.~17, and expression (95) reduces to
\begin{equation}\label{2param10}
\sigma=\frac{e^2}{\hbar}
 F_\sigma\left(\frac{\xi_A}{\cal L},\;T\right).
\end{equation}
In the most interesting case of an infinite system $({\cal L}=L_\varphi )$, we obtain
\begin{equation}\label{2param11}        
\sigma=\frac{e^2}{\hbar}
 F_\sigma\left(\frac{\delta A}{T^{1/z\nu}},\;T\right).
\end{equation}
instead of formula (77).

It is precisely in this form that the scaling relation (97) was used for an analysis of experimental data [78].

Owing to the second argument of the function $F_\sigma$, the behavior of the family of $\sigma(T)$ curves near the quantum phase transition changes. Let us fix the magnitude of the control parameter $A$ in such a manner as to fulfill the condition $\delta A{=}0$. This means that from the entire set of $T\sigma (T)$ curves we chose a separatrix corresponding to the temperature dependence of the conductivity of the boundary state. As is seen from relationship (98), the conductivity remains temperature-dependent, although the first argument of the function $F_\sigma$  remains unaltered, being identically equal to zero. We have already noted this temperature dependence in the analysis of the two-parametric flow diagram in Fig.~16. Given the mechanism of the formation of the dephasing length, the temperature dependence of the separatrix can be calculated analytically [78].

Thus, the existence of a sloped separatrix is a signature of two-parametric scaling.

{\section{4. Experimental}

For the sake of convenience, we shall divide the variety of experiments performed into three groups based on the type and specific features of the material and control parameter. Let these be ultrathin films with a thickness $b$ serving as a control parameter; materials of a variable composition which can be varied in this way or that, and high-temperature superconductors. Such groups will be described in Sections 4.1--4.3. Section 4.4 is devoted to experiments in which the breakdown of superconductivity leads to the formation of a `bad' metal with a negative derivative $\partial R/\partial T$ of the resistance $R$ at low temperatures. Finally, in Section 4.5 we shall consider data on the current--voltage characteristics and nonlinear phenomena in the vicinity of a superconductor--insulator transition.

\subsection{4.1 Ultrathin films on cold substrates}

The general scheme of experiments on ultrathin films deposited on cold substrates is as follows. A substrate with preliminarily applied contacts is placed into the measuring cell of a cryostat; the film is deposited in several small steps, and after each new deposition, the temperature dependence of the resistance is measured. Thus, a whole series of films is obtained and measured in a single experiment without warming the cell to temperatures substantially exceeding liquid-helium temperature. In particular, it is precisely such a procedure that was used in experiments whose results are given in Fig.~3.

Since the interval of effective thicknesses, in which such experiments are performed, varies from 4--5\,\AA\ to several dozen \AA, we can be sure that the study concerns just a two-dimensional object: $d{=}2$. A second advantage of this arrangement of the experiment lies in the fact that we can reach a constancy of all random factors affecting the resistance and clarify the effect of precisely the film thickness on the temperature dependence of its resistance. However, in such an experiment we should avoid the coalescence of atoms into droplets during film deposition, i.e., we should avoid the formation of a granular film. To this end, the cold substrate is most frequently precoated with a layer of amorphous Ge with a thickness $b_0\approx 5\;$\AA, which wets the substrate, remaining by itself amorphous at low temperatures, i.e., it does not impose its lattice period to the overlying film [81].

The important role played by the layer of amorphous Ge is beyond all doubt; it was illustrated, for instance, in Fig.~3. At the same time, the processes that are responsible for the transport properties of ultrathin films and the mechanisms of the influence of the film thickness b are not yet completely clear. Usually, it is assumed that the film thickness $b$ determines the effective mean free path of electrons owing to the diffuse scattering of electrons by the film surface, whereas the Ge layer does not exert a direct effect on the film conductivity. It is, however, possible that, since the thicknesses $b_0$ and $b$ are comparable, the Ge layer affects the electronic spectrum or the effective concentration of electrons in the film under study. There also exist other possible variants of the influence of the Ge layer on electron transport in the main film [82]. At the same time, this is of no consequence, in a sense, since the very fact of the effect of the thickness $b$ is beyond any doubt and quantitatively this effect can be characterized by the resistance of the film rather than by the film thickness. This quantitative characteristic, resistance per square, allows comparing films of different materials.

{\bf Ultrathin bismuth films}. These films appear to have been studied most thoroughly. The phase transition obtained on these films is illustrated in Fig.~18a. The thinnest films behave as insulators; their resistance increases exponentially with decreasing temperature. In the thickest films, a superconducting transition occurs, and its temperature $T_{\rm c}$ lowers with decreasing thickness of the film. The film thicknesses indicated alongside the curves are the average values calculated from the amount of the deposited material and known density of the metal.

The direct vicinity of the phase transition can be examined in some more detail using Fig.~18b, which displays $R(T\,)$ curves for a dense series of weakly differing states. Generally speaking, some additional information is required to assume that the resistance of the states lying inside the interval marked by arrows on the right-hand side of Fig.~18b tends to infinity as $T\rightarrow 0$, i.e., that these states can be considered to be insulating and the insulator--superconductor transition to be unsplit (cf. the experimental data for the three-dimensional  Nb--Si system shown in Fig.~1). To investigate this question, the temperature dependence of the conductance of those states that exhibited no superconducting transition was studied. It turned out [85, 86] that the conductance $y$ for the thinnest films changes according to the Arrhenius law:
\begin{equation}\label{expT}        
y=y_0\exp(-T_0^I/T),
\end{equation}
and that with increasing film thickness $b$ this dependence is replaced first by the Shklovsky--Efros law:
\begin{equation}\label{expT2}
y=y_0\exp[(-T_0^I/T)^{1/2}],
\end{equation}
and then by a logarithmic dependence:
\begin{equation}\label{lnT}     
y=y_0-\ln(T/T_0^I).
\end{equation}

All three formulas (99)--(101) contain a parameter $T_0^{\,\rm I}$ with the dimensionality of temperature (superscript 'I' indicates that here we are dealing with the insulator side). This makes it possible to bring all experimental points together to a single $y(T/T_0^{\,\rm I})$ curve with the aid of a simple procedure. For the $y(T/T_0^{\,\rm I})$ curve represented in the $(\ln T,g)$-axes, the change in $T_0^{\,\rm I}$ is reduced to a shift of the curve along the abscissa axis. Therefore, starting from the $y(T/T_0^{\,\rm I})$ curve for the thinnest film and merging each following curve with the preceding one by means of a parallel translation, we can, by gradually enlarging the range of variability of the argument  $T/T_0^{\,\rm I}$, construct a curve of the universal function $y(x)$, in which all experimental points lie and which is described to a good accuracy in the different segments by formulas (99)--(101).

The parameter $T_0^{\,\rm I}$ proves to be a monotonic function of the film thickness $b$ or the film resistance $R^{\,*}$ measured at a certain fixed temperature $T^{\,*}$ that exceeds the superconducting transition temperature. In an analogous way we can also proceed on the superconductor's side of the transition, selecting there the parameter $T_0^{\,\rm S}$ in such a way that all experimental $y_b(T\,)$ curves become merged into a single curve $y(T/T_0^{\,\rm S})$. By building up both functions, $T_0^{\,\rm I}(R^{\,*})$ and $T_0^{\,\rm S}(R^{\,*})$, we shall see that they have a singularity at one and the same value $R^{\,*}=R_{\rm c}^{\,*}$ corresponding to the resistance of the separatrix at a temperature $T^{\,*}$ (Fig.~18c). This means that we approach one and the same value of the critical thickness $b_{\rm c}$ from both sides, i.e., the phase transition is unsplit.

\begin{figure}
\includegraphics{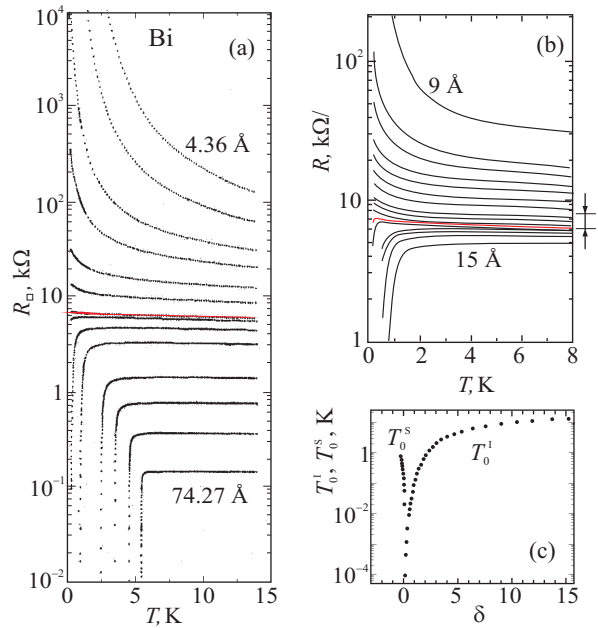}
 \caption{(a) Evolution of the temperature dependence of the resistance of amorphous Bi films deposited on a Ge underlayer of thickness $b_0=6\,$\AA\  with film thickness $b$ increasing from above downward [83]. (b) Central part of an analogous series of $R(T,b{=}{\rm const})$ curves on an enhanced scale and with a smaller step in $b$ on the interval of thicknesses $b$ from 9 to 15\,\AA; the films were deposited over a Ge underlayer 10\,\AA\  thick onto the (100) surface of an SrTiO$_3$ single crystal [84]. (c)~Characteristic scaling parameters $T_0^{\,\rm I}$ and $T_0^{\,\rm S}$ on each side of the quantum phase transition; the quantity $\delta$  plotted on the abscissa represents normalized difference between the film resistance $R^{\,*}$ and the critical value $R_{\rm c}^{\,*}$ at $T^{\,*}=14\,$K: $\delta=(R^{\,*}-R_{\rm c}^{\,*})/R_{\rm c}^{\,*}$ [85].
}
 \label{Al+BiScal}
\end{figure}

The above-described empirical procedure for determining $T_0^{\,\rm I}(R^{\,*})$ is valid for the entire insulator's domain rather than for the critical vicinity of the quantum transition. In the critical vicinity, the theory poses limitations both on the form of the $y(T\,)$ functions and on the scaling variable. According to the theory [70, 71] constructed on the basis of the general theory of quantum transitions and the model of dirty bosons, the separatrix that separates the $R(T\,)$ curves with a superconducting transition (Fig.~18a,b) from the curves in which there is no transition, should pass horizontally in two-dimensional systems, and on both sides of the separatrix the sign of the derivative along the curves should remain constant, although different in the regions of the superconductor and insulator (see the end part of Section 3.2). In essence, it is precisely the fulfilment of these two conditions that symbolizes the first level of agreement with the theory and makes it possible to carry out the scaling procedure and to determine critical exponents in accordance with formula (77). In connection with the experiments discussed, it is expedient to write out the latter formula as
\begin{equation}\label{2Scaling}        
 R(\delta,T)=R_cF(\delta x/T^{1/\nu z}),\qquad \delta x=|b-b_c|.
\end{equation}
As follows from Fig.~18a,b, the results of experiments performed on Bi films seem ideal from the viewpoint of these conditions. The separatrix $R_{b_{\rm c}}(T\,)\equiv R(T,b=b_{\rm c})$ is indeed horizontal, as is predicted by the theory [70, 71]. The processing of the results using formula (102) indeed makes it possible to bring together all the experimental points from the curves given in Fig.~18b to two branches of a single scaling curve (Fig.~19a) and to determine the product $\nu z$ of the critical exponents.

The representation of the temperature dependence of resistance $R(T\,)$ near the transition point on an extended scale (Fig.~19b) shows that even in bismuth the separatrix is only approximately horizontal, and in other materials the situation is even worse. In Al and Pb films, the transition manifests itself but the separatrix is sloped. For example, in Al films the resistance along the separatrix changes by a factor of at least 1.5, from 20 to 30\,k$\Omega$, as the temperature decreases from 15 to 1\,K. This diminishes the accuracy of the scaling procedure.

\begin{figure}
\includegraphics{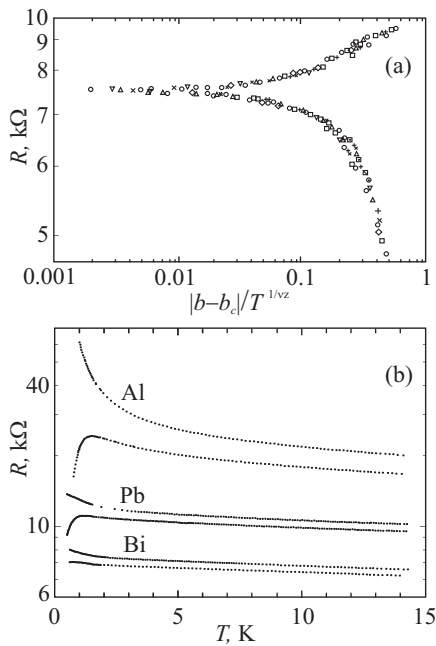}
 \caption{(a) Resistance of ultrathin Bi films [curves $R(T\,)$ from Fig.~18b] as a function of the scaling variable in the temperature range from 0.14 to 0.50\,K. Different data symbols correspond to different temperatures [84]. (b) Temperature dependence of the resistance of the last insulating and first superconducting films in the sequences of Bi, Pb, and Al films with a gradually increasing thickness $b.$ In the case of Al films, the slope of the separatrix is seen clearly [85].}
 \label{Al+BiScal}
\end{figure}

However, the agreement is worse even for Bi films at the following level of comparison with the results of the theory. According to the scaling theory of `dirty bosons', the metallic state corresponding to the separatrix must possess a universal value of the resistance, $R_{\rm un}$, which is on the order of the resistance $R_{\rm Q}$ given by formula (87). The resistance $R_{\rm un}$ must be insensitive to the microscopic features of a concrete system and must depend only on the universality class of the transition. The universality of $R_{\rm un}$ allows, in principle, calculating the proportionality factor $c_{\rm u}=R_{\rm un}/R_{\rm Q}$ by taking advantage of some comparatively simple model. Some different numerical values of $c_{\rm u}$ were given in Section 3.2 after formula (87). Whatever the true value of $R_{\rm un}$, it must be reproduced in different experiments. However, the values of $R_{\rm un}$ obtained in various laboratories using different materials and distinct intermediate underlayers (e.g., amorphous Si or Sb instead of Ge) had approximately a twofold spread, from 6 to 12\,k$\Omega$. The universality of the critical resistance $R_{\rm un}$ is confirmed only to an order of magnitude.

As has already been noted, the bosonic scenario for the quantum phase transition assumes the existence of electron pairs on both sides of the superconductor--insulator transition. This means that the modulus $\Delta$  of the order parameter at the transition, i.e., the energy of pairing or the width of the superconducting gap, must not become zero; the superconductivity must be destroyed as a result of strengthening fluctuations of the order-parameter phase. This was checked on films of several metals (Pb and Sn in Ref.~[87], Bi in Refs~[88, 89]) using tunnel experiments which make it possible to measure the density of states as a function of energy near the Fermi level. The first example of such measurements is given in Fig.~20. From the value of the tunneling voltage $V_0$, at which the normalized tunneling conductivity $G_{\rm N}$ is equal to unity, it is possible to estimate the superconducting gap: $\Delta \approx eV_0$. When approaching the transition, i.e., as the film thickness $b$ approaches a critical value $b_{\rm c}$, the superconducting gap $\Delta$  behaves as usual: it tends to zero, remaining proportional to the transition temperature, $\Delta \propto T_{\rm c}$.

At the same time, according to Fig.~20, a certain uncommonness in the behavior of the system is nevertheless observed: the density of states, which is proportional to $G_{\rm N}$, does not vanish inside the gap. This is apparently due to the fact that the method of measuring density of states with the aid of a tunneling junction is integral, and over the area of the contact there are a large number of vortices of both signs, inside which the superconductivity is destroyed and the electrons are normal [90]. As can be seen from Fig.~20, the averaged density of states at the Fermi level (in the middle of the gap) becomes greater as the state of the system gets nearer to the transition point.

Since the moving vortices cause fluctuations of the order-parameter phase, it is clear that as the transition is approached the amplitude of these fluctuations grows rapidly, simultaneously with a decrease in the gap width. However, it is difficult to say what occurs earlier, i.e., whether this is the growth of the fluctuations in the phase or the decrease in the gap width. In any case, no signs of the retention of the gap (modulus of the order parameter) has been found in this experiment at the transition in the absence of a magnetic field.

An increase in the number of vortices indicates an increase in the number of normal electrons near the vortex axes (in their cores). Until we deal with ultrathin films, it makes no sense to discuss whether the electrons in the vortex core are localized or quasifree relative to the motion along the vortex axis, whereas in the case of thicker films and three-dimensional systems such a question appears to be appropriate.

Let us now go over to the description of experiments with another control parameter --- a magnetic field.
\begin{figure}
\includegraphics{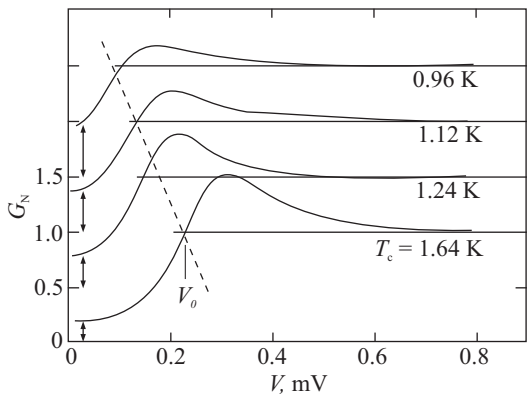}
 \caption{Normalized differential tunneling conductivity for four ultrathin Bi films of different thicknesses $b$ on a Ge substrate [89]. Alongside the curves, the superconducting transition temperatures determined from the $R_b(T\,)$ curves are indicated. For clarity, the curves are moved upward, each by 0.5, relative to the preceding one; the figures given at the ordinate axis correspond to the lower curve recorded for the film with the superconducting transition temperature $T_{\rm c}=1.64\,$K; $V_0$ is the voltage across the tunneling junction, at which $G_{\rm N}=1$. The arrows show the residual density of states in the middle of the gap. The dashed line is drawn to make more visual the gradual decrease in the gap width with approaching the transition point.} \label{Valles0}
\end{figure}

The superconducting transition in a film with a thickness $b>b_{\rm c}$ can be destroyed by a magnetic field. In this case, we obtain a superconductor--insulator transition with a magnetic field as the control parameter. The thicker the film, the less its normal resistance and the stronger the critical field. The application of a magnetic field gives the possibility of conducting several fundamentally new experiments on ultrathin Bi films and, in particular, to compare the different ways of approaching a superconductor--insulator transition.

Figure 21, which was taken from Ref.~[89], demonstrates the magnetic-field-induced behavior of an ultrathin Bi film with a thickness $b>b_{\rm c}$, which, without a field, appears to be superconducting with a transition temperature of 1.64\,K. All $R_B(T\,)$ curves can be more or less unambiguously divided into two groups: those that demonstrate a tendency toward the establishment of a superconducting state with decreasing temperature $(R\rightarrow 0$ as $T\rightarrow 0$), and those for which it is possible to assume that $R\rightarrow \infty$  as $T\rightarrow 0$. The strength of the field $B_{\rm c}$ in which a separatrix is obtained that separates these two groups of curves with different asymptotic behaviors is considered as critical. In Fig.~21a, according to this definition, one has $B_{\rm c}\approx 2.5\,$T.

Let us compare the $R_B(T\,)$ resistance curves given in Fig.~21a with analogous $R_b(T\,)$ curves plotted in Fig.~18a,b. In the last figure, the superconducting transition temperature decreases as the critical thickness $b_{\rm c}$ is approached, whereas in the case of similar $R_B(T\,)$ curves shown in Fig.~21a with the magnetic field in the role of the control parameter {\it no visible shift in the temperature of the superconducting transition onset is observed}, but at temperatures less than that of the transition onset, an increase in dissipation is observed with increasing the magnetic field strength.

\begin{figure}[t]
\includegraphics{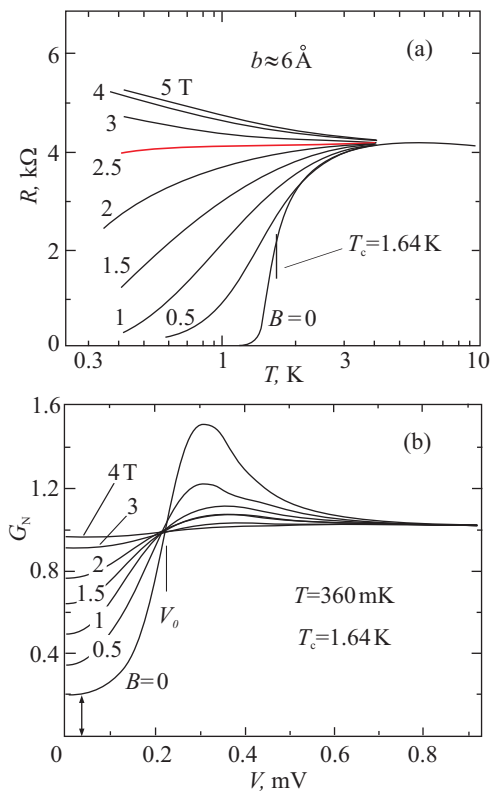}
 \caption{(a) Temperature dependences of the resistance of an ultrathin Bi film in magnetic fields of up to 5\,T [89]; the temperature of the superconducting transition in the zero field is arbitrarily determined as the value of $T$ at which the resistance decreases to half the normal value. (b)~Normalized differential tunneling conductivity of the same film, which demonstrates the evolution of its density of states in a magnetic field in the vicinity of the Fermi level [89]. The arrow indicates the residual density of states inside the gap at $B=0$ (cf. Fig.~20). } \label{BiValles}
\end{figure}

The comparison of the behavior of resistance curves can be supplemented by the comparison of tunnel characteristics (cf. Figs~20 and 21b). It is seen from Fig.~21b that an increase in the field strength leads to an increase in the density of states in the center of the gap. This, in general, is natural: the strengthening of a field should lead to an increase in the density of vortices. According to the theoretical calculation of the density of states averaged over a cell of the vortex lattice for the `conventional' (far from the superconductor--insulator transition) type-II superconductor [90, 91], in the middle of the gap we have $G_{\rm N}\approx B/B_{\rm c2}$. Near the transition, however, there is a finite density of states in the zero field; in Fig.~21b, it amounts to approximately 0.2, just as in Fig.~20 for the same film. At the same time, the density of states is $G_{\rm N}\approx 0.9$ in a field $B_{\rm c}\sim 2.5\,$T, when the superconductivity should seemingly be destroyed; it becomes close to unity in a field of approximately 4\,T, at which the resistance curve $R_{4{\rm T}}(T\,)$ demonstrates a clear tendency toward growth with decreasing temperature (Fig.~21a).

One more specific feature of tunnel characteristics seen in Fig.~21b is that all of them intersect the straight line $G_{\rm N}=1$ at the same point corresponding to the voltage $V=V_0$ across the junction. It is usually assumed that this point of intersection determines the magnitude of the gap. Thus, the tunneling experiment shows that the magnetic field in ultrathin superconducting bismuth films {\it does not suppress the gap $\Delta$, as usually occurs in superconductors, but leads to a growth of the density of states inside the gap and to a decrease in the coherent peak.}

This behavior of the superconducting gap with strengthening magnetic field, discovered in Ref.~[89], is a serious argument in favor of the idea that the modulus of an order parameter can remain finite on the insulator side of the transition as well, taking account, instead of the superconducting gap, of the energy of the pairwise superconducting correlations of the localized electrons. As the magnetic field aligns electron spins due to the Zeeman effect, the correlation energy has to decrease, and the hopping conductivity to increase. For films with a thickness $b>b_{\rm c}$, we can therefore expect the appearance of a negative magnetoresistance in strong fields at low temperatures. In the series of the $R_B(T\,)$ curves in Fig.~21a this could manifest itself in the low-temperature intersection of curves recorded in strong magnetic fields. However, as can be seen from Fig.~21a, no such intersection occurs in bismuth films in fields of up to 5\,T, although, it cannot be excluded that fields below 5\,T are insufficiently strong.

By detecting transitions on films of different thicknesses and tracing the dependence of the critical resistance $R_{\rm c}$ at transitions in the presence of a magnetic field on the thickness $b$ using one and the same series of films, it is possible to verify the universality of the critical resistance in one experiment [92]. From the theoretical viewpoint, the value of $R_{\rm c}$ at transitions in a field can differ from those at transitions in the absence of a field, but it should nevertheless be universal as well. However, it turned out that $R_{\rm c}$ at transitions in a magnetic field substantially varies with a change in the thickness of the film, despite the fact that all the transitions a fortiori relate to one and the same class of universality, and the poorly controlled experimental factors are identical. With an increase in $b$, a decrease is revealed in the normal resistance of the film, a strengthening of the critical field, and a decrease in the critical resistance $R_{\rm c}$, although, according to Fig.~21, the normal resistance of the bismuth film and its critical resistance $R_{\rm c}$ in a magnetic field are virtually coincident. This is, however, not the case in beryllium.

{\bf Ultrathin beryllium films.} Beryllium can be deposited onto a cold polished surface in the form of a continuous film even in the case of a very small effective thickness and even in the absence of an amorphous Ge underlayer [93]. Then, the superconducting transition temperature of amorphous Be films can reach 10\,K, although the transition temperature of crystalline Be is less than 30\,mK. [In this sense, Be behaves similarly to Bi: an amorphous Bi film deposited onto a cold substrate can become superconducting at a temperature above 5\,K (see Fig.~18a), whereas crystalline Bi does not exhibit superconductivity at all.]

\begin{figure}
\includegraphics{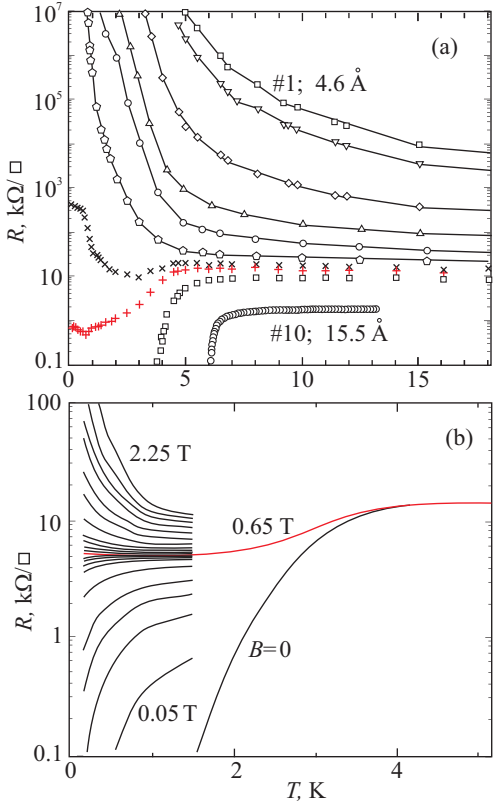}
\caption{Superconductor--insulator transition in ultrathin Be films. (a) Variation of the temperature dependence of the resistance of amorphous Be films upon increasing their thickness $b$; the numbers (1--10) alongside the films correspond to the number of sequential depositions; near curves 1 and 10, the effective thickness of the corresponding films is indicated [94]. (b) Temperature dependence of the resistance of ultrathin Be film with an effective thickness $b$ intermediate between the thicknesses of films 8 and 9 in magnetic fields from zero to 2\,T [95, 96].} \label{Be1}
\end{figure}

As is seen from Fig.~22, the ultrathin Be films also display superconductor--insulator transition. Both the film thickness $b$ and the magnetic field $B$ can serve here as the control parameter. The family of resistance curves in a zero field at different thicknesses differs only a little from the appropriate curves for Bi (the nonmonotonic behavior of two curves in the immediate vicinity of the transition point is probably due to some factors related to experimental conditions, for example, an overheating of the sample or the presence in it of a certain characteristic small size limiting the coherence length [see inequality (64)]. These two families of curves differ significantly only in the values of the critical resistance.

The family of resistance curves in a magnetic field (Fig.~22b) deserves more attentive consideration. If we limit ourselves to temperatures $T{<}1.6\,$K, we can easily single out a clearly pronounced horizontal separatrix in the field $B\approx 0.65\,$T. However, this separatrix has an additional rise for $T{>}1.6\,$K, which apparently also refers to superconductivity. Indeed, for $T{>}3.8\,$K the resistance $R$ is independent of the magnetic field or depends on it very weakly, but for $T{<}3.8\,$K a strong field dependence appears. This dependence is naturally explained by the influence of the field on the superconducting fluctuations or on the equilibrium superconducting state. Therefore, we should consider the temperature $T_{{\rm c}B}{\approx}3.8\,$K corresponding to the onset of transition as the representative temperature of the superconducting transition in the film under consideration; if we utilize the traditional method and select the temperature $T_{\rm c}$ at which the resistance in the zero field decreases by a factor of two (or, for example, by 10\%) relative to the normal resistance, then we should obtain a value of approximately 2.5\,K.

The separatrix of the family of curves presented in Fig.~22b is characterized by two different values of the resistance, namely, by the normal resistance $R_{\rm N}=10.7\;{\rm k}\Omega /\Box$, and by the quantum critical resistance $R_{\rm c}=4.4\;{\rm k}\Omega /\Box$. Other ultrathin Be films behave analogously. As the film thickness increases, $R_{\rm N}$ gradually decreases to 5.6\,k$\Omega /\Box$  and the critical resistance $R_{\rm c}$ grows to 7.8\,k$\Omega /\Box$ [95]. The critical resistance is $R_{\rm c}<R_{\rm N}$ in some films, and $R_{\rm c}>R_{\rm N}$ in other (thicker) ones. The thermodynamic superconducting transition in the absence of a field, with a $T_{\rm c}$ of about 3\,K, and the quantum phase transition in the magnetic field are in no way connected with one another. In this case, the quantum transition corresponds well to the one-parametric scaling scheme: the separatrix is strictly horizontal and all the data fall in a single curve upon their processing with formula (102). However, the duality of the 2e-bosons--vortices system is not realized: the resistance $R_{\rm c}$ depends on the film thickness.

\begin{figure}
\includegraphics{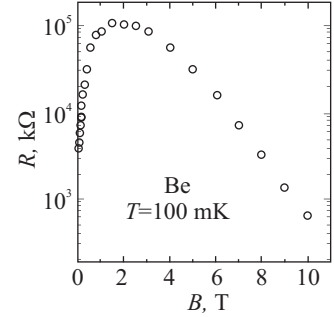}
\caption{Field dependence of the resistance of a Be film with a thickness slightly less than critical [99]. } \label{Be-R(b)}
\end{figure}

According to what was said in Sections 1.4, 2.2, 2.4, and 2.5, we can expect the occurrence of a transition to a Bose insulator in Be with the localization of electron pairs at the transition to the insulating state and their subsequent decomposition and delocalization of electrons in a strong magnetic field. This process must be accompanied by emergence of a significant negative magnetoresistance similar to that observed in granular films (see Figs~4 and 5). This process was not revealed in Bi films (see, however, the comments on Fig.~21a), but it was observed on a giant scale in InO and TiN films (see Section 4.2). In beryllium, the giant negative magnetoresistance was indeed observed [97--99], but only in the high-resistance films lying in the diagram in the insulator region, although comparatively close to the superconductor--insulator transition. Figure 23 gives an example of the field dependence of the resistance measured at a temperature of 100\,mK. The initial resistance of the film, which was equal to about 4\,M$\Omega$, first increased by almost two orders of magnitude with increasing field strength and then decreased by almost three orders of magnitude.

Thus, superconductivity in beryllium is observed in low-resistance (thicker) films, while giant negative magnetoresistance occurs in high-resistance (thinner) films, so that the relation between and the common nature of these phenomena are not evident {\it a priori}. However, a thorough analysis of indirect data (see Ref.~[100]) suggests that the negative magnetoresistance of the high-resistance films is due to precisely the decomposition of localized pairs in the Bose insulator that was formed in weak magnetic fields.

Thus, the experiments on ultrathin films, which are undoubtedly two-dimensional systems, showed that:

(a) the separatrix separating the resistance curves $R(T\,)$ at different thicknesses, which refer to the superconductor and the insulator regions, is horizontal in Bi and Be films, although it has generally a finite slope, ${\rm d}R/{\rm d}T\neq 0$, as $T\rightarrow 0$ in films of other metals (Fig.~19b);

(b) the critical resistance $R_{\rm c}$ takes on various values in films of different materials, on different substrates, etc., although these values differ by no more than a factor of two either way from $R_{\rm Q}=\hbar /(2e)^2\approx 6.45\,$k$\Omega$; in transitions occurring in a magnetic field, $R_{\rm c}$ varies with varying film thickness in films of one and the same series [92];

(c) in transitions occurring in a magnetic field used as the control parameter, the critical resistance $R_{\rm c}$ can differ from the film resistance $R_{\rm N}$ in the normal state (Fig.~22b);

(d) if the role of the control parameter is played by the film thickness $b$, then both the transition temperature (see Figs~18, 22a) and the gap width (see Fig.~20) decrease as the transition point is approached; in this case, the gap `becomes overgrown' gradually: a finite density of states appears in it, which increases with the approaching transition (see Fig.~20);

(e) if the film thickness $b$ is close to the critical value $b_{\rm c}$, then the magnetic field does not shift the onset of the resistive transition toward lower temperatures (Fig.~21a) and does not suppress the gap or, to be exact, does not decrease the value of the characteristic energy $\Delta$ in the spectrum, but gradually increases the density of states inside the gap (Fig.~21b); it should be noted, however, that the last result was obtained only on Bi films.

\subsection{4.2 Variable-composition materials}

Alloys and compounds whose electrical properties are determined by deviations from the stoichiometric composition or by special type of defects compose a special type of materials in which, in principle, superconductor--insulator transitions can be observed. Usually, these are films with a thickness on the interval
\begin{equation}\label{b}   
  100\,\mbox{\AA}\lesssim b\lesssim 2000\,\mbox{\AA}.
\end{equation}
Due to the thickness limitation from below, the state of the film is low-sensitive to the state of the boundary and electron scattering by the boundary. In thicker films, problems are expected with the homogeneity of the concentration distribution of the constituent elements or vacancies and with the formation of granules --- hence, the limitation on $b$ from above. The dimensionality of the electron systems in such films should be interpreted with caution: the electron mean free path $l$ in them is usually less than $b$; the London penetration depth $\lambda$ determining the diameter of vortices is greater than $b$, and the superconductive coherence length $\zeta$  is comparable to $b$.

It is precisely such properties that are characteristic of amorphous In--O films, the description of experiments on which we turn to now.

{\bf Amorphous In--O films.} Upon the deposition of high-purity $\rm In_2O_3$ on an $\rm SiO_2$ substrate using electron-beam sputtering in a vacuum, an amorphous ${\rm InO}_x$ film is formed without crystalline inclusions with a certain oxygen deficit $q=1.5-x$ [101, 102]. The concentration $q$ of vacancies, which act as donors, depends on the residual oxygen pressure in the vacuum chamber during sputtering. In small limits, $q$ can be additionally changed by means of a soft annealing at a temperature no higher than $50\,^\circ$~C; the annealing in a vacuum increases $q$, while the annealing in air, on the contrary, decreases it. The oxygen deficit, in turn, determines the concentration n of electrons that do not participate in chemical bonds between In and O atoms. These electrons can be either localized under the action of the random potential of the amorphous material or be delocalized if $n$ is sufficiently large, $n>n_{\rm c}$. At low temperatures, the system of delocalized electrons becomes superconducting.
\begin{figure}
\includegraphics{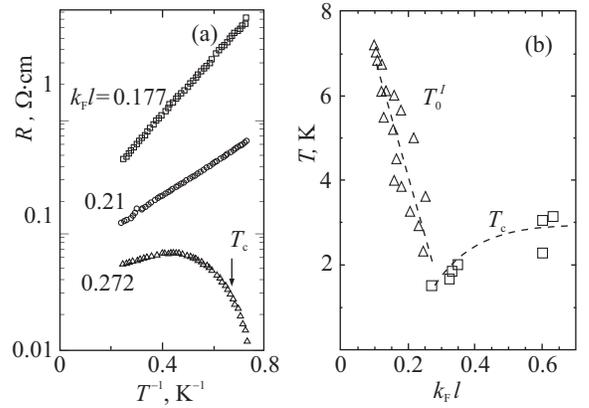}
\caption{(a) Temperature dependence of the resistance of an amorphous In--O film 2000\,\AA\  thick in three states: immediately after deposition (upper curve) and after two subsequent heat treatments [101]; the arrow above the lower curve indicates the superconducting transition temperature $T_{\rm c}$ at which the resistance is half the maximum. (b) Dependence of the activation energy $T_0^{\,\rm I}$ of insulating films and of the transition temperature $T_{\rm c}$ of superconducting films on the parameter $k_{\rm F}l$ [101].} \label{ShahaOv}
\end{figure}

Figure 24a displays the temperature dependence of the resistance of an amorphous In--O film in three different states [101]. For the quantitative characterization of the film states, the product $k_{\rm F}l$ was employed, which was determined at room temperature from the data on the film resistance and the Hall effect in the films. This product, in general, takes into account both the electron concentration $n=k_{\rm F}^{\,3}/3\pi ^2$ and, through the electron mean free path $l$, the degree of disorder. When the state of the metal is close to the localization threshold, the parameter $k_{\rm F}l$ becomes less than unity. In this region, $k_{\rm F}l$ still can be used for the characterization of the film, although $l$ cannot yet be considered as the mean free path.

\begin{figure}
\includegraphics{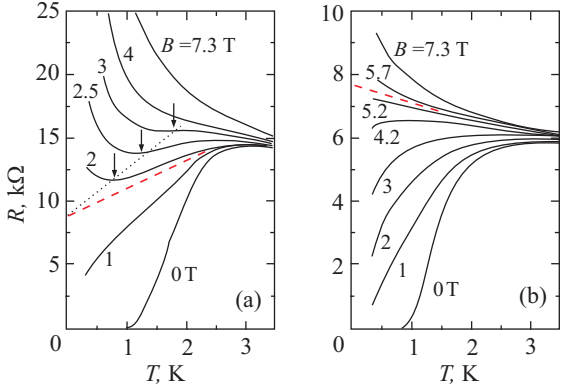}
\caption{Temperature dependence of resistance of two amorphous In--O films 200\,\AA\  thick in different magnetic fields. The separatrices of both families are shown by dashed lines. In figure (a), the critical value $R_{\rm c}$ (intercept of the separatrix with the ordinate axis) is determined by the extrapolation of the positions of the minima marked by the arrows in the $R(T\,)$ curves (dotted line) [105, 106].}
\label{G1}
\end{figure}

The fact that the two measured functions $R(T\,)$ plotted in Fig.~24a on the $(1/T,\ln R)$ coordinates are straight lines with different slopes means that in the appropriate states the resistance changes according to the law $R=R_0\exp (T_0^{\,\rm I}/T\,)$ with different activation energies $T_0^{\,\rm I}$. In the lower curve, a superconducting transition is observed, whose temperature $T_{\rm c}$ is indicated by an arrow. As can be seen from Fig.~24b, no gap is seen on the abscissa axis between the $T_0^{\,\rm I}(k_{\rm F}l\,)$ and $T_{\rm c}(k_{\rm F}l\,)$ functions. This means that the transition is unsplit (cf. the data for the Nb--Si system in Fig.~1, and for amorphous bismuth in Fig.~18). Although these graphs correspond rather to the phase diagram given in Fig.~2c, the existing accuracy does not make it possible to exclude the variant displayed in Fig.~2b.

Basic experiments on the quantum phase transition in In--O films were conducted in a magnetic field. Their results can be presented in two forms: as a series of $R_B(T\,)$ curves recorded in different magnetic fields, or as a set of isotherms $R_T(B)$. If the series of $R_B(T\,)$ curves possesses a horizontal separatrix $R_{B_{\rm c}}(T\,)=R_{\rm c}$, i.e., the transition can be described within the framework of one-parametric scaling, then the isotherms $R_T(B)$ intersect at one point with an abscissa $B=B_{\rm c}$. In some series of experiments with weak critical fields $B_{\rm c}$, it is precisely this that is the case [103, 104, 106]. In other experiments, for example, in the absence of a magnetic field (see Fig.~24), as well as in the case of strong critical fields $B_{\rm c}$, the separatrix of the families of curves $R(T\,)$ has finite slope, which can be of different signs (Fig.~25). Earlier, the positive slope of the separatrix, $\partial R/\partial T>0$, as in Fig.~25a, was considered to be the indication of the presence of granules and macroscopic inclusions. After the appearance of paper [59], grounds appeared to consider that this separatrix can, on the contrary, indicate the absence of macroscopic characteristic lengths in the random potential. In any case, it can be asserted that the schemes of one-parametric scaling are insufficient for describing these experiments.

\begin{figure}
\includegraphics{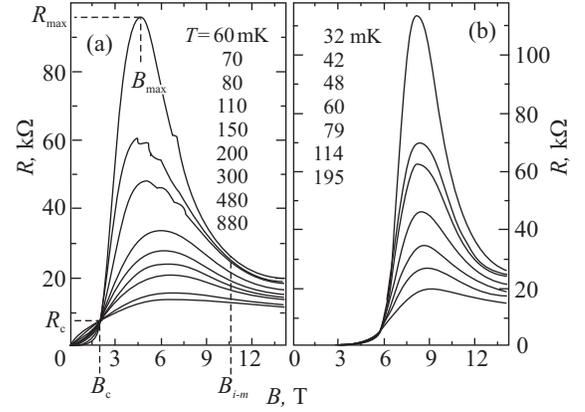}
\caption{Families of $R(B)$ isotherms for two close states of an amorphous In--O film 200\,\AA\  thick in fields (a) perpendicular and (b) parallel to the film [108]. The position of the point of intersection of the isotherms in figure (a) determines the critical values of $R_{\rm c}$ and $B_{\rm c}$; the meaning of the field $B_{\rm I-M}$ is explained in Fig.~27 and in the main text.} \label{G2}
\end{figure}
\begin{figure}
\includegraphics{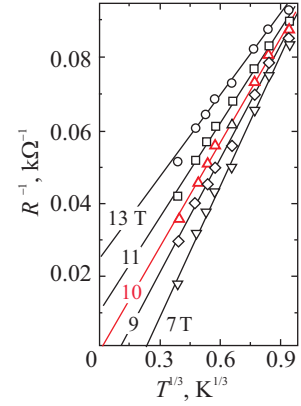}
 \caption{Extrapolation of the temperature dependence of the film conductivity to $T=0$ in various magnetic fields [processing of the curves presented in Fig.~26a using formula (104)] [108].}
\end{figure}

It should be noted that in the curves given in Fig.~25 there are no signs of the suppression of the superconducting transition temperature by a magnetic field, just as in the case of ultrathin films. The decrease in the resistance in a magnetic field of 1\,T starts virtually at the same temperature as in a zero field (cf. Fig.~21a).

The most interesting and most important feature of the $R(B)$ isotherms for the In--O film is the presence of a maximum [104] and negative magnetoresistance in strong magnetic fields. Figure 26 displays two families of $R(B)$ isotherms in magnetic fields, namely, in a field perpendicular to the film (Fig.~26a) and in a field parallel to it (Fig.~26b). The states of the film in these two experiments are close to each other, although they are not identical. All isotherms in each of these states intersect almost at one point, whose coordinates determine the critical values of the resistance $R_{\rm c}$ and the field strength $B_{\rm c}$ (Fig.~26a). As can be seen from a comparison of both families, the $R(B)$ dependences differ qualitatively only in weak fields, $B<B_{\rm c}$, where the resistance is determined by the motion of vortices and, therefore, strongly depends on the field direction. In strong fields, $B>B_{\rm c}$, the difference is only quantitative.

The appearance of a negative magnetoresistance seems to be quite natural within the framework of the assumption of the localization of electron pairs with opposite spins, since the magnetic field aligning the spins destroys pair correlations [107, 108]. The fact that the effect is observed for all field directions [108] confirms the suggested interpretation. An additional confirmation comes from an analysis of the temperature dependence of the film resistance in strong magnetic fields.

The observed increase in the resistance with a temperature decrease in the field B=5\,T, i.e., near the maximum of the $R(B)$ dependence, is described, albeit with low accuracy, by the activation dependence (99) with an activation energy of 0.13\,K [107]. The dependence in stronger fields can be described by none of the formulas (99)--(101). They can, however, be described (Fig.~27) with the aid of the formula for the conductivity $\sigma$ in the critical vicinity of the metal--insulator transition in three-dimensional space:
\begin{equation}\label{T1-3}        
 \sigma=s_1+s_2T^{1/3},\qquad s_2>0,
\end{equation}
where the parameter $s_1$ reverses sign at the metal--insulator transition (see Ref.~[7]). Where $s_1>0$, this parameter makes sense of the conductivity at $T=0$: $s_1=\sigma (T{=}0)>0$. According to the standard interpretation of the temperature dependence of the conductivity of three-dimensional systems in the vicinity of the metal--insulator transition, it follows from the results displayed in Fig.~27 that in the electron system of the In--O film the quantum superconductor--insulator transition in the field $B_{\rm c}$ is followed, upon a further increase in the field induction, by an insulator--metal transition in a field $B_{\rm I-M}{=}10\,$T. The assumption of the 3-dimensional nature of the system is reasonable, since the mean free path of normal electrons is {\it a fortiori} less than the film thickness $b=200\,$\AA.

\begin{figure}
\includegraphics{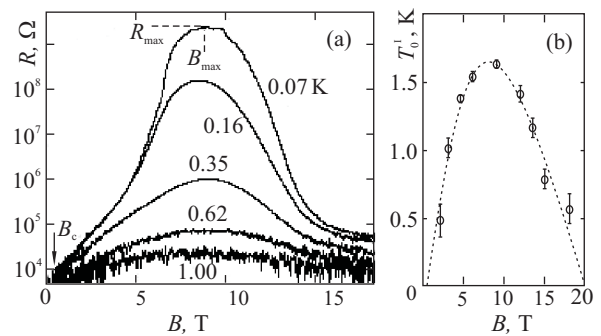}
\caption{(a) Family of five $R(B)$ isotherms of an amorphous In--O film with thicknesses of 200--300\,\AA\  in state 1 (see text) in a perpendicular magnetic field [109]. (b) Variation of the activation energy $T_0^{\,\rm I}$ with strengthening magnetic field, obtained from an analysis of isotherms; on both edges of the interval of magnetic fields, the accuracy of determining $T_0^{\,\rm I}$ decreases because of the closeness to the phase transition points [109].}.
 \label{Shahar1}
\end{figure}

A detailed study of the vicinity of the quantum superconductor--insulator transition in In--O films in a magnetic field was performed by Sambandamurthy et al. [109]. The authors of Ref.~[109] revealed a state in which the resistance at a temperature of 70\,mK increased by more than five orders of magnitude in comparison with the critical resistance $R_{\rm c}\approx 5\,$k$\Omega$ (see Fig.~28; cf. Fig.~26a). In comparison with the state presented in Figs~26 and 27, the newly found state falls more deeply into the insulator region with a field change and proves to lie outside the critical region of the metal--insulator transition. Correspondingly, the temperature dependence of the resistance bear an activation nature, in accordance with formula (99). The activation energy $T_0^{\,\rm I}$ depends on the magnetic field $B$, reaching a maximum at the same field $B_{\rm max}$ as the $R(B)$ dependence itself (Fig.~28b). With strengthening field in the region of $B>B_{\rm max}$, the activation energy decreases gradually, so that in a field of about 20\,T we can expect that $T_0^{\,\rm I}$ will become zero, i.e., the insulator will pass into a metal for this state of film, as well.

Even in the maxima of the $R(B)$ functions, the values of the activation energy $T_0^{\,\rm I}$ are small; they lie in the temperature range of 0.5--2~K, like the temperatures $T_{\rm c}$ of the superconducting transition [109]. As can be seen from Fig.~24, the activation energy is greater on the insulator side; it lies on the interval of 2--7\,K. The existence of an activation energy indicates the presence of a gap in the spectrum. It follows from all available experimental data that this gap is connected with a superconducting interaction, although there is no superconductivity itself at these values of the control parameters. By analogy with the gap in high-temperature superconductors, we shall call this gap a pseudogap (see also the note in the end part of Section 4.3).

For the sake of convenience, let us divide the possible states in In--O films into groups. To those states that in a zero magnetic field are located on the insulator side, we ascribe index 0; to the states lying in the superconducting region, for which the resistance in the zero field at $T=0$ is equal to zero, we ascribe indices 1--4 in such a manner that the deeper the state is located in the superconducting region, the greater the index. A convenient measure of the proximity of states to the transition point is the critical field $B_{\rm c}$. State 1, whose properties are demonstrated in Fig.~28, is nearest to the transition point ($B_{\rm c}=0.45\,$T); in state 2, the critical field is $B_{\rm c}=2\,$T (Fig.~26a), and state 3 (Fig.~29), on the contrary, lies the most deeply in the superconducting region ($B_{\rm c}=7.2\,$T). The nearer the state to the transition point, the higher the resistance peak in the $R(B)$ curve. For state 1 with its $B_{\rm c}=0.45\,$T, the value of $R_{\rm max}$  at $T=70$\,mK exceeded $R_{\rm c}$ by five orders of magnitude [109]; for state 2 with $B_{\rm c}=2\,$T, the ratio $R_{\rm max}/R_{\rm c}$ was about 10, and for state 3 with $B_{\rm c}=7.2\,$T, this ratio was only 1.35 [110, 111]. With a further increase in the charge carrier concentration and a shift deeper into the normal region, the peak of the magnetoresistance disappears completely (state 4).

Simultaneously, there occurs a narrowing of the magnetic field interval $\Delta B=B_{\rm I-M}-B_{\rm c}$: $\Delta B=20\,$T for state 1 (Fig.~28b), 8 T for state 2 (Fig.~26a), and 0 for state 3. The last follows from the results of the extrapolation displayed in Fig.~29 (curve with circles). A procedure similar to that demonstrated in Fig.~27 showed that at any values of the magnetic field strength, including $B_{\rm max}$, the parameter $s_1$ is positive, i.e., a finite conductivity should be retained at $T=0$ and any $B>B_{\rm c}$. This means that the {\it superconductor--insulator transition transforms into the superconductor--normal metal transition} with increasing concentration of charge carriers in the amorphous In--O film.

\begin{figure}
\includegraphics{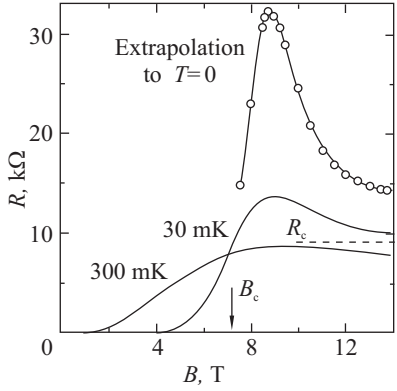}
\caption{Magnetoresistance of an amorphous In--O film 200\,\AA\  thick whose representative point lies relatively deep in the superconducting region [107]. The upper curve is drawn through points (shown by circles) obtained by the extrapolation of the experimental values of $R(T\,)$ at various $B$ to $T=0$ using formula (104).}
 \label{G4}
\end{figure}

The peak in the $R(B)$ dependence is retained near the superconductor--insulator transition in the insulator region, as well, i.e., in the samples in state 0 [109, 110]. A qualitative difference from the curves shown in Fig.~26 lies only in the fact that the resistance at $B=0$ has a finite and by no means small value (a similar curve for Be films was given in Fig.~23). According to measurements [110], the resistance variation with temperature in fields close to $B_{\rm max}$ obeys the activation law (99); in a high field $B\approx 15\,$T, the pseudogap was closed and an insulator with a finite density of states at the Fermi level was formed, in which the resistance obeyed the Mott law: $R\propto \exp {(T_0/T\,)^{1/4}}$.

The above-described results of measurements performed on amorphous In--O films can be represented in a common phase diagram. Attempts to construct such a diagram were made at various stages of the studies [104, 107, 109]. The variant presented in Fig.~30 does not appear to be final, either. However, it is suitable in that it provides a possibility of involving and comparing all data that are available to date.

\begin{figure}
\includegraphics{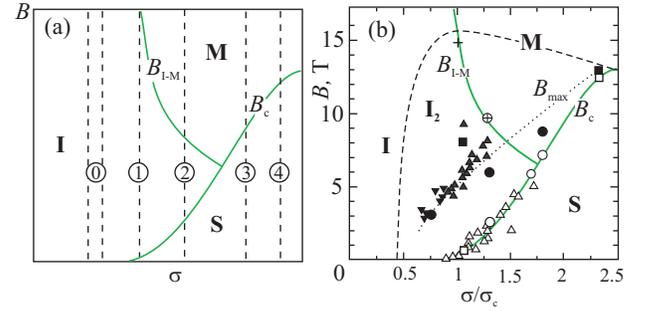}
\caption{(a) Phase diagram in the $(\sigma ,B)$ plane, which shows the mutual arrangement of superconducting, insulating, and metallic phases in amorphous In--O films at low temperatures. The diagram is presented schematically, i.e., without values of the quantities laid off along the coordinate axes and without experimental points, although with dashed straight lines representing different types of samples (see the main text). (b)~The same diagram, but with experimental points. The dashed curve separates the region of the existence of the Bose glass (continuous to the left of the curve $B_{\rm I-M}$, and in the form of fine inclusions to the right of this curve). All dark symbols refer to the curve $B_{\rm max}$; white symbols, to the curve $B_{\rm c}$, and the symbols with interior crosses, to the curve $B_{\rm I-M}$. The meaning of the different symbols is explained in the main text.}
 \label{diagram}
\end{figure}

In order to help the reader in comparing different experiments performed on various In--O samples, the diagram is given first in a schematic form in Fig.~30a. As the axes, the control parameters $\sigma$  and $B$ were chosen. The conductivity $\sigma$  is analogous in its meaning to the parameter $k_{\rm F}l$, which was used earlier in Ref.~[101]. Concrete samples in the diagram are marked by vertical straight lines.

The total of all these measurements can be formulated as follows. The greatest peak of magnetoresistance (an increase and the subsequent decrease) is observed in those states that are close to the transition point in a zero magnetic field and which are located in this case to the right of the transition, on the superconductor side (samples of type 1 and 2, Fig.~30a). However, this property is apparently not universal; in Be, as we saw in Section 4.1 (see Figs~22 and 23), the peak of magnetoresistance lies to the left of the transition, while in Bi it is not at all present.

The complete phase diagram is displayed in Fig.~30b. In order to facilitate the comparison of the data of different experiments, the $\sigma$-axis is represented in a dimensionless form. As the basis for constructing this diagram, the data from Ref.~[109] were taken: white triangles denote $B_{\rm c}$ values; the dark triangles correspond to $B_{\rm max}$; the base-down triangles mark the states with a superconductivity, and the base-up triangles correspond to states exhibiting no superconductivity. The straight cross marked the field strength equal to 14--15\,T, at which, according to Fig.~28, the basic decrease in the resistance of a sample in the appropriate state stops and the activation energy becomes poorly determined [109]. The data from Refs~[106, 107, 110] are shown by circles: white circles designate $B_{\rm c}$; dark circles correspond to $B_{\rm max}$, and the circle with an interior straight cross stands for the field of the insulator--metal transition. Analogously, we used white and dark squares for the data from Ref. [111]. The solid curves separate the superconducting phase from the nonsuperconducting ones (line $B_{\rm c}$), and the insulator from the metal (line $B_{\rm I-M}$. The dotted curve $B_{\rm max}$ passes through all dark symbols. The $B_{\rm c}$ curve is drawn through the white symbols, and the $B_{\rm I-M}$ curve passes through the symbols with interior crosses. The dashed curve M separates the region of the existence of localized pairs. On the left-hand side of Fig.~30b, this curve separates the regions of the Fermi glass and Bose glass; on the right-hand side, this curve separates the region in which the inclusions of a Bose glass in metal are present.

If we extrapolate the $B_{\rm I-M}(\sigma )$ curve to the superconducting region, then the mutual arrangement of the points of the superconductor--insulator and metal--insulator transitions at $B=0$ will prove to correspond to the diagram given in Fig.~2c. This agrees with the conclusions made above on the basis of Fig.~24.

{\bf Polycrystalline TiN films}. Polycrystalline TiN films are produced by magnetron sputtering of a target from pure Ti in a nitrogen plasma. The resistivity of the films depends on the nitrogen pressure during sputtering, apparently, since the pressure determines the excess concentration of nitrogen in the resultant film, so that the subscript $x$ in the TiN$_x$ formula can reach a value of 1.3 [112, 113]. The standard thickness of the films on which the experiments were performed was about 50\,\AA. In air, the films are very stable at room temperature. Their resistance should be considered as a check rather than control parameter; the temperature of the superconducting transition in a zero field can serve as another check parameter for evaluating the proximity of the state to the superconductor--insulator transition.

Figure 31a shows the resistance curves $R(T\,)$ which include the onset of the superconducting transition for four TiN films [114]. The vertical bars in the curves indicate the $T_{\rm max}$  temperatures at which the resistance reaches maximum (onset of the transition). Furthermore, the transition temperatures $T_{\rm c}$ are indicated alongside the curves; they were calculated under the assumption that the conductivity $\sigma =1/R$ consists of the normal part
\begin{equation}\label{sigmaN}
  \sigma_n=a+bT^{1/3}
\end{equation}
and the contribution $\Delta\sigma_{\rm s}$ caused by the superconducting fluctuations (Aslamazov--Larkin correction [56]):
\begin{equation}\label{Aslam}       
\begin{array}{l}
  \displaystyle\sigma=\sigma_n+\Delta\sigma_s=a+bT^{1/3}+
  \frac{e^2}{16\hbar}\left[\ln\left(\frac{T}{T_c}\right)\right]^{-1},
  \\ \Delta\sigma_s\ll\sigma_n.
\end{array}
\end{equation}
A comparison of the $T_{\rm max}$  and $T_{\rm c}$ temperatures in Fig.~31a with the superconducting transition temperature $T_{\rm c0}=4.7\,$K of the massive TiN sample shows that the superconductivity in these films is strongly suppressed by disorder, so that their states are indeed located at the edge of the superconducting region.

In formula (106), an expression was taken for the normal conductivity that is valid in the critical vicinity of the normal metal--insulator transition in the three-dimensional region [7], and an expression valid in the two-dimensional system was taken for the contribution of superconducting fluctuations to the conductivity. The use of different dimensionalities is justified by the fact that the mean free path of normal electrons is $l\ll b\approx 50\,$\AA , and the coherence length $\zeta$ in TiN and, all the more, the London penetration length $\lambda$ determining the transverse size of vortices, are much more than the film thickness, i.e., $\zeta ,\lambda \gg b$. Figure 31b depicts the conductivity $\sigma ^{\,*}=\sigma -\Delta \sigma _{\rm s}$ which, upon fulfillment of the right-hand inequality in formulas (106), coincides with $\sigma _{\rm n}$. Formally, the accuracy of this representation and the extrapolation are small, since the inequality in formulas (106) is violated for sure near the transition point and $\sigma ^{\,*}$ proves to be the difference of two large quantities, one of which contains a free parameter $T_{\rm c}$. However, the very selection of the representation (105) for $\sigma _{\rm n}$ sharply limits the interval of possible values of the parameter $T_{\rm c}$; moreover, the term $\Delta \sigma _{\rm s}$ diminishes rapidly with increasing temperature, so that the $\sigma ^{\,*}$ curve on the right-hand side of the graphs in Fig.~31b practically coincides with the directly measured quantity $1/R$: $\sigma ^{\,*}\approx \sigma \equiv 1/R$. All this substantially elevates the reliability of the conductivity extrapolation being performed [114, 115].
\begin{figure}
\includegraphics{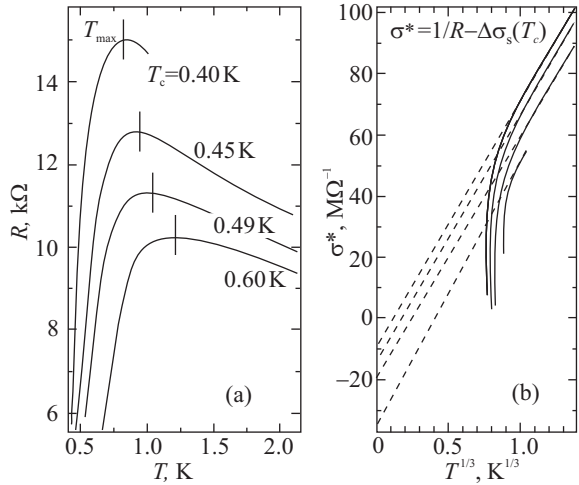}
\caption{((a) Temperature dependence of the resistance of four TiN films in which the superconductivity is partially suppressed by disorder [114]. The meaning of the temperatures $T_{\rm max}$  and $T_{\rm c}$ indicated alongside each curve is explained in the main text. For all the curves, one has $T_{\rm max} \approx 2T_{\rm c}$. (b) Temperature dependence of the normal part of the conductivity, $\sigma ^{\,*}$, for the same four films, obtained after subtracting the Aslamazov--Larkin correction with an optimum value of $T_{\rm c}$ from the total conductivity; the dashed curves have been obtained by the extrapolation of the $\sigma ^{\,*}(T\,)$ dependence for $T\gtrsim1$\,K [114]}
\label{Bat-TiN1}
\end{figure}
\begin{figure}
\includegraphics{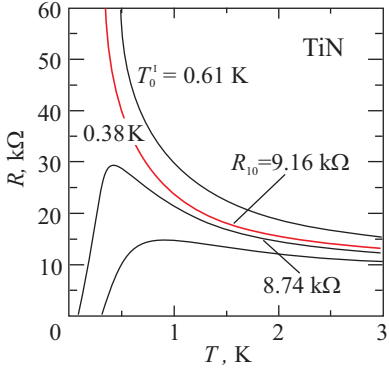}
\caption{Superconductor--insulator transition in a zero magnetic field in a family of TiN films. Film resistances at $T=10\,$K are given for the two middle curves (these can be assumed to be a control parameter); for the two upper curves, the activation energies $T_0^{\,\rm I}$ are indicated [116].} \label{Bat-TiN2}
\end{figure}

The extrapolation of the straight lines given by formula (105) to the temperature $T=0$ yields negative values of the parameter $a$. Consequently, if there were no superconducting transition, the state with which we are dealing would be insulating. On the whole, we obtain a phase diagram predicted by Larkin [8] for three-dimensional systems; it is schematically depicted in Fig.~2c. The states of four films that have been studied in Ref.~[114] are found on the left-hand side of the critical region of the nonexistent normal metal--insulator transition; with changing $T$, their representative points move vertically along the $x={\rm const}$ lines.

A superconductor--insulator transition for another series of TiN films [116] is demonstrated in Fig.~32. Here, we should first of all note the sharpness of this transition. The resistance of the superconducting film and the film that becomes insulating differ by only 5\% at $T=10\,$K. An analysis of the curve that is nearest to the transition point, which lies in the insulating region, shows that the film resistance changes according to the activation law (99) with the activation energy $T_0^{\,\rm I}=0.38\,$K. Using the above analysis of curves in Fig.~31 and the phase diagram in Fig.~2c as the base, we can assert with confidence that the boundary state at the transition, which corresponds to the separatrix of the family of curves in Fig.~31, is not metallic, either. This means that the scheme of one-parametric scaling is not applicable to TiN.

All characteristic features of the superconductor--insulator transition in a magnetic field that were observed in TiN~films [113, 114] are very similar to those that were observed in other materials. The sets of $R(T\,)$ curves obtained in weak fields [113] are similar to those observed for ultrathin films (Figs~21a, 22b) or for In--O films (see Fig.~25). The reproducibility and the origin of the double reentrant transition which was observed in some TiN films in Ref.~[113] have not been explained so far (similar double reentrant transitions were also observed on the Josephson junction arrays; see Fig.~44 in Section 5.1 and the accompanying text). The $R(T\,)$ curves in strong magnetic fields [114] resemble analogous curves for In--O in Fig.~26, not only qualitatively, but even quantitatively. The peak of magnetoresistance is observed in the states on both sides of the transition point in a zero field, with a somewhat larger amplitude in the region of the insulator [117]: the resistance in strong fields decreased by three orders of magnitude at a temperature of 60\,mK.

\subsection{4.3 High-temperature superconductors}

Superconductor--insulator transitions have repeatedly been observed in all basic families of cuprate high-temperature superconductors. The structure of high-temperature superconductors belonging to these families represents a stack of cuprate CuO$_2$ planes containing mobile carriers. The coupling between the planes is weak; therefore, the high-temperature superconductors in the normal state demonstrate, as a rule, a strong anisotropy of conductivity. In high-temperature superconductors there are control parameters which change the degree of doping of the cuprate planes at zero temperature and the probability of carrier scattering in these planes and of tunneling between the planes. At the same time, the systems of carriers in high-temperature superconductors possess some specific features that are not covered with the theoretical concepts discussed in Section 2. The aim of this section is just to reveal and to discuss these features.

\begin{figure}[b]
\includegraphics{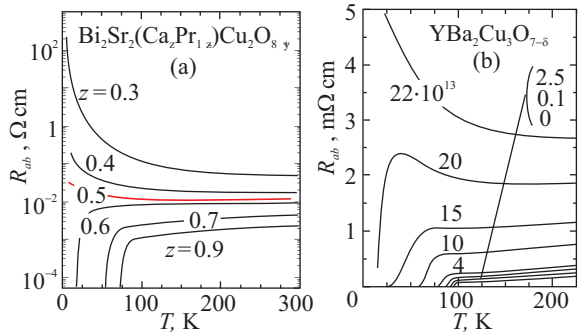}
\caption{a) Temperature dependence of the resistance of ${\rm Bi_2Sr_2(Ca}_z{\rm Pr}_{1-z}){\rm Cu_2O}_{8+y}$ samples at various Ca concentrations (indicated alongside the curves) [118]. (b) Temperature dependence of the resistance of a YBCO film after a stepped change in the dose of irradiation with Ne$^{\,+}$ ions [122]. The doses are indicated near the curves as the number of ions per cm$^2$. }
 \label{HTSC-BiY}
\end{figure}

Both the superconductivity and, apparently, the conductivity proper exist in high-temperature superconductors owing to their specific atomic structure determining the appearance of a periodic potential in which the electron system is embedded. This potential must mainly be retained with a change in the control parameters. Therefore, the most common control parameter is the concentration of substitutional impurities or vacancies. Figure 33a displays the curves of the temperature dependence of resistance $R_{ab}$ along the basal plane for samples of different compositions for the compound ${\rm Bi_2Sr_2CaCu_2O}_{8+y}$ (abbreviated as BSCCO), in which Pr atoms substitute for part of the Ca atoms located between the cuprate planes: ${\rm Ca}{\rightarrow}({\rm Ca}_z{\rm Pr}_{1-z})$ [118]. At a critical concentration of Ca atoms, $z_{\rm c}{=}0.52$, a superconductor--insulator transition occurs. An analogous effect is observed upon the substitution of some other rare-earth atoms [119] and Y [120, 121] for Ca. Since both the rare-earth elements and Y are substitutional impurities in BSCCO, their substitution for Ca atoms does not change the crystal structure of the compound.

It is obvious that the conductivity is always realized against the background of a certain disorder, which commonly exists as a result of a nonstoichiometry. However, the corresponding random potential must be small in comparison with the crystal potential, which must ensure the retention of the initial electronic structure. In Fig.~33a, the control parameter appears to act on the system via a smooth change in the average parameters of the structure rather than through a change in the level of local disorder. The opposite limiting case is demonstrated by the experiment conducted in Ref.~[122], whose results are given in Fig.~33b. An epitaxial ${\rm YBa_2Cu_3O}_{7-\delta }$ (YBCO) film about 2000\,\AA\  thick was placed into a beam of 1-MeV ${\rm Ne}^+$ ions which passed through the film, producing defective regions in the form of cylinders with a diameter of about 8\,\AA. Small doses $\tilde {f}$ of irradiation led to a reduction in the superconducting transition temperature $T_{\rm c}$ and to an increase in the residual part of resistance, $R_0(\tilde {f})$, in the temperature dependence $R(T\,)$ of resistance for $T>T_{\rm c}$:
 $$R(T)=R_0(\tilde{f})+R_T(T)$$
At large doses $\tilde f$, the superconductivity disappeared, and the resistance increased with decreasing $T$, according to the Shklovsky--Efros law (100). The irradiation caused the breakdown of the crystalline potential. Therefore, the transition observed is, in fact, a consequence of the destruction of the `living environment' of the electron system itself.

Let us examine the evolution of the normal and superconducting properties caused by a change in the chemical composition using the example of ${\rm La}_{2-x}{\rm Sr}_x\rm CuO_4$ (LSCO) compounds. Superconductivity in LSCO exists if the degree of doping x lies on the interval
\begin{equation}\label{LSCO}
0.04\lesssim x\lesssim0.26.
\end{equation}
At the optimum degree of doping, $x_{\rm opt}\approx 0.16$, the superconducting transition temperature reaches approximately 40\,K [123].

With overdoping, $x>x_{\rm opt}$, the normal state is the usual metal, in which the anisotropy of resistance changes only a little with a change in the temperature, while in the superconducting state the superconducting planes are strongly coupled [124]. Therefore, the quantum phase transition at $T{=}0$ and $x$ belonging to the right-hand edge of interval (107) is a 3D superconductor--normal metal transition [125]. However, the signs of the derivatives of the longitudinal ($\rho _{ab}$) and transverse ($\rho _c$) resistivities become different in the region $x_{\rm opt}>x\gtrsim0.04$: in the direction perpendicular to the CuO$_2$ planes, the resistivity $\rho_c$ grows rapidly with decreasing temperature, while the resistivity $\rho_{ab}$ along the layers diminishes [124, 126]. As a result, the anisotropy of resistivity at temperatures slightly greater than the transition temperature, $T\gtrapprox T_{\rm c}$, can exceed three orders of magnitude. With a further reduction in $x$, the derivative $\partial \rho /\partial T$ proves to be negative in both directions, so that the nonsuperconducting state becomes similar to a usual insulator [127, 128].

\begin{figure}[b]
\includegraphics{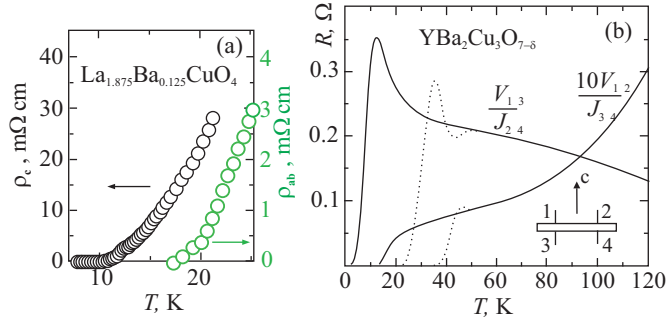}
\caption{(a) Temperature dependence of the transverse ($\rho _c$, left-hand scale) and longitudinal ($\rho _{ab}$, right-hand scale) resistivities of an LaBaCuO single crystal in a zero magnetic field; the vertical scales on the left-hand and right-hand ordinate axes differ by four orders of magnitude [129]. (b) Temperature dependence of the $V_{12}/J_{34}$ ratio, which in a YBCO single crystal with a reduced oxygen concentration can be approximately regarded as the $R_{ab}$ resistance along cuprate planes (current $J\!\parallel \![ab])$, and of the $V_{13}/J_{24}$ ratio, which is approximately equal to $R_c$ (current $J\!\parallel \!c\perp [ab]$), in the absence of a magnetic field (dotted curves) and in magnetic field of 5\,T (solid curves). The sample had the shape of a plate with dimensions of $2\times 1\times 0.05$\,mm. The inset shows the arrangement of the contacts (1--4) and the direction of the c-axis [132].}
 \label{LaFrid}
\end{figure}

With decreasing a degree of doping $x$, a transformation of the superconducting state itself occurs. The bonds between the cuprate planes weaken strongly, so that the planes transform into quasiindependent two-dimensional systems, and the quantum transition at $x$ belonging to the left-hand edge of interval (107) is converted into a 2D superconductor--insulator transition [125, 129]. Nevertheless, the arising superconducting state is global and three-dimensional, although the process of the establishment of this state can be extended over a certain temperature range. In Ref.~[129], which contains an analysis of different experiments with ${\rm La}_{2-x}{\rm Ba}_x{\rm CuO_4}$ ($x{=}0.125$), a whole series of phase transitions was experimentally examined in this compound with decreasing temperature. First, charge structures --- stripes --- appear in the CuO$_2$ planes and the coupling between the cuprate planes weakens, after which an antiferromagnetic ordering of magnetic moments localized on copper ions occurs. Then, a 2D superconducting transition occurs in the cuprate planes, but dissipation is retained because of the presence of fluctuation vortices, and only with a further decrease in temperature does a BKT transition manifests itself, and a coherent superconducting state is established.

If the disorder disrupts the identity of cuprate planes which in fact become superconducting above all, then the temperature of the appearance of the superconducting current can depend on the direction of this current (the `Fridel effect' [130, 131]). As can be seen from Fig.~34a, the resistivity $\rho _{ab}$ becomes zero at $T\approx 18$\,K, whereas $\rho _c$ does so only at $T\approx 10$\,K [129]. An analogous effect was observed in the underdoped YBCO crystal (Fig.~34b [132]), although the anisotropy of conductivity in YBCO is substantially less.

It is understandable from the above that the occurrence of quantum phase transitions in high-temperature superconductors depends on a whole number of side factors which can change and complicate the entire picture of the phenomenon, e.g., the strong anisotropy of the crystal structure, the difference in the mechanisms of electron transport along and across the cuprate planes, the magnetic ordering of the localized spins, etc. We shall not go deeply into this boundless region, and limit ourselves only to brief notes concerning the influence of the magnetic field.

By varying the chemical composition, we can make the superconducting transition temperature small, i.e., bring the system to a state close to the transition in a zero magnetic field, and then destroy superconductivity with the aid of a magnetic field and investigate the transition in the magnetic field as was done, for example, in ultrathin Bi films (see Fig.~21). When this program is realized for the basic families of high-temperature superconductors, the families of the thus-obtained curves are visually very similar to those given above, but an increase in the resistance in the normal phase occurs according to a logarithmic rather than an activation law [133, 134].

In Sections 4.1 and 4.2, much attention was given to the negative magnetoresistance in strong fields, which indicates the destruction of fluctuation-related incoherent Cooper pairs or the destruction of pair correlations between localized carriers. Negative magnetoresistance in strong fields was also repeatedly observed in high-temperature superconductors. In YBCO [135] and BSCCO [135, 136], a negative derivative $\partial \rho /\partial B$ was observed for the transverse resistivity $\rho _c$, while the in-plane resistivity $\rho _{ab}$ remained positive or equal to zero. This is explained apparently by the specific character of transverse magnetotransport in these families. However, a negative magnetoresistance in LSCO is also observed in $\rho _{ab}$. The curves (taken from Ref.~[134]) in Fig.~35a are very similar to those that were discussed above in connection with the experiments on InO and TiN. In this case, the negative magnetoresistance is probably explained by the destruction of fluctuation-related quasilocalized superconducting pairs, which, according to Galitski and Larkin [59], must lead to an increase in the conductivity along two-dimensional layers. An analogous effect was also observed in the magnetoresistance of the electronic high-temperature superconductor ${\rm Nd}_{2-x}{\rm Ce}_x\rm CuO_4$. We shall return to a semiquantitative analysis of these data in Section 4.4.

  \begin{figure}
\includegraphics{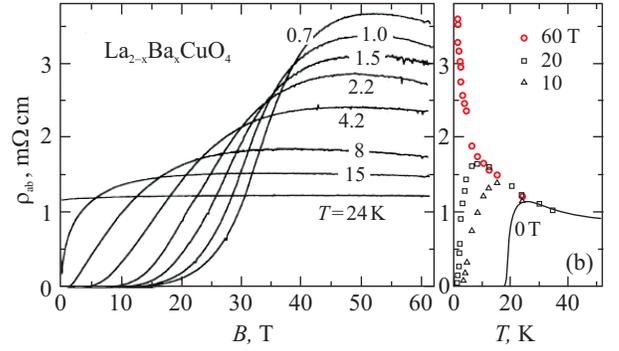}
\caption{(a) Longitudinal magnetoresistance $\rho _{ab}$ of an LSCO single crystal, measured in a pulsed magnetic field at various temperatures [134]. (b) Temperature dependence $\rho _{ab}(T\,)$ in a field $B=0$ (solid curve) and $B=10$, 20, and 60\,T (data points) [134].}
 \label{Boebin}
  \end{figure}

The pair correlations of localized carriers lead to the emergence of a gap or, at least, to a decrease in the density of states at the Fermi level. Superconducting pair correlations in a system of delocalized carriers simultaneously bring about superconductivity and the appearance of a superconducting gap. Therefore, the decrease in the density of states at the Fermi level that is caused by a superconducting interaction but is not accompanied by establishing superconductivity can naturally be called the pseudogap. This term already exists, and it appeared precisely in connection with high-temperature superconductivity. Usually, the pseudogap implies an anisotropic rearrangement of the density of states for $T>T_{\rm c}$, which is caused by antiferromagnetic fluctuations, fluctuations of charge-density waves, or by structural rearrangements accompanied by phase separation [137]. Amorphous or fine-crystalline materials, which were discussed in Sections 4.1 and 4.2, demonstrate the simplest isotropic variant of a pseudogap emerging only due to the random potential without the participation of the periodic crystal field.
  \begin{figure}[h]
\includegraphics{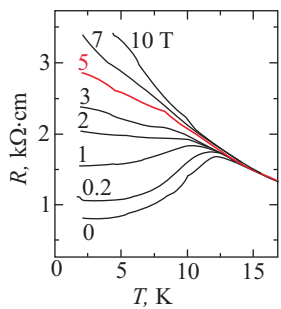}
\caption{Temperature dependence of the longitudinal magnetoresistance $R_{ab}$ of a two-dimensional organic superconductor $\kappa$-(BEDT-TTF)$_2$Cu[N(CN)$_2$]Cl at a small pressure in a magnetic field perpendicular to the two-dimensional layers [140].}
 \label{Org}
  \end{figure}
All that was said in this section relative to the specific features of the high-temperature superconductors that are connected with superconductor--insulator transitions also refers to organic superconductors. Organic crystals are usually strongly anisotropic and possess two-dimensional or even quasi one-dimensional conducting structure. In these anisotropic structures, different states of the electron system are competing, e.g., superconducting states, ferro- and antiferromagnetic states, or states with waves of charge or spin density, etc. [138]. Quite unusual sequences of phase transitions can be observed in this case, for example, a transition in a zero magnetic field to the superconducting state at a certain temperature $T_{\rm c1}$ with the subsequent reverse transition for $T_{\rm c2}<T_{\rm c1}$ to a high-resistance normal state [139]. To isolate a pure superconductor--insulator transition under these conditions is quite difficult. As an example, Fig.~36 depicts the evolution of the curves for the temperature-dependent longitudinal resistance of a two-dimensional organic superconductor $\kappa$-(BEDT-TTF)$_2$Cu[N(CN)$_2$]Cl in a magnetic field. It is evident that the superconducting state is established in a zero magnetic field only partially, and that in strong fields no exponential increase in the resistance occurs with decreasing temperature.

\subsection{4.4 Crossover from superconductor-metal to superconductor--insulator transitions}

In Section 1.2, an algorithm was formulated that makes it possible to distinguish between a superconductor--insulator transition and a superconductor--normal metal transition. According to this algorithm, it is necessary to extrapolate the temperature dependence of the conductivity $\sigma (T\,)$ to $T=0$ on the nonsuperconducting side; the type of transition is determined by the sign of the extrapolated value $\lim \sigma (T\rightarrow 0)$. Such a complex procedure is required since in a three-dimensional metal near the metal--insulator transition there is a region of `bad' metal with a conductivity $\sigma (0)$ smaller than the Mott limit (see, e.g., Ref.~[7]):
 $$\sigma(0)=\frac{e^2}{\hbar}\frac1\xi<\sigma_M\equiv\frac{e^2}{\hbar}k_F.$$
The temperature-dependent part of the conductivity of the `bad' metal is determined by the quantum correction and has a positive derivative $\partial \sigma /\partial T>0$, just like the hopping conductivity in the insulator. In other words, the sign of the derivatives $\partial \sigma /\partial T$ and, naturally, $\partial R/\partial T$ changes in the depth of the metallic region for $\sigma (0)>\sigma _{\rm M}$ rather than at the metal--insulator transition point.

When we discussed in the Introduction which of the transitions, superconductor--insulator or normal metal--insulator, occurs earlier with a change in the disorder or electron concentration and, superimposing the appropriate schematic phase diagrams (see Fig.~2), moved them relative to one another, we did not take into account that on the abscissa axis of the normal metal--insulator diagram there is one additional characteristic point, $x_{\rm M}$, at which $\sigma (0)=\sigma _{\rm M}$. If we take this circumstance into account, then two variants should appear in the diagram in Fig.~2a, depending on the location of the point$ x_{\rm M}$. If the point of the quantum transition M--S is located between the points I--M and $x_{\rm M}$, then the superconductor transforms at this point into a bad metal and the derivative $\partial R/\partial T$ is negative at this point, but the resistance tends to a finite value as $T\rightarrow 0$. The behavior of the resistance during such quantum transition is very similar to that observed in the transitions in InO or TiN, but the increase in the resistance with decreasing temperature on the nonsuperconducting side of the transition, and the maximum of the magnetoresistance will change by only several percent rather than by several orders of magnitude. The evolution of conductivity in the amorphous Nb--Si alloy shown in Fig.~1 corresponds to precisely such a case. An analogous behavior of conductivity appears to be observed in the Mo--Ge [141, 142] and Mo--Si [143, 144] films, in ultrathin Ta films [145], and in the high-temperature superconductor NdCeCuO [146--148].

  \begin{figure}
\includegraphics{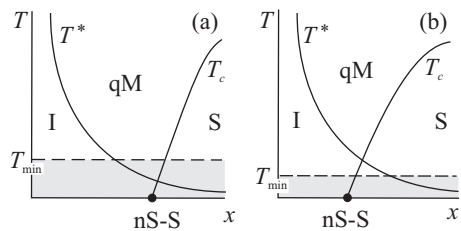}
\caption{Two variants of the phase diagram for two-dimensional systems. Curve $T^{\,*}(x)$ is plotted with formula (108); the dashed straight line $T=T_{\rm min}$  specifies the temperature range $T>T_{\rm min}$ in which all realistic experiments are performed; qM denotes quasimetal, and nS, nonsuperconductor (analogous diagrams for three-dimensional systems were given in Fig.~2). }
 \label{diagr-2D}
  \end{figure}

Before going over to concrete examples, it is necessary to make a refinement. The diagrams in Fig.~2 relate, strictly speaking, only to three-dimensional systems, while all experiments [141--145] were carried out on thin films. In the two-dimensional systems of normal noninteracting electrons there is formally no metallic state at absolute zero; at any arbitrarily small disorder, a temperature decrease leads, first, to a changeover from the logarithmic increase in the resistance to an exponential increase at
\begin{equation}\label{108}
 T^*\approx\varepsilon_F\exp(-2k_Fl)
\end{equation}
and then to the electron localization at a temperature $T=0$. However, it is evident from formula (108) that already at $k_{\rm F}l\simeq$ 2--3 the temperature $T^{\,*}$ becomes inaccessibly low, and in the temperature range with $T>T^{\,*}$ the conductivity of a two-dimensional system is described by the classical formula with a relatively small quantum correction. Therefore, in two-dimensional systems everything depends on the location of the intersection of the curve $T^{\,*}(x)$ and the curve of superconducting transitions $T_{\rm c}(x)$ (Fig.~37). In each experimental facility and each laboratory, a minimum accessible temperature $T_{\rm min}$  exists. The unattainable region is depicted in the diagrams in Fig.~37 by gray. The transitions discussed in this section are realized when the point of intersection of the $T^{\,*}(x)$ and $T_{\rm c}(x)$ curves is located in the unattainable region (Fig.~37a); after the breakdown of superconductivity, the resistance changes logarithmically in accordance with formula (101). A real superconductor--insulator transition occurs when the point of intersection is located above the level of $T_{\rm min}$, as in the diagram shown in Fig.~37b.

Let us illustrate the aforesaid by a concrete example. Figure 38 displays the temperature dependence of the resistance of amorphous ultrathin Ta films of different thicknesses $b$ [145]. All films with $b\geqslant b_{\rm c}=3.1\,$\AA\  are superconducting. If we select $T_{\rm min}\sim 0.5$~K as the lower temperature boundary, then the resistance of the films with a thickness ranging $3.1>d>2.1$~nm at a temperature of down to $T_{\rm min}$  varies logarithmically, and in a film with $d=$1.9~nm there occurs a crossover to the exponential increase in the resistance. But if we decrease $T_{\rm min}$  to 10~mK, then the interval of thicknesses of films with a logarithmic dependence of resistance will narrow, but hardly disappear. {\it A fortiori} we cannot expect an exponential increase in the resistance in the superconducting region at temperatures higher than the superconducting transition temperature. Therefore, after fixing an appropriate value of a control parameter, we can destroy superconductivity by a field and implement a transition to the state of a bad metal with a negative derivative of the resistance with respect to temperature, $\partial R/\partial T<0$. The main difference between this state and the insulating state is a relatively slow logarithmic increase of resistance upon a decrease in temperature. Such an increase is observed, for instance, in the film of the amorphous $\rm Nb_{0.15}Si_{0.85}$ alloy [149] in a magnetic field of 2~T, whereas a superconducting transition in a zero field occurs at $T_{\rm c}\approx 0.23$~mK.

  \begin{figure}
\includegraphics{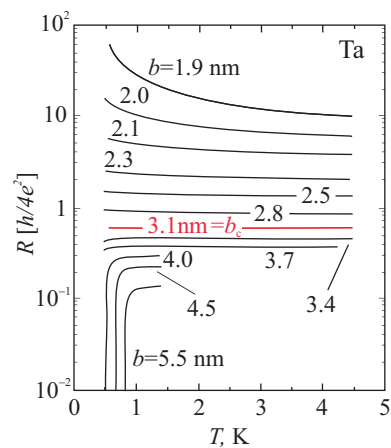}
\caption{Temperature dependence of the resistance of amorphous ultrathin Ta films of different thicknesses [145]. The superconducting transition temperature decreases with thickness $b$ and becomes zero at $b=3.1$\,nm.}
 \label{Yoon}
  \end{figure}

When the evolution of the states takes place in accordance with the variant presented in Fig.~37b, we can expect the appearance in the nonsuperconducting phase of not only a negative derivative of the resistance with respect to temperature, $\partial R/\partial T<0$, but also a negative magnetoresistance, $\partial R/\partial B<0$, in strong magnetic fields. However, the effect should be small compared to that observed in InO or TiN, since only a weak localization caused by quantum corrections to the conductivity occurs on the nonsuperconducting side of the transition. Such negative magnetoresistance in strong fields was indeed observed in at least two materials: in amorphous Mo--Si films [143, 144], and in textured ${\rm Nd}_{2-x}{\rm Ce}_x\rm CuO_4$ films [146--148] (Fig.~39).

The temperature dependence of the resistance of the latter material in different magnetic fields are given in Fig.~39a,b. As can be seen, the uncommon behavior of the resistance, which makes it possible to discuss these experiments in connection with superconductor--insulator transitions, was only observed at low temperatures, $T\ll T_{\rm c}$. This is, first and foremost, the negative derivative $\partial R/\partial T<0$ at low temperatures in the fields in which the superconductivity has already been destroyed. Figure 39b, where the low-temperature part of these curves is given on an enlarged scale, illustrates a second feature, namely, the intersection of the $R(T\,)$ curves in fields of 5 and 7\,T. At a temperature lower than the point of intersection of these curves, the increase in the field strength leads to a decrease in the resistance.
\begin{figure}
\includegraphics{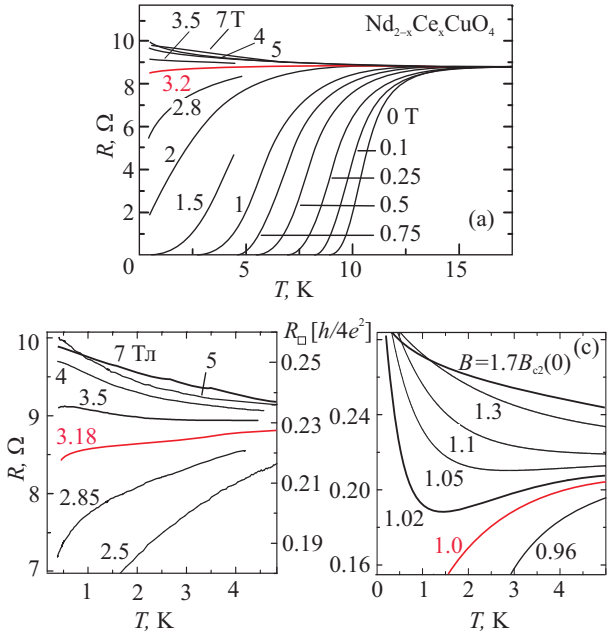}
\caption{(a) Temperature dependence of the resistance of an epitaxial
${\rm Nd}_{2-x}{\rm Ce}_x\rm CuO_4$ film 1000\,\AA\  thick [148]. (b) Part of Fig.~39a on an enlarged scale; the negative derivative of the resistance with respect to temperature is seen ($\partial R/\partial T<0$) in the fields exceeding the critical one, as is the intersection of the curves in the fields of 5 and 7~T. (c) $R(T\,)$ functions calculated with formulas (48) and (109) [148].}
 \label{NdCe}
\end{figure}

Since the relative changes in the resistance caused by variations of the temperature and field strength in ${\rm Nd}_{2-x}{\rm Ce}_x\rm CuO_4$ films are small, the experimental curves can be compared with the results of theoretical calculations [59] performed within the framework of the perturbation theory. The purpose of this comparison is twofold. First, to explain to which extent the high-temperature superconductor with a superconductivity destroyed by a magnetic field behaves in the region of strong superconducting fluctuations similar to a conventional superconductor. The second purpose is to answer the question of whether the superconducting fluctuations in the dirty limit at low temperatures can, to an order of magnitude, describe the observed negative magnetoresistance and whether they are the forerunners of localization of superconducting pairs.

The comparison was carried out in Ref.~[148] for a film in which the superconducting transition in the zero magnetic field occurred at $T_{\rm c}\approx 12$\,K. The conductivity was calculated using the formula
\begin{equation}\label{Comparison1}     
  R^{-1}=
  \sigma_0+\delta\sigma(B,T)-\alpha\frac{e^2}{2\pi\hbar}\ln(T/\widetilde{T}),
\end{equation}
where the term $\delta\sigma(B,T\,)$ defined by formula (48) takes into account the superconducting fluctuations, and the last term, which is called the Aronov--Altshuler correction and which allows for electron--electron interaction in the diffusion channel, is not connected with the superconducting interaction. The value of $T_{\rm c0}$ that enters into formula (48) and the value of the classical conductivity $\sigma _0$ were taken from the experiment; the value of $\widetilde T{=}20$\,K determines the temperature at which the Aronov--Altshuler correction is zero, and the coefficient $\alpha =1/2$ was selected so as to obtain the agreement of the calculated results with the experimental curve in the field of 7~T. The resultant set of curves shown in Fig.~39c possesses features inherent in the family of experimental curves displayed in Fig.~39b: the curves break out into those that are bent downward, and those that are bent upward, whereas the magnetoresistance is negative in strong fields at low temperatures. Notice that we obtained correct scales of the variation of the resistance depending on the temperature and field strength, and also the `correct' region of the appearance of negative magnetoresistance.

This group of materials with the intermediate type of transition also includes the two-dimensional superconducting electron system at the interface between two layered oxides, $\rm LaAlO_3$ and $\rm SrTiO_3$, which are both insulators. The (100) surface of single-crystal $\rm SrTiO_3$ terminated by a $\rm TiO_2$ layer was coated with an $\rm LaAlO_3$ film with a thickness of more than four unit cells [150]. The density of two-dimensional electron gas at the thus-created interface could be changed by applying a voltage across the gate deposited onto the back part of the $\rm SrTiO_3$ crystal.

\begin{figure}
\includegraphics{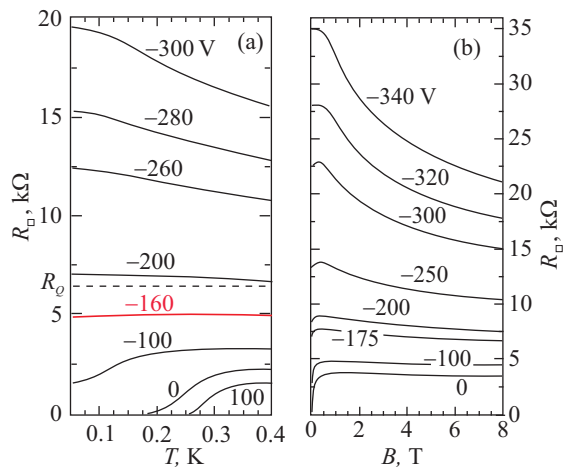}
\caption{(a) Temperature dependence of the resistance of a two-dimensional electron gas in an $\rm LaAlO_3\!-\!SrTiO_3$ heterostructure at various concentrations of the electron gas [150]. To facilitate the comparison of these curves with the magnetoresistance curves, the gate voltage determining the electron concentration is indicated alongside each of them. $R_{\rm Q}=2\pi\hbar/4e^{\,2}\approx 6.45\,{\rm k}\Omega$. (b) Magnetoresistance in the $\rm LaAlO_3\!-\!SrTiO_3$ heterostructure at various concentrations of the electron gas at a temperature $T=30$\,mK [150].}
 \label{LaAlO}
\end{figure}

Figure 40a displays eight curves obtained at different gate voltages from the set of 35~curves published in Ref.~[150]. These curves demonstrate the full set of the possible behaviors of the films, i.e., a sharp increase in the resistance with a decrease in temperature [the flattening of the curve corresponding to the gate voltage equal to $-300$\,V at low temperatures is explained by a size effect (see inequality (64) and Ref.~[64])], a comparatively slow logarithmic increase in $\delta R\propto -\ln T$, the emergence of superconducting fluctuations, and, finally, the superconducting transition whose temperature increases with increasing two-dimensional electron density at the interface.

Like in ultrathin films, the superconducting system discussed here is undoubtedly two-dimensional. This means that the transition to the superconducting dissipationless state in the zero field occurs in two stages: Cooper pairs are formed at a temperature $T=T_{\rm c0}$, but the dissipation remains finite due to the motion of vortices, diminishing with decreasing temperature, to the temperature of the BKT transition, $T_{\rm c}<T_{\rm c0}$ (see Section 3.2). Although the resistance in the vicinity of $T_{\rm c}$ is several orders of magnitude less than the normal resistance $R_{\rm N}$ of the film, it remains reliably measurable. This makes it possible, by using the formula [151]
\begin{equation}\label{GaNel}       
    R(T)\propto\exp\left(\frac{b_R}{(T-T_c)^{1/2}}\right),
\end{equation}
where $b_R$ depends on the difference $T_{\rm c0}-T_{\rm c}$ and the dynamic parameters of the vortex system, to determine $T_{\rm c}$ in each separate state and thus to find the dependence of $T_{\rm c}$ on the control parameter, which in this case is the electron concentration n. According to formula (65) (see also Fig.~14), this makes it possible to determine the product $z\nu$ of the critical exponents. Such a procedure, which was followed in Ref.~[150], yielded a value of $z\nu =2/3$ for an electron system in the heterostructure $\rm LaAlO_3-SrTiO_3$ in the zero field:
$$T_c\propto(n-n_c)^{2/3}.$$

The procedure described is an alternative to the one usually utilized, in which the resistance is represented as the function of a scaling variable (61) (see, e.g., Fig.~19a). It would be of interest to compare the values of $z\nu$  obtained by these two methods. However, we are not aware of such experiments at present.

The dependence plotted in Fig.~40b demonstrates a negative magnetoresistance in this two-dimensional system, which is similar to that observed in In--O [109, 110] and Be (see Fig.~23) [99]: positive in weak fields, and negative in strong fields, but substantially lower in magnitude. Both the absolute and relative values of the negative magnetoresistance increase when moving away from the superconducting region. In this respect, the electron system in the $\rm LaAlO_3\!-\!SrTiO_3$ heterostructure resembles the ultrathin Be films.

On the other hand, the magnetoresistance of the $\rm LaAlO_3\!-\!SrTiO_3$ heterostructure behaves just as it does upon the destruction of weak localization by a magnetic field. As a result of a strong disorder and frequent events of elastic scattering, the areas of the closed diffusion trajectories are very small, which shifts the process of destroying the weak localization to the strong-field region. This inference is applied to all examples of negative magnetoresistance in this section.

\subsection{4.5 Current--voltage characteristics and nonlinear phenomena }

A quantum superconductor--insulator transition occurs between two opposite extremely nonlinear states of a medium: an increase in current in a superconductor to a certain limit does not lead to the appearance of a voltage, and the increase in voltage in the insulator at $T=0$ does not lead to the appearance of a current until the potential of the electric field creates the possibility of transitions between localized states. Let us simulate both these states using single tunnel junctions. As a model for the superconducting state, we take a tunnel junction with two superconducting sides. If this junction is in the Josephson regime, the current $J$ through it can increase at a zero voltage $(V=0)$ up to a critical value $J_{\rm c}$; then, the voltage moves jumpwise into a linear characteristic $J=V/R$ (curve S in Fig.~41). A similar junction between a normal metal and a superconductor can serve as a model for an insulator. The existence of one superconducting bank emphasizes that we are dealing with an insulator that was formed with the participation of a superconducting interaction. At a voltage $V\approx\Delta /e$, the current through the junction, which first was equal to zero ($J=0$), also gradually moves into a linear characteristic (curve I in Fig.~41).

  \begin{figure}[h]
\includegraphics{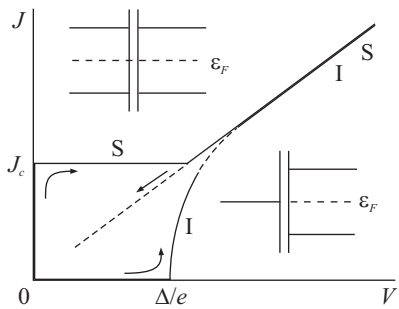}
\caption{Superconductor--superconductor (in the upper part of the figure) and normal metal--superconductor (in the right-hand part of the figure) tunnel junctions, and a schematic of their zero-temperature current--voltage characteristics.}
 \label{SNI}
\end{figure}

Both characteristics are realized in the represented form at a current (characteristic S) or voltage (characteristic I) that is specified from outside. With a decrease in the external action, a hysteresis usually arises, and the representative point approaches the origin along the dashed line (shown in Fig.~41 by arrows). Hystereses also arise when studying the current--voltage characteristics of the substance in the state near the superconductor--insulator transition, rather than the characteristics of the junctions.

The curves S and I in Fig.~41 describe the current--voltage characteristics of the corresponding junctions very roughly, since many important factors are ignored here. However, these over-simplified representations demonstrate one important feature of the nonlinear properties of these states: if the axes $J$ and $V$ are interchanged, the curves S and I pass into one another (this feature was already mentioned in Section 3.2).

In the physics of metals and semiconductors, the density of states on the sides of a tunnel junction is measured with the aid of differential current--voltage characteristics $\partial J/\partial V(V)$. Above, we have already considered similar experimental curves (see Figs~20 and 21). In the superconducting junctions and materials, because of the presence of supercurrents, it is necessary to assign the current rather than voltage, when studying current--voltage characteristics. Therefore, it is usually the function $\partial V/\partial J(J)$ that is measured, whose interpretation is somewhat more complex, in spite of the above-noted symmetry.

  \begin{figure}
\includegraphics{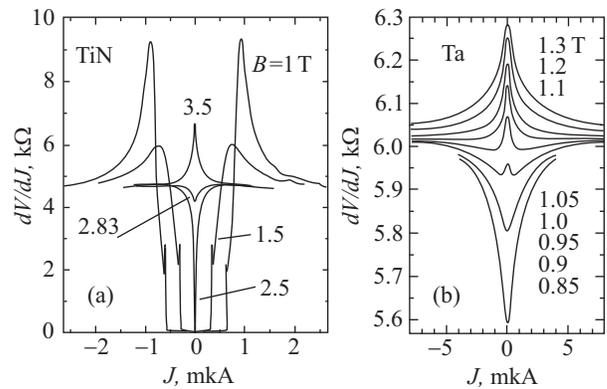}
\caption{Differential resistance $\partial V/\partial J$ of films near the superconductor--insulator transition in various magnetic fields as a function of a dc current through the film. The measurement temperature was less than 0.1\,K. (a) TiN film 5\,nm thick with a superconducting transition temperature $T_{\rm c}\approx 1$\,K [152]; signs of hysteresis phenomena are seen in the curves at $B=1$ and 1.5\,T. (b) Ta films 5\,nm thick with $T_{\rm c}\approx 0.67$\,K [154].}
 \label{SpecificVAC}
\end{figure}

Figure 42a, which was taken from Ref.~[152], displays the results of measurements of the function $\partial V/\partial J(J)$ in a TiN sample with a superconducting transition temperature $T_{\rm c}$ of about 1~K and a critical field $B_{\rm c}$ of approximately 2.9~T. In the field of 1~T, a current smaller than a certain critical value, $J<J_{\rm c}\approx 0.6\;\mu$\,A, flows through the sample without resistance. The current $J>J_{\rm c}$ destroys superconductivity. As the current increases, the response curve ${\rm d}V/{\rm d}J(J\,)$, having passed through a maximum, acquires a value corresponding to the resistance of the normal state, which amounts to approximately 4.6\,k$\Omega$. It should be emphasized that the response curve ${\rm d}V/{\rm d}J(J\,)$ does not relate to the density of states, and the presence of a maximum is connected with a redistribution of current over the section of the film, which is accompanied by a gradual decrease in the proportion of the supercurrent (the film width is 50\,$\mu$m, which is substantially more than the London penetration depth).

With strengthening field, the width of the interval of the superconducting currents decreases (curves at B=1.5 and 2.5~T) and in the vicinity of the critical field $B_{\rm c}$ the minimum of ${\rm d}V/{\rm d}J(J\,)$ near $J=0$ transforms into a maximum. The same transformation of the ${\rm d}V/{\rm d}J$ curves in the vicinity of the zero current in fields of order $B_{\rm c}$ was also observed in other materials in the neighborhood of superconductor--insulator transitions, e.g., in InO [153] and Ta [154] (Fig.~42b).

The right-hand part ($J>0$) of the $\partial V/\partial J(J\,)$ plots in Fig.~42b strongly resembles the fan of the $R(T\,)$ curves arising with a change in the magnetic field strength used as the control parameter (cf., for example, Fig.~25). In Ref.~[153], both series of curves were obtained using one and the same InO sample. A comparison showed that if we make a transformation $T\propto J^{\;0.4}$, then the curves are superimposed on each other rather well. This made it possible to explain the evolution of the current--voltage characteristics similar to those that are shown in Fig.~42b by an overheating of the electron system relative to the ambient temperature, via constructing one series of curves with the aid of calculations based on another series.

The electron temperature $T_{\rm e}$ is determined by the balance between the Joule heat $VJ=J^{\,2}R(T_{\rm e})$, which is liberated in the sample, and the energy flux $Q$ from the electrons to the phonons:
\begin{equation}\label{Nonline1}        
  Q=\alpha(T_e^5-T^5)=J^2R(T_e),
\end{equation}
where $\alpha$ is the proportionality factor, the phonon temperature is assumed to be equal to the ambient temperature $T$, and the resistance is assumed to be dependent only on $T_{\rm e}$. Using the experimentally obtained functions $R(T_{\rm e})$ and equation (111), which implicitly assigns $T_{\rm e}(J\,)$, it is possible to calculate the current--voltage characteristics $\partial V/\partial J$ as functions of the argument $\alpha ^{\,-1/2}J$ containing $\alpha$ as an adjustable parameter, and to compare them with the experimental curves.

Interest in the current--voltage characteristics near the superconductor--insulator transition was stimulated, in particular, by the fact that the scaling relationships for the conductivity or for the resistance near the transition point can be generalized by including dependence on the electric field strength $E$ [71]. For two-dimensional superconductors, the generalized expression for the resistance is as follows [71, 141]:
\begin{equation}\label{R-BTE}       
 R(B,T,E)=
 R_cF\left(\frac{\delta x}{T^{1/z\nu}},\frac{\delta x}{E^{1/(z+1)\nu}}\right).
\end{equation}
(The above-considered expressions (77) and (102) are obtained from expression (112) if the field strength $E$ is assumed to be small and fixed.) It appeared very interesting to apply two independent scaling procedures to determining two different combinations of critical exponents: $z\nu$  and $(z+1)\nu$. However, the experiments [153] showed that an increase in the field strength $E$ immediately leads to a deviation of the electron temperature $T_{\rm e}$ from the ambient temperature $T$. Since it is precisely $T_{\rm e}$ that should be used in formula (112) as the temperature, it is hardly possible to investigate the dependence of function (112) on the second argument at the constant first argument.

Apparently, it is precisely the overheating of the electron system and the deviation of its temperature from the ambient temperature that also explain the hysteresis phenomena that were observed first in InO [155] and then in TiN [116] deep in the insulator region at very low temperatures.

\begin{figure}
\includegraphics{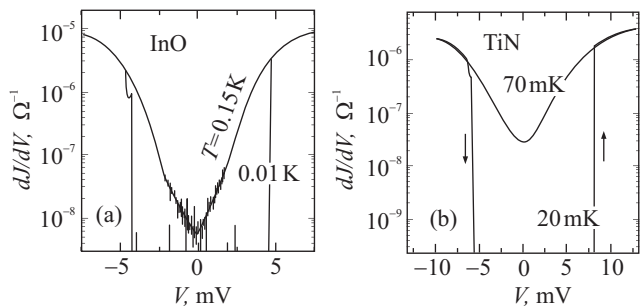}
\caption{Differential conductivity ${\rm d}J/{\rm d}V$ of an InO film in a magnetic field of 2\,T at two temperatures (indicated alongside the curves) [155]. (b)~Differential conductivity ${\rm d}J/{\rm d}V$ of a TiN film in a magnetic field of 0.9\,T [116].The difference in the voltages at which the upward and downward jumps occur indicates the existence of a hysteresis.}
 \label{Superins}
\end{figure}

The InO sample, whose ${\rm d}J/{\rm d}V(V\,)$ curves are presented in Fig.~43a, was superconducting in the zero magnetic field, but its critical field amounted only to 0.4~T. This means that the sample was very close to the superconductor--insulator transition (sample of type 1 according to the classification of Fig.~30a; white square in Fig.~30b). The desired current--voltage characteristics were recorded in a magnetic field of 2~T, i.e., deep in the insulator region. Therefore, it is the voltage $V$ across the sample rather than the current $J$ that was the regulated variable here. The ${\rm d}J/{\rm d}V(V\,)$ curves are strongly temperature-dependent: at the low-temperature characteristic at a voltage $V\approx 5$\,mV, there is a jump between the upper and lower branches; the lower branch cannot be fixed at all, since the signal decreases by more than three orders of magnitude. At a temperature $T=150$\,mK, which should be considered `high' in this case, the jump is observed no longer, although the instability and the telegraphic noise at small $V$ are retained.

Analogous curves were obtained for TiN (Fig.~43b) in the sample whose representative point was also located very close to the quantum transition point, but on the insulator side. Its resistance in the zero field changed according to the activation law (99) with an activation energy $T_0^{\,\rm I}=0.25$\,K. At a temperature of 20~mK, jumps also occur between the two branches, and again the lower branch of the characteristic is located below the noise level. A hysteresis is highly visible in Fig.~43b, i.e., a difference in the voltages at which the jumps upward and downward occur. The position of the jumps depends on the magnetic field applied to the sample.

It was initially assumed that the jumps indicate the transition of the system of localized carriers to a highly correlated state. Later on, another interpretation was suggested by Altshuler et al. [156]. It was shown in Ref.~[156] that in the case of an exponential dependence (99) of resistance on the temperature the deviation of the electron temperature $T_{\rm e}$ from the phonon temperature $T_{\rm ph}$ leads to an S-like current--voltage characteristic $J(V\,)$ and to a bistability. The concrete predictions made in Ref.~[156] were confirmed by measurements on InO [157].

\section{5. Related systems }

\subsection{ 5.1 Regular arrays of Josephson junctions}

Formally, even a single Josephson junction is a device in which it is possible to accomplish a superconductor--insulator transition. Indeed, let us turn to the curve S in Fig.~41, which schematically depicts the current--voltage characteristic of a Josephson junction. With the flow of a Josephson dc current $J\le J_{\rm c}$ through the junction, the potential difference between the banks of the junction is equal to zero, so that a superconducting state is realized in the junction. However, the Josephson current can be suppressed in some way, although preserving the superconductivity of the banks. Then, the current--voltage characteristic of the junction will follow curve I in the same Fig.~41 with a current jump at a voltage $V=2\Delta /e$, and this can be considered to be the realization of an insulating state.

The Josephson current can be suppressed, for example, by changing the coupling of the junction with the dissipative environment [21, 158] or by changing the environment itself. To do this, an experimentalist has at his disposal a whole series of parameters, e.g., the Josephson energy $E_{\rm J}$ and the Coulomb energy $E_{\rm C}$ of the junction itself:
\begin{equation}\label{arr1}        
 E_{\rm J}=\frac{\pi}{4}\left(\frac{\hbar/e^2}{R_n}\right)\Delta,\qquad E_C=e^2/2C
\end{equation}
($R_{\rm n}$ and $C$ are the normal resistance and the capacitance of the tunnel junction, respectively), and also the shunt resistance $R_{\rm sh}$ whereby he can simulate the external source of dissipation. By varying these parameters, it is possible to make a `nonconducting' junction from a `superconducting' junction and even to construct a phase diagram for the states of a single junction [159].

The development of experimental methods made it possible to create one- and two-dimensional periodic arrays from identical Josephson junctions. On this basis, there arose a separate branch of the physics of superconductivity, with a rich variety of physical phenomena (see, e.g., the review [160]). We here only briefly consider the ideas and the results that have a direct relation to the subject of this review and are concerned only with the systems whose properties allow comparison with the properties of continuous films.

Let us begin with two-dimensional systems. Let us imagine a square array with the number of cells on the order of $200\times 50$, in whose nodes the islands of an aluminium film are located, which are connected between themselves through Josephson tunneling junctions ${\rm Al\!-\!AlO}_x\!-\!{\rm Al}$, placed in the middles of the edges of a mesh. The typical sizes, taken from Ref.~[161], are as follows: the area of a unit cell $s_{\rm cell}\approx 4\;\mu$m$^2$; the area of an island $s_{\rm isl}\approx 1\;\mu$m$^2$; the area of a tunnel junction $s_{\rm tun}\approx 0.01\;\mu$m$^2$, and its Coulomb energy $E_{\rm C}$, on the order of 1~K. The arrays of another research group were several times less in area $s_{\rm cell}$ of the cell and approximately the same for the values of $s_{\rm tun}$ and $E_{\rm C} [162]$.

Special experiments showed that it is possible to ensure that the spread in the parameters would not exceed 5\%. This array, in essence, is similar to a granular superconducting film in which all granules are strictly identical and have an identical temperature $T_{\rm c}$ of the superconducting transition, superconducting gap $\Delta$, number of nearest neighbors, etc.

\begin{figure}
\includegraphics{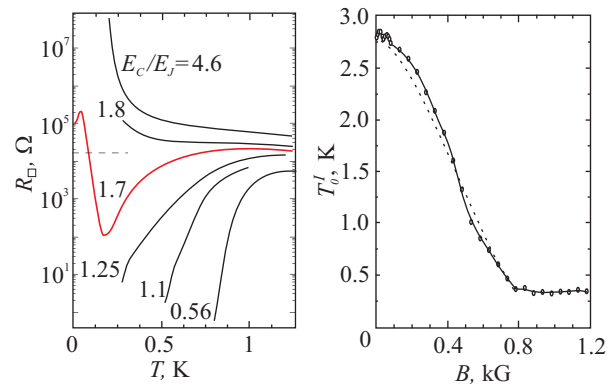}
\caption{(а) Zero-field temperature dependence of the resistance of six square Josephson junctions arrays consisting of $190\times60$ cells each, differing in the energy ratios
$x=E_C/E_J$, [161]. The horizontal dashed line marks the value of universal resistance
$R_{\rm un}=c_uR_Q=16.4$~k$\Omega$, where $R_Q$ is defined by formula (87),and the value
$c_u=8/\pi$ is taken from Ref. [73]; (b) a decrease in the activation energy $T_0^I$ in the Arrhenuis law (114) with strengthening magnetic field for a Josephson array in an insulating state [162]. The dotted curve corresponds to the calculated quantity $0.25E_C+\Delta(B)$}
 \label{Arr25}
\end{figure}

Figure 44a displays temperature dependence of the normalized resistance of six such arrays of identical size, differing in the energy ratios $x{=}E_{\rm C}/E_{\rm J}$. The superconductor--insulator transition demonstrated by this set of curves is very similar to those that occur in continuous films. In principle, this is rather natural, if we take into account that the model of a granular superconductor [45] discussed in Section 2.2 is also entirely applicable to Josephson junction arrays. The arrays, from the viewpoint of this theory, are simultaneously both simpler and more complex objects than a continuous granular film. The simplicity consists in the parameters of all granules--cells being identical and measurable independently; all nonzero constants $B_{i\,j}$ and $J_{i\,j}$ entering into Hamiltonian (19) are also identical and are determined by the capacitance $C$ and normal resistance $R_{\rm n}$ of a junction, respectively. An additional complexity lies in the fact that the array is a multiply connected object and its unit cell is by no means determined only by the parameters (113) of the junction itself; the energies $E_{\rm C}$ and $E_{\rm J}$, on the one hand, and the areas of the cell $s_{\rm cell}$ and island $s_{\rm isl}$, on the other hand, are completely independent. The energies $E_{\rm C}$ and $E_{\rm J}$ can be sufficiently small ($E_C\sim E_J\lesssim\Delta$), but the areas $s_{\rm cell}$ and $s_{\rm isl}$ can remain comparatively large. Accordingly, the theory of transport phenomena in such arrays [163, 164] does not reduce to the theory of granular superconductors.

We analyzed the temperature dependence of resistance at low temperatures in two arrays which behave, judging from the temperature dependence of their resistance shown in Fig.~44a, as insulators. It was found that this dependence follows an activation law
\begin{equation}\label{arr2}        
 R_\Box\propto\exp(T_0^I/T),\qquad T_0^I=\Delta+0,25E_C.
\end{equation}
Relation (114) for $T_0^{\,\rm I}$ is by no means a numerical coincidence. This relation was observed independently by three experimental groups [161, 162, 165] in different square arrays. This means that even in arrays--insulators the aluminium islands remain superconducting and serve as containers for Cooper pairs. For an electron to tunnel from one island to another, first, there should occur a destruction of the pair, which requires an energy $\Delta$  per electron, and, second, there should occur a redistribution of effective charges in all tunnel capacitances (which requires an additional energy $E_{\rm C}/4)$.

In a magnetic field, the coefficient $T_0^{\,\rm I}$ in the Arrhenius law (114) decreases to $0.25E_{\rm C}$; the magnetic field destroys the superconductivity in aluminium islands and makes the gap $\Delta$ vanish (Fig.~44b). Thus, negative magnetoresistance can exist in insulating arrays, as well. In contrast to continuous films (e.g., InO; see Figs~27, 30), these arrays behave as insulators in strong fields, as well, since the normal electrons, because of the Coulomb blockade, remain localized in the islands. A similar behavior was also observed in granular superconductors (see Fig.~5).

Let us now return to Fig.~44a, according to which the critical value of the control parameter is $x_{\rm c}\approx 1.7$. The $R(T\,)$ curve at $x$ near this value has additional features, namely, the decrease in the resistivity related to the development of a superconducting state is replaced by an increase in the resistance by three orders of magnitude at temperatures below 200~mK, and then, for $T<40$\,mK, the resistance continues decreasing. Such a behavior is called a double reentrant transition. Double reentrant transitions were also observed in continuous films, e.g., TiN [113]. However, there is no complete clarity of this issue to date; it is assumed that such transitions are due to an inhomogeneous granular structure. In this sense, the experiments on arrays have one advantage: the structure of arrays can be controlled much better. However, it is unclear to which extent the double reentrant transitions are reproducible on different arrays.

Recall that the reentrant behavior near the transition point in a granular superconductor was predicted in Ref.~[45] (see Fig.~8 and comments on it in Section 2.3 and Refs~[46--48]).

The difference between the arrays and continuous films is especially substantial in very weak magnetic fields perpendicular to the array plane. The field is concentrated in the regular periodically repetitive holes of the array. Since the holes are surrounded by superconducting rings, the magnetic flux through them is quantized, so that an integer number of vortices passes through each hole, each vortex containing one magnetic flux quantum $\Phi _0=2\pi \hbar /2e$. Therefore, for measuring the field $B$ in the arrays, the concept of frustration $\,f$, the average number of magnetic flux quanta per array's cell, is used:
\begin{equation}\label{frustration}     
 f=B/W\Phi_0,
\end{equation}
where $W$ is the number of cells per cm$^2$. The characteristic value of the field B depends on the dimensions of the cell but, generally speaking, it is very small: the frustration $\,f=1$ usually corresponds to a field induction $B$ in the range from $\approx 4$\,G [161, 166] to $\approx 40$\,G [162, 167].

\begin{figure}[b]
\includegraphics{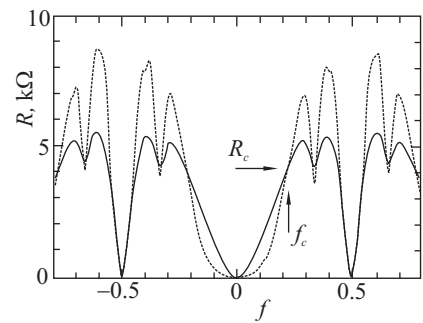}
\caption{Resistance of a Josephson junction array with a normal resistance $R_\Box=10.5$\,k$\Omega$  and parameter $x=0.9$ as a function of frustration $f$ at a temperature of 180\,mK (solid curve) and 80\,mK (dashed curve) [166]. The arrows indicate the critical values of the resistance $R_{\rm c}$ and frustration $f_{\rm c}$ determined as the coordinates of the point of intersection of the isotherms.}
 \label{Arr3}
\end{figure}

The measurements of the transport properties of arrays in a magnetic field [161, 162, 166, 167] showed that the resistance $R(\,f\,)$ has minima at those values of frustration that are described by rational fractions: $f=f_{nm}=n/m$, where $n$ and $m$ are integers, and is periodic in $f$ in the sense that in the vicinity of the values $f$ and $f+1$ the function $R(f)$ behaves alike. Figure 45 presents, as an example, $R(f)$ isotherms for two temperatures, 0.08 and 0.18~K, on a square Josephson junction array ($x=0.9$, i.e., $x<x_{\rm c}$). It is seen that the array resides in the superconducting state not only in the zero field at $f=0$, but also at $f_{12}=0.5$, when vortices exist in each second cell of the array. When the lattice of vortices is commensurate with the lattice of holes, then the lattice of vortices is rigidly pinned, and the magnetic field is stationary located outside of the superconducting film and in no way influences the superconductivity of the array. Small changes in a magnetic field disrupt the commensurability of the lattices and the vortices become mobile, which leads to dissipation accompanying the flow of current through the lattice of junctions.

The $R(f)$ curve (see Fig.~45) also exhibits minima at $f=f_{13}=1/3$ and $f=f_{23}=2/3$. Their depth depends on the quality and number of periods of Josephson lattice and on the temperature. Under favorable circumstances, the resistance at these points can also reach zero. The heights of the local maxima of the resistance also depend on the same factors: one can see clearly from Fig.~45 that a temperature decrease leads to their growth.

Thus, a change in the magnetic field gives rise to a chain of phase transitions between superconducting states at $f=f_{nm}$ (certainly, with sufficiently small n and m) and insulating states in the case of the incommensurability of the lattices of vortices and junctions. For this to occur, magnetic fields are required that are several thousand times weaker than those that cause analogous chains of transitions under the conditions of the quantum Hall effect (see, e.g., the review [7]).

It is easily seen that the picture represented strongly provokes the introduction of the idea of the vortex--electron pair duality: the vortices are localized at $f=f_{nm}$ and the pairs ensure superconductivity; the pairs exist for sure in the insulating state and are localized for sure in the islands. It only remains to suppose that the vortices can be superconducting and that the transition to the insulating state is certainly caused by their delocalization. The duality, however, implies that the vortex system allows a representation in the form of a gas of quasiparticles. It should be noted that the array gives more grounds for such representation than a continuous film. When moving in the film, a vortex is always in the dissipative medium, while moving over an array, it mostly exists outside of the film (see, in particular, Ref.~[168] in which for arrays of a special form it was possible to derive a dual transformation exactly).

\begin{figure}[t]
\includegraphics{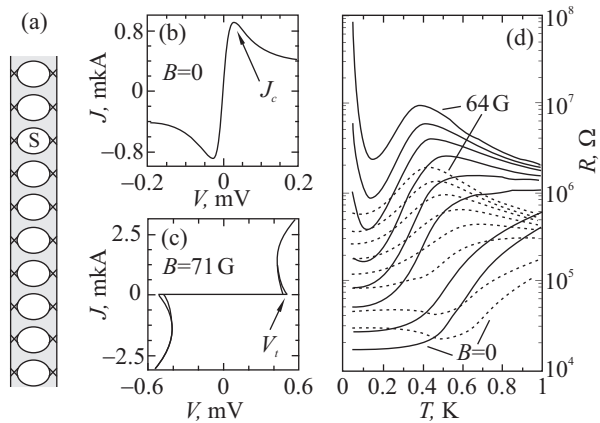}
\caption{(a) Schematic of a one-dimensional chain of paired Josephson junctions with cells each containing a hole with an area $S$ and two Josephson junctions along the perimeter of the hole. (b) Current--voltage characteristic of a chain with a length of 255~cells in the superconducting state in a zero magnetic field at $T=50$\,mK; $J_{\rm c}$ is the critical superconducting current [169]. (c) Same, in a magnetic field $B=71$\,G, i.e., in the insulating state (the critical field of the superconductor--insulator transition is approximately 62~G [169]); $V_{\rm t}$ is the threshold voltage. (d)~Temperature dependence of the resistance of two one-dimensional chains of paired Josephson junctions of various lengths (255~cells, solid curves; and 63~cells, dashed curves) in different magnetic fields. The field values from bottom to top: $B=0$, 27, 47, 53, 57, 60, 62, and 64~G [169].} \label{Havi1D}
\end{figure}

Naturally, the same technique makes it possible to prepare one-dimensional arrays of Josephson junctions. Transitions in such systems were studied in Ref.~[169]. The sample depicted in Fig.~46a has the form of a strip consisting of aluminium islands, each connected with its neighbors to the left and to the right through two parallel tunnel junctions $\rm Al/Al_2O_3/Al$. These two junctions implement a Josephson coupling between the adjacent elements of the one-dimensional system. In this case, the effective Coulomb binding energy $E_{\rm C}= e^{\,2}/2C$ is determined by the total capacitance of two parallel junctions, and the effective Josephson energy $E_{\rm J}$ can be varied using a magnetic field, since it depends on the magnetic flux $BS$ passing through a hole with an area $S=0.12\;\mu $m$^2$, along whose perimeter the junctions are located:
\begin{equation}\label{1D-Haviland}     
\begin{array}{c}
 E_J=E_{J0}|\cos(\pi BS/\Phi_0)|,\\
\Phi_0=\rule{0pt}{5mm}2\pi\hbar/2e=20.7\;G\,{\rm \mu}^2.
\end{array}
\end{equation}
The theoretical model describing this system is discussed in detail in the review [65] as the simplest example of a quantum phase transition.

In Ref.~[169], three identical chains of different lengths (containing 255, 127, and 63~junctions) were measured. Figures 46b and 46c display the current--voltage characteristics of the longest chain. In the zero magnetic field, the superconducting current $J_{\rm c}$ reaches approximately 0.8\,$\mu$A (Fig.~46b). $J_{\rm c}$ diminishes with strengthening field, to become zero at a field of about 62~G. Then, the current--voltage characteristic changes radically: a section with $J(V\,)=0$ ($|V|<V_{\rm t}$) appears in it (Fig.~46c). The threshold voltage $V_{\rm t}$ increases with strengthening field, reaching a maximum at $B=86$\,G. This is that induction of the field at which, according to formula (116), the energy $E_{\rm J}$ becomes zero.

Figure 46d displays the temperature dependence of the resistance of two chains of different lengths in magnetic fields that include the field of the superconductor--insulator transition. All the curves referring to the short chain flatten out at low temperatures:
$$R(T)\approx{\rm \,const}\quad\mbox{при}\quad T<T_{\rm short},$$
where $T_{\rm short}$ decreases gradually from 0.3~K at $B=0$ to 0.15~K at $B=64$\,G. This should be expected according to the scaling hypothesis, since the argument of the arbitrary function in Eqn~(63) for $L<L_\varphi$ is $\xi /L$ rather than $\xi /L_\varphi$, and $\xi$  is temperature-independent in the case of one-parametric scaling. On the contrary, the resistance increases in the long chain at the lowest temperatures and in the fields $B\gtrsim60$\,G. This means that inequality (64) for the long chain at temperatures down to the lowest ones used in measurements was fulfilled, at least, on the insulator side. The last limitation is connected with the fact that on the superconducting side the $R(T\,)$ curves for the long chain also come to a constant level for $T<T_{\rm long}$. However, $T_{\rm long}<T_{\rm short}$ in all the fields.

\subsection{5.2 Superconductor--insulator type transitions in an atomic trap}

The terminology used in the consideration of quantum phase transitions recently appeared in atomic physics in connection with experiments on the Bose condensation of a gas of ultracold atoms. The authors of Ref.~[170] discussed the possibility of establishing conditions for atoms, which resemble those under which electrons exist in solids and which lead to quantum phase transitions. Soon after, this experiment was realized in Ref.~[171].

The rarefied gas of ${}^{87}{\rm Rb}$ atoms was subjected to laser cooling and placed into a magnetic trap in which the atoms were retained because of the presence of a magnetic moment in them. The total number of bosonic atoms in the trap, ${\cal N}\approx 2\cdot 10^5$, was much fewer than that in $1\;\mu$m$^3$ of the substance, but it was sufficient for the statistical laws to be applicable to them. A three-dimensional crystal lattice was imitated with the aid of three standing linearly polarized optical waves with a wavelength $\lambda \approx 852$~nm that were orthogonal to each other and had mutually orthogonal polarizations. The neutral atoms in the field of an electromagnetic wave acquire an electric dipole moment proportional to the field strength. The force acting on the atom is determined by the product of the dipole moment by the field gradient. The potential for the atoms, which is proportional to the sum of the intensities of all three waves, takes the form of the potential of a simple cubic lattice:
\begin{equation}\label{trap1}       
 U(x,y,z)=
 U_0\left(\sin^2(kx)+\sin^2(ky)+\sin^2(kz)\right),
\end{equation}
with $k=2\pi/\lambda$; $U_0$ is the depth of the potential well in one standing wave of the laser field. This depth can be varied by changing the intensity of the laser waves. It can be conveniently characterized by comparing it with the kinetic energy $E_{\rm k}=\hbar ^{\,2}k^{\,2}/2m$ of an atom, which is assigned by the laser wavelength $\lambda$ and by the atomic mass $m$. In the experiment under consideration, the well depths could be changed from zero to $22E_{\rm k}$. On the whole, the trap contained about 150,000 sites with an average number of atoms of approximately 2.5 per site in the center of the trap.

Since the potential for the bosonic atoms, created by standing laser waves, does not contain disorder, the system designed in Ref.~[171] corresponds to the model of bosons on a lattice of sites [55], which was discussed in Section 2.4. The corresponding phase diagram is given in Fig.~9a. The growth of the hopping frequency $J$ occurs with a decrease in the well depth of the periodic potential (117). According to the predictions of the theory [55], at small $J$, i.e., at a large amplitude $U_0$ of the periodic potential, all bosons are localized in the wells and together form a Mott insulator; as $U_0$ decreases to a critical value, the bosons are delocalized and pass into the Bose condensate.

For determining the degree of the coherence of atomic wave functions, a testing laser was utilized. In order to eliminate the influence of the structure-forming periodic potential, it was sharply switched off, so that the atomic wave functions began to evolve in the free space. Since the temperature was very low and the kinetic energy of the atoms was very small, the evolution proved to be comparatively slow, and it was possible to fix the interference pattern that appeared as a result of diffraction of the laser beam by the atomic system with that degree of coherence of wave functions that was formed against the background of the periodic potential.

\begin{figure}[h]
\includegraphics{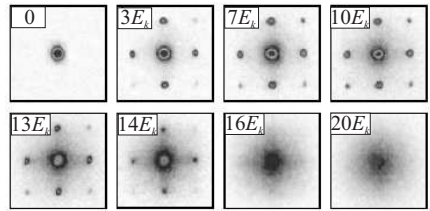}
\caption{. Diffraction of a testing laser beam in a system of ultracold ${}^{87}{\rm Rb}$ atoms depending on the amplitude of the periodic potential created by standing laser waves; the wave amplitude is indicated in the upper left-hand corner of each pattern [171].}
 \label{ColdAtoms}
\end{figure}

The results of measurements are represented in eight interference patterns (Fig.~47); the amplitude $U_0$ of a periodic potential in which the pattern was formed is indicated in $E_{\rm k}$ units in the upper left-hand corner of each pattern. In the absence of a periodic potential ($U_0=0$), the interference pattern is a result of the diffraction of the laser beam by the unstructured Bose condensate. As long as the amplitude of the periodic potential is small, [$U_0=(3\!-\!10)E_{\rm k}$], all bosonic atoms remain delocalized, but the Bose condensate formed of them exhibits a spatial density modulation. Since the totality of all the atoms comes forward as a single quantum object, the relaxation of the system after switching off the external periodic potential occurs slowly. Therefore, the modulation picture takes the form of a usual Laue diffraction pattern, and the intensity of the side interference maxima grows with increasing amplitude of the periodic potential. In this case, the representative point is located in the phase diagram in Fig.~9a sufficiently far to the right, in the region of superfluidity. According to the notation used in Ref.~[55], the system is superfluid. However, when the minima of the periodic potential become sufficiently deep ($U_0\gtrsim13E_k$), there occurs a localization of the bosonic atoms: the representative point in the phase diagram of Fig.~9a approaches the ordinate axis. The wave functions of the localized bosons are incoherent, and the system rapidly relaxes after switching off the potential. Therefore, the interference structure fades, giving way to the incoherent background $[U_0=(14\!-\!20)E_{\rm k}]$.

It is thus far unclear what problems, besides mere demonstration, can be solved with the conducting of such experiments, but for sure the further development of this avenue will not be long delayed. In any case, experiments have already appeared with ultracold atoms, in which a disorder-tuned Anderson transition is investigated [172, 173].

\section{Concluding discussion}

In this review, we attempted to describe and to compare different theoretical approaches to the issue of superconductor--insulator transitions, conclusions and predictions within the frameworks of different models, and also to enumerate and to systematize experimental facts. In this section, we shall try to summarize available data, by refining the statement of the problem and formulating what may be considered solidly established and what requires additional study. Here, we also describe some comparatively new results which can play a key role in further studies.

\subsection{6.1 Scenarios of the transition}

There is no doubt that the very existence of a quantum superconductor--insulator transition has been established for sure and that, together with the characteristics of the material (such as the film thickness, disorder, charge-carrier concentration), a magnetic field can also play the role of a control parameter. The question is rather why the transitions in various materials occur in different scenarios and what factors determine which of the scenarios is realized.

In the Introduction we have already discussed the division of the transitions into two basic types, fermionic and bosonic, depending on what occurs at the transition point: whether the modulus of the order parameter becomes zero or the amplitude of the fluctuations of its phase reaches a critical value. The critical values of the conductance examined experimentally in the two-dimensional case, which are on the order of 10\,k$\Omega$, cannot apparently help in the selection of a scenario. The values (87) obtained through calculations within the framework of the bosonic model, which is based on the 2e-bosons--vortices duality [70, 73, 75], have the same order of magnitude. However, the logarithmic estimate of the critical conductance in the fermionic scenario, derived in Ref.~[8] on the base of the results obtained in Refs~[11, 40]:
\begin{equation}\label{y_c}     
   y_c=\left(\frac{1}{2\pi}\ln(1/T_{c0}\tau)\right)^2,
\end{equation}
also gives a close value of  $R_{\rm un}$ if we make a reasonable assumption that $\ln(1/T_{c0}\tau)\gtrsim5$.

The basic experimental evidences in favor of the bosonic model are the negative magnetoresistance in strong magnetic fields and the presence of a pseudogap. Although the negative magnetoresistance is also predicted [59] within the framework of the BCS scheme with allowance for superconducting fluctuations in the magnetic field at low temperatures $T\approx 0$, the giant magnitude of the magnetoresistance peak in InO, TiN, and ultrathin Be films makes it necessary to give preference to the explanations that proceed from the bosonic model and to assert that, at least in these materials, on the nonsuperconducting side of the transition there indeed exist equilibrium electron pairs in localized states. The presence of a pseudogap can be established directly, primarily, from the tunnel current--voltage characteristics. It should be noted, however, that the measurements available are undoubtedly insufficient not only to perform a classification of materials on their basis, but even to reliably interpret the characteristics themselves.

In order to speak about the bosonic model, it is probably not necessary that both facts be simultaneously established experimentally, i.e., the presence of a negative magnetoresistance, and the presence of a pseudogap. In this respect, the experimental data on ultrathin Bi films are demonstrative. The negative magnetoresistance in Bi films is virtually absent. However, the current--voltage characteristics of the tunneling contacts on Bi films exhibit some specific features, namely, a finite differential conductivity at a zero bias $V$, which indicates a finite density of states inside the superconducting gap, and the intersection of all current--voltage characteristics at one point in the line $G_{\rm N}=1$ upon variation of the magnetic field, which indicates that the gap is independent of the magnetic field [89]. These features were demonstrated in Figs~20 and 21. In particular, when the control parameter is the magnetic field, the modulus of the order parameter at the transition point does not, apparently, become zero.

Giant negative magnetoresistance arises when a decrease in the binding energy of a pair as a result of the paramagnetic effect leads to delocalization and even to an insulator--metal transition, as in $\rm In\!-\!O$ films [108]. But if the decrease in and the switching-off of pair correlations does not lead to delocalization, then its influence on the transport can be insignificant. This effect was demonstrated theoretically in numerical calculations by the Monte Carlo method: at specific relationships between the parameters, the pairing strongly influences the probability of the localization [64]. As can be seen from Fig.~13, the Anderson localization exists only in the presence of attraction between the electrons in a certain interval of values of the parameter W/t characterizing disorder.

Thus, $\rm In\!-\!O$ and amorphous Bi represent, apparently, two different variants of a bosonic scenario. However, the number of variants is probably not limited to these two cases. Recall that the maximum of the peak of magnetoresistance in $\rm In\!-\!O$ is located on the superconducting side of the transition point in the zero field, while in Be films, on the side of the insulator. We shall return to this issue in Section 6.3.

Characteristically, the transitions in the majority of the materials that were considered in this review stimulate the discussion of precisely the bosonic scenario. The explanation for this can probably be perceived from the diagram in Fig.~2a. If it is a decrease in the efficiency of the superconductive interaction that is the main response of the system to a change in the control parameters, then the system will most probably go from the superconductive state to the metallic, rather than insulating, state. Therefore, transitions in the fermionic scenario should primarily be sought among the materials in which the breakdown of superconductivity yields a `bad' metal; the corresponding examples were given in Section 4.4.

On the other hand, the events in both scenarios are very similar in the immediate proximity to the transition point: the initially uniform system becomes macroscopically inhomogeneous according to the BCS theory (see Sections 2.1 and 2.6); negative magnetoresistance appears in the dirty limit (see Section 2.5), and the Cooper pairs appear in two-dimensional superconductors at a temperature that exceeds the transition temperature $T_{\rm c}\equiv T_{\rm BKT}$ (see Section 1.5).

\subsection{6.2 Role of macroscopic inhomogeneities }

Initially, the theory of transitions in granular superconductors was developing separately and superconductor--insulator transitions in granular and uniformly disordered systems were considering as different phenomena. Gradually, however, it became clear that, first, the physical properties of systems of these two types (their transport properties, magnetoresistance, etc.) near the transition points are very similar, and, second, macroscopic inhomogeneities as a certain kind of granularity spontaneously appear in uniformly disordered systems near the transition point. This `electronic--structural instability' can arise for two reasons: as a result of a strong disorder [62, 63] or electron--electron interaction [44]. As in systems of normal electrons, these two principally different factors lead to analogous consequences.

\begin{figure}[h]
\includegraphics{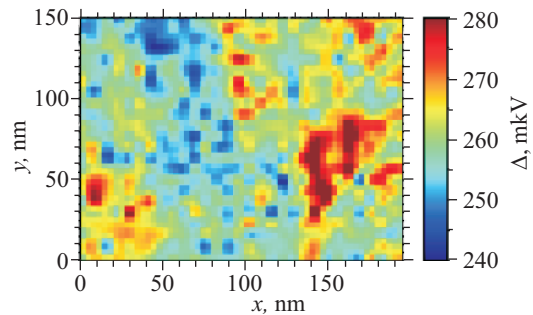}
\caption{Spatial fluctuations of the superconducting gap $\Delta$  in a TiN film [174] }
 \label{TiN-map}
\end{figure}

With the development of the technology of low-temperature tunnel spectroscopy, the possibility appeared of experimentally studying induced macroscopic inhomogeneities. The authors of Ref.~[174] could simultaneously and independently measure the resistance and current--voltage characteristic in a TiN film 5~nm thick with the aid of the Pt/Ir tip of a scanning tunneling microscope mounted in a dilution refrigerator. As a result of strong disorder, the film was in a state close to the superconductor--insulator transition: the superconducting transition in it occurred on the temperature interval of 2--1.3~K, whereas the superconducting transition temperature in the bulk material is $T_{\rm c}^{\,\rm bulk}=4.7$~K [113].

The differential conductivity of the tunnel junction, measured at a temperature of 50~mK, has a usual form: it reveals a superconducting gap inside which the density of states decreases to zero. However, these measurements demonstrate two specific features. First, the average width of the gap, $\tilde\Delta$, was approximately $265\;\mu$eV, in contrast to the value of the gap $\Delta ^{\,\rm bulk}=730\;\mu$eV. The second feature is seen from Fig.~48, which demonstrates the result of scanning of the film surface. The superconducting state proved to be spatially inhomogeneous (see also the earlier study [175]).

A quantitative comparison with the theory has not yet been done, although the material in Ref.~[174] for such a comparison in fact already exists: according to this study, the stronger the disorder, i.e., the nearer the sample to the quantum transition point, the greater the ratio $\tilde {\Delta \;}\!\!/T_{\rm c}$ in it. This result, which confirms the theory developed in Refs~[44, 63], is very important even in the qualitative form.

Notice that the electron system behaves differently near the metal--insulator transition: the electron wave functions become fractal [176].

\subsection{6.3 Localized pairs }

Although the existence of localized superconducting pairs can at present be considered as a recognized fact, the conditions that favor their appearance, their internal structure, and the wave function have not virtually been discussed.

The formation of localized pairs is favored or, on the contrary, prevented by the statistical properties of the random potential. Let us explain what we mean by the example of amorphous In--O in which the strongest shift of the state into the depth of the insulator region was observed under the effect of a magnetic field, with the subsequent appearance of the strongest negative magnetoresistance (see Fig.~28). The structural element of this material is the $\rm In_2O_3$ molecule inside which all valence electrons participate in the formation of covalent bonds and, therefore, are strongly coupled. The chemical composition of the real amorphous substance is described by the formula ${\rm In_2O}_{3-y}$. The fraction $y/2$ of structural units has an oxygen vacancy, and two valence electrons in the immediate neighborhood of each vacancy prove to be weakly connected with the ion core and are easily delocalized, leaving pairwise correlated wells in the random potential. In the case of two other materials with giant negative magnetoresistance (TiN and amorphous Be), there is probably also an analogous `quasichemical' influence on the structure of the random potential.

The possibility of superconducting interaction between localized carriers is merely postulated in many theoretical models. For example, the attraction between the electrons at a separate lattice site was introduced in Hamiltonian (51) without discussing the problem of its origin. When asking this question, it is useful to glance at the problem of Cooper pairing due to the exchange of phonons from another angle, by examining the transition from an insulator to a normal metal on the basis of the wave functions of electrons in a strongly disordered medium [177].

For the realization of a coherent electronic state, condition (3) determining the minimum size (4) of a superconducting particle should be satisfied. In a volume with smaller characteristic dimensions, the superconductivity is already absent, but as long as the spacing between the electron energy levels remains smaller than the energy $\hbar \omega _{\rm D}$ of a short-wave phonon:
\begin{equation}
\label{deltaE}      
  \delta\varepsilon\ll\hbar\omega_{\rm D},
\end{equation}
the superconducting interaction manifests itself in the form of the parity effect [see formulas (5)--(7)]. This means that with switching on a superconducting interaction the phonon attraction mechanism can decrease the energy of the pair of electrons localized at the same site only if condition (119) is satisfied.

For a localized electron, inequality (119) can prove to be too rigid: using formula (3) for the estimation and expressing $\delta\varepsilon$ through a Bohr radius $a_{\rm B}$ of the localized state, we shall obtain a hardly feasible inequality
\begin{equation}\label{aB}
  (g_0a_B^3)^{-1}\ll\hbar\omega_D.
\end{equation}

However, Feigel'man et al. [177] pointed to the fact that limitation (119) can be softened by the proximity to the metal--insulator transition. Indeed, when approaching the metal--insulator transition from the side of the insulator, the localization length $L_{\rm loc}$ grows from $a_{\rm B}$ to infinity. Therefore, returning to the phase diagrams shown in Fig.~2, we can say in the language of these diagrams that to the left of the point $x_{\rm I-M}$ there is an interval of values of the control parameter,
\begin{equation}\label{interval}             
x_L<x<x_{\rm I-M},
\end{equation}
in which the wave functions of normal electrons are localized, but nevertheless are subject to the action of superconducting interaction; the left-hand edge $x_L$ of this interval is determined by the equality $\delta \varepsilon =\hbar \omega _{\rm D}$.

The fractal nature of the wave functions of the localized electrons near the metal--insulator transition can extend this interval. The fractal dimensionality of wave functions is $D_{\rm f}<3$; according to the numerical calculations [178], this quantity is $D_{\rm f}=1.30\pm 0.05$ near a standard 3D Anderson transition. The fractal nature of the wave function increases its significant dimension $L_{\rm loc}$, preserving the volume in which the modulus squared of the wave function differs from zero.
\begin{figure}
\includegraphics{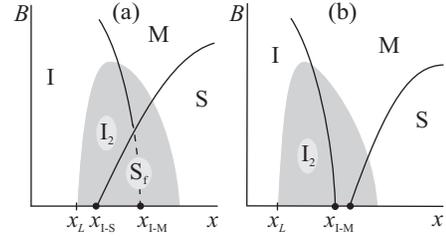}
 \caption{Two variants of an insulator (I)--metal (M)--superconductor (S) phase diagram on the $(x,B)$ plane at $T=0$. Gray regions are those in which the superconducting interaction occurs between electrons with fractal wave functions. Diagram (a) appears to be realized in InO and TiN films, and diagram (b) in Be films.}
 \label{diagr-xB}
\end{figure}

The mutual arrangement of superconductor--insulator and metal--insulator transitions under a change of the control parameter proves, thus, to be one additional essential factor, besides the `quasichemical' one, that is essential for the formation of localized pairs. This arrangement, as we know, can be different. Two of the possible phase diagrams on the plane $(x,B)$ at $T=0$ are represented in Fig.~49. These diagrams differ in the mutual arrangement of the line of the superconducting transitions and the line of the Anderson transitions, which divide the regions of metal (M), insulator (I), and superconductor (S). The region in which the pairing is possible but is affected by the fractality of wave functions is marked out with gray.

Both these phase diagrams can seemingly be realized in practice: the diagram displayed in Fig.~49a is realized in InO and TiN (it can easily be checked that it is precisely this diagram that is depicted in Fig.~30), while the diagram represented in Fig.~49b applies to Be films.

Formally, the above-developed ideas about the localized pairs and the negative magnetoresistance connected with their destruction are applicable to the gray part of region I designated in Fig.~49 as $\rm I_2$. However, the peak of magnetoresistance also exists in the gray part of the metallic region (see, e.g., the experimental data on the magnetoresistance of InO in Fig.~29). The pairing in this region probably occurs according to the strengthened variant of superconducting fluctuations described in Ref.~[59]. Finally, an extremely interesting region $\rm S_f$ exists as well. It was called in Ref.~[177] the region of fractal superconductivity. Its study is only beginning.

\begin{figure}[b]
\includegraphics{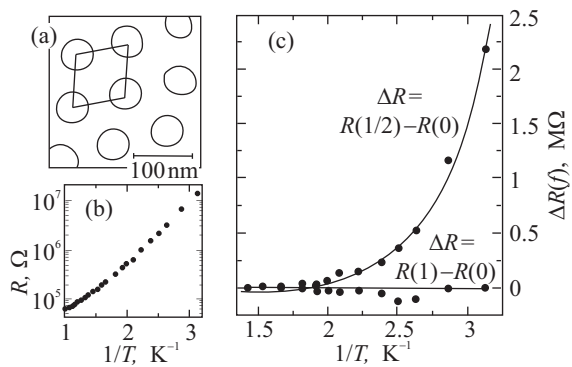}
 \caption{Behavior of a perforated Bi film on the nonsuperconducting side of a quantum phase transition: (a) periodic lattice of holes on the substrate for a Bi film and its unit cell shown by a rhombus; (b)~temperature dependence of the film resistance indicating that the film is in an insulating state, and (c) temperature dependence of the addition $\Delta R$ to the resistance of the film in the presence of a field ($f=1/2$ and $f=1$) as compared to the resistance without a field [179].}
 \label{BiHoles}
\end{figure}

From the viewpoint of an experimental study of the wave function of localized pairs, of great interest are experiments [179] on ultrathin Bi films on anodized aluminium oxide substrates with holes of radius $r_{\rm hole}=23$~nm, which form a periodic lattice with a period of 95~nm (Fig.~50a). The film deposited onto such a substrate also had a periodic lattice of holes. The process of film application and step-by-step testing was described in detail in Section 4.1. For conjugating the film with the substrate, a layer of amorphous Ge coated with an additional Sb~layer 1~nm thick was used. For a control, a substrate without holes was placed nearby, onto which the deposition was produced in parallel and which was also tested after each thickening of the Bi~film.

The sets of $R(T\,)$ curves for the films produced on two substrates are very similar, both to each other and to those that were repeatedly demonstrated above (for example, see Fig.~18). On the thinnest (both continuous and perforated) films, the resistance at low temperatures changes according to the Arrhenius law (99). For an analysis, one such state was chosen (on a substrate with a lattice of holes) not very distant from the transition point. This state is insulating in the sense that the film resistance grows exponentially with a decrease in the temperature (Fig.~50b). However, magnetoresistance oscillations determined by frustration (115) appear in this film in weak magnetic fields: the resistance oscillates with a period $\Delta f=1$ (the concept of frustration was discussed in detail in Section 5.1). The most probable explanation of the frustration dependence of resistance lies in the fact that the magnetic field in the film is structured and is expelled to the holes. According to classical electrodynamics, this means that persistent currents flow around the holes, and it follows from the periodicity of oscillations and quantization conditions (10) and (115) that these currents are formed by electron pairs with a charge $2e$. It turns out that on the scales of $r_{\rm hole}$ superconducting currents exist, whereas on the scale of the sample dimension there are neither superconducting currents nor conductivity at all.

By analogy with the Bohr radius $a_{\rm B}$ of a localized electron, let us designate the attenuation length of the wave function of an isolated localized electron pair as $a_{\rm 2B}$. Because of the overlap of the wave functions of pairs, the attenuation occurs on the scale
\begin{equation}\label{a2B}     
 L_{\rm loc}\geqslant a_{2B},
\end{equation}
which is specified by relationship (54) and is determined by the deviation of the control parameter from the critical value (analogously, the hopping conductivity near the metal--insulator transition is determined by the correlation length $L_{\rm loc}$ rather than by $a_{\rm B}$). The experiment performed in Ref.~[179] makes it possible to estimate the limitation from below on the attenuation length of the wave functions of localized pairs in a concrete film at concrete values of the control parameters, which are given in Fig.~50:
\begin{equation}\label{summ1}
  r_{\rm hole}<L_{\rm loc}<\infty.
\end{equation}
No theoretical explanation of this `local Meissner effect' in a macroscopic insulator exists so far. In particular, it is not clear how inequality (123) is correlated with the penetration depth.
Inequality (122) makes it possible to qualitatively understand the nature of positive magnetoresistance on the left-hand slope of the magnetoresistance peak in $\rm In\!-\!O$ on the field interval
\begin{equation}\label{B-B}     
  B_c>B>B_{\rm max}.
\end{equation}
We have not yet discussed this segment of the magnetoresistance curves $R(B)$ (see Figs~26 and 28).

It is assumed that the conductivity in the field interval (124) is determined by diffusion and hoppings of the localized pairs. Therefore, the decrease in $L_{\rm loc}$ with strengthening field on this interval is accompanied by a decrease in the hopping probability and by an increase in the resistance. In this case, however, there is also an opposite effect of the field action on $L_{\rm loc}$: an increase in the field strength leads to a decrease in the binding energy and a growth in $a_{\rm 2B}$ and, therefore, to an increase in $L_{\rm loc}$. The presence of two opposite effects appears to lead to the expansion of interval (124); its right-hand edge is determined by the field strength at which $L_{\rm loc}\approx a_{\rm 2B}$, so that the first factor is levelled off.

\subsection{6.4 Pseudogap }

The concept of a pseudogap in the vicinity of a superconductor--insulator transition was mentioned above in Section 4.3 in connection with the localization of electron pairs. Since this term is not commonly accepted, let us formulate a definition which will be utilized here. We shall call the pseudogap a minimum, caused by the superconducting interaction, in the density $g(\varepsilon )$ of single-particle states at the Fermi level in the system that is not in a coherent dissipationless state. This definition, first, involves the long- and well-known minimum $g(\varepsilon )$ in the fluctuation regime of conventional superconductors for $T>T_{\rm c}$ [59], and, second, the entire region of the states of a two-dimensional superconductor. In an ideal two-dimensional superconductor in a zero magnetic field, this is the range of temperatures (11) in which, along with Cooper pairs, there coexist vortices causing dissipation. The finite temperature range exists both in the presence of disorder and in a magnetic field. This region can be represented with the aid of Fig.~15: it is located between two surfaces, from which the upper one is stretched onto the dashed curves passing through the point $T_{\rm c0}$, and the lower one is stretched onto the solid curves passing through the point $T_{\rm c}$.

Formally, these involve all the cases of existence of nonlocalized electron pairs in a dissipative medium with a suppressed macroscopic coherence. A fundamentally new possibility of the existence of a pseudogap can be due to the effect of localized pairs on the function $g(\varepsilon )$ or, speaking more carefully, the pairing effect on fractal electron wave functions.

Until recently, no experimental measurements of the $g(\varepsilon )$ function or the pseudogap in it in the vicinity of superconductor--insulator transitions were available. However, such studies have appeared recently owing to the use of low-temperature scanning tunneling microscopy. The unique potentials of this technique and at the same time the related problems are clearly seen by the example of Ref.~[180] in which TiN films were investigated.

\begin{figure}
\includegraphics{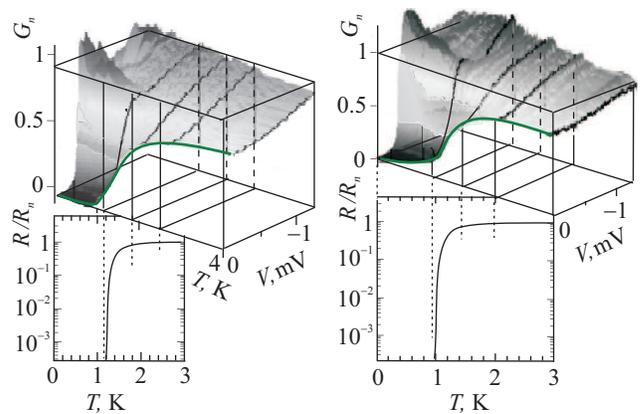}
\caption{Density of states near the Fermi level as a function of temperature for two different TiN films ($G_{\rm n}$ is the normalized differential conductivity). Transverse curves at the surfaces correspond to $G_{\rm n}(V)$ graphs at four temperatures relating as 1\,:\,1.5\,:\,2\,:\,3. For comparison, the resistive curves of the superconducting transitions are given for both films on the corresponding scales [180].}
 \label{TiN-2}
\end{figure}

The measurements were carried out utilizing two TiN films 5~nm in thickness. The resistance was measured at each temperature in parallel with the current--voltage characteristic. This made it possible to compare the evolution of the density of states $g(\varepsilon )$ with the resistive curve of the transition (Fig.~51).

The results of the comparison are as follows. At the lowest temperatures, the curve of the density of states looks the way it usually does  in superconductors: it exhibits a dip to zero in the region of $\varepsilon_{\rm F}\pm\Delta$, and two coherent peaks on the sides. With the appearance of dissipation [in the vicinity of the BKT transition (see, for comparison, Fig. 6)], the coherent peaks disappear, and the minimum in the vicinity of $\varepsilon _{\rm F}$ becomes less deep. The Cooper pairs in this region move in a gas of vortices and antivortices causing fluctuations of the order-parameter phase. The binding energy of pairs does exist, and the coherence is absent.

Then the minimum of the function $g(\varepsilon )$ becomes smeared, but it is retained even at comparatively high temperatures. The problem here lies in the fact that it is difficult to distinguish whether this minimum indicates the presence of localized pairs or is caused by superconducting interaction in the Cooper channel, i.e., by conventional superconducting fluctuations or even by the Aronov--Altshuler correction [33] to $g(\varepsilon )$ (caused by electron--electron interaction in the diffusion channel) which has no relation to superconductivity, at all. As is known, this correction increases with strengthening disorder and transforms into a Coulomb gap at the normal metal--insulator transition (see Fig.~31 and the related discussion concerning the location of a virtual metal--insulator transition in TiN).

Thus, since the superconducting transitions are broadened near the quantum superconductor--insulator transition as a result of strengthening disorder, tunneling spectroscopy made it possible to reliably observe a `conventional' pseudogap in the zero magnetic field. It can be supposed that when it is possible to combine tunneling spectroscopy with a strong magnetic field, this will help in revealing and isolating the effect from the localized pairs, as well.

\subsection{6.5 Scaling }

The basic collection of experimental data was compared with the results of scaling models for two-dimensional systems. The extent of agreement was discussed in detail in Sections 4.1 and 4.2 (for the successive stages of the comparison, see the end part of Section 3.2). To summarize the discussion, the following can be said.

No resistance $R_{\rm un}$ that is universal for all the systems exists. However, the theory, apparently, does not insist on its existence [75]. The problem can, rather, be formulated as follows: does there exist a special resistance $R_{\rm c}$ connected precisely with the quantum phase transition or is this the same resistance $R_{\rm N}$ that characterizes the boundary state (separatrix) at a high temperature? Some answers to this question come from the experiments on Be, in which $R_{\rm c}$ does exist and $R_{\rm c}\ne R_{\rm N}$, and the passage from one limit to another in the temperature dependence of resistance of the boundary state takes the form of a step (see Fig.~22). On the other hand, the separatrix in Bi everywhere has a small slope $\partial R/\partial T$, so that $R_{\rm c}\approx R_{\rm N}$ (see Fig.~18). This can be considered as a random coincidence, and the inclined separatrices in the case of Al films (Fig.~19b) or $\rm In\!-\!O$ films (Fig.~25) can be seen as a smooth passage from $R_{\rm N}$ to $R_{\rm c}$. Since an inclined separatrix does not make it possible to continue the procedure of scaling with the employment of a scaling variable, the question of $R_{\rm c}$ acquires a special importance: if the resistance $R_{\rm c}$ is not a universal quantity, it is important to understand how it depends on the properties of the corresponding quantum boundary state and whether it is possible to directly affect $R_{\rm c}$.

Formally, an inclined separatrix means that one should apply two-parametric scaling. This is especially necessary if the separatrix exhibits a tendency toward an increase in the slope up to infinity with decreasing temperature, as in the case of TiN (see Fig.~32) or high-temperature superconductors (see Figs~33 and 35). However, the schemes of two-parametric scaling still have not been applied to superconductor--insulator transitions, to the best of our knowledge.

We are grateful to S.M. Apenko, I.S. Burmistrov, A. Gold, V.V. Lebedev, Z. Ovadyahu, N. Trivedi, M.T. Feigel'man, A.M. Finkel'stein, A. Frydman, and D.V. Shovkun for numerous discussions. This work was supported in part by the grants received from the Russian Foundation for Basic Research, the RF Ministry of Education and Science, and the A von Humboldt Foundation.
potential.

\newpage


\end{document}